\documentclass[goettingen,print]{thesismp}
\usepackage{txfonts}
\usepackage{natbib}
\usepackage{fancybox}
\usepackage{multirow} 
\usepackage{mathrsfs}
\usepackage{wrapfig}
\usepackage{subfig}
\usepackage{setspace}
\usepackage{verbatim} 
\usepackage[figuresright]{rotating}

\title{Observations, analysis and interpretation with non-LTE of chromospheric structures of the Sun}
\author{Bruno S\'anchez-Andrade Nu\~no}
\town{Oviedo, Asturias, Spanien}

\refereea{Prof. Dr. F. Kneer}
\refereeb{Prof. Dr. W. Kollatschny}
\submitteddate{Januar}
\submittedyear{2008}
\examinationdate{15 Februar, 2008}
\publicationyear{2008}
\isbn{ 978-3-936586-81-7 }

\def\thesubsubsection {\thesubsection .\alph{subsubsection}}

\begin{document}
\setstretch{1}
\maketitle
\newpage
\begin{minipage}{\linewidth} 
  \vspace{5cm}
   \begin{flushright}
\emph{ For my parents Conchita and Julio, \\ my sister Deva\dots \\ }
  \vspace{.5cm}
 \emph{\dots and all those who shall learn \\ something from this work.}
\end{flushright} 
\end{minipage}
\thispagestyle{empty}

\tableofcontents

\chapter*{Summary\markboth{Summary}{Summary}}
\addcontentsline{toc}{chapter}{Summary}

This thesis is based on observations performed at the \emph{Vacuum Tower Telescope} at the \emph{Observatorio del Teide}, Tenerife, Canary Islands. We have used an infrared spectropolarimeter (Tenerife Infrared Polarimeter -- TIP) and a Fabry-Perot spectrometer (``G\"ottingen'' Fabry-Perot Interferometer -- G-FPI). Observations were obtained during several campaigns from 2004 to 2006. We have applied methods to reduce the atmospheric distortions both during the observations and afterwards in the case of the G-FPI data using image processing techniques. 

We have studied chromospheric dynamics inside the solar disc. The G-FPI provides means to obtain very high spatial, spectral and temporal resolution. We observe at several wavelengths across the H$\alpha$ line. With different post-processing techniques, we achieve spatial resolutions better than $0\farcs5$. We present results from the comparison of the different image reconstruction methods.  A time series of 55~min duration was taken from AR\,10875 at $\vartheta\approx36\degr$. From the wealth of structures we selected areas of interest to further study in detail some ongoing processes. We apply non-LTE inversion techniques to infer physical properties of a recurrent surge. We have studied the occurrence of simultaneous sympathetic mini-flares. Using temporal frequency filtering on the time series we observe waves along fibrils. We study the implications of their interpretations as wave solutions from a linear approximation of magneto-hydrodynamics. We conlude that a linear theory of wave propagation in straight magnetic flux tubes is not sufficient. 

Furthermore, emission above the solar limb is investigated. Using infrared spectroscopic measurements in the \ion{He}{i} 10830\,\AA\ multiplet
 we have studied the spicules outside solar disc. The analysis shows the variation of the  off-limb emission profiles as a function of the distance to the visible solar limb. The 
ratio between the intensities of the blue and the red components of this triplet $({\cal R}=I_{\rm blue}/I_{\rm red})$ is an observational signature of the optical thickness along the light path, which is related to the intensity of the coronal irradiation. The observable ${\cal R}$ as a function of the distance to the visible limb is given. We have compared the observational ${\cal R}$ with the intensity ratio obtained from \citet{Centeno06}, using detailed radiative transfer calculations in semi-empirical models of the solar atmosphere assuming spherical geometry
. The agreement is purely qualitative. We argue that this is a consequence of the limited extension of current models. With the observational results as constraints, future models should be extended outwards to reproduce our observations. To complete our analysis of spicules we report observational properties from high-resolution filtergrams in the H$\alpha$ spectral line taken with the G-FPI. We find that spicules can reach heights of 8 Mm above the limb. We show that spicules outside the limb continue as dark fibrils inside the disc.
\nopagebreak 

One and a half centuries after the hand-drawings by Secchi, the chromosphere is still a  source of unforeseen and exciting new discoveries.

\chapter{Introduction}

This thesis deals with the chromosphere of the Sun. To give some insight to the readers which are not familiar with the topics of this work we introduce in Section {\ref{intro:sun}} the main characteristics of the Sun with a short general description. This will elucidate the position of the chromosphere in the solar structure and its role for the outer solar atmosphere. In the subsequent Section \ref{intro:chromo}, those aspects of the chromosphere which are treated in the present work are specified. Finally Section \ref{ref:out} indicates the structure of this thesis work.

\section{The Sun\label{intro:sun}}
\begin{flushright}
\emph{It is just a ball of burning gas\\ \dots right?\\
\vspace{1cm}}
\end{flushright}

The Sun is the central object of the Solar System, which also contains planets and many other bodies such as planetoids (small planets), comets, meteoroids and dust particles. However, the Sun on its own harbors 99.8\% of the total mass of the system, so all other objects orbit around it. 

The Sun itself orbits the center of our Galaxy, the \emph{Milky Way}, with a speed of $217$ km/s. The period of revolution is $\sim230$ million years (the last time the Sun was on this part of the Galaxy was the time the Dinosaurs appeared). Compared to the population of stars in our galaxy, the Sun is a middle-aged, middle-sized, common type star. In astrophysicist's language it is of spectral type \emph{G2} and of luminosity class \emph{V}, located on the main sequence of stars in the Hertzsprung-Russell diagram. According to our understandings derived from models, it has been on the main sequence for $5\,000$ million years and it will remain there for another $5\,000$ million years before starting the giant phase.

The Sun is the closest star to us, the next one being $250\, 000$ times further away, but still light from the Sun's surface takes around 8 minutes to reach the Earth. It is the only star from where we get enough energy to study its spectrum in great detail and with short temporal cadence. With indirect methods, we can produce images of the surface structuring on other nearby starts. But on the Sun, with current telescopes and techniques, we resolve structures down to 100 km size on its surface, which represents approximately the resolution limit in this thesis work. We can also investigate the structure of its atmosphere and the effects of its magnetism. Actually, we are embedded in the solar wind that has its origin in the outer solar atmosphere, the corona of the Sun. Thus, we can make \emph{in-situ} measurements.  With special techniques and models, we can reconstruct the properties of its interior.
\pagebreak
\begin{wrapfigure}[14]{r}{0.5\textwidth}
\vspace{0.5cm}
\begin{center}
\includegraphics[width=0.5\textwidth]{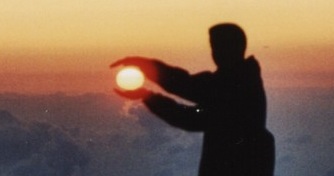}
\caption{The apparent size of the Sun on the sky is $\sim32 ' $, a little bit larger than one half degree. }
\label{fig:foto:sol}
\end{center}
\end{wrapfigure}

The Sun is the most brilliant object in the sky, 12 orders of magnitude brighter than the second brightest object, the full Moon, which actually only reflects the sunlight.   Its light warms the surface of the Earth and is used by plants to grow. Its radiation is the input for the climate. The solar wind separates us from the interstellar medium. The magnetism of the Sun protects us from cosmic high-energy radiation and it influences the climate on Earth. Violent events in the solar ultraviolet radiation and the solar wind can also disrupt radio communications.

The  Sun possesses a complex structure. Essentially, it can be described as a giant conglomerate of Hydrogen and Helium ($\sim74$\% and $\sim24$\% of the mass, respectively) and traces of many other chemical elements. Due to its big mass the self-gravitation keeps the structure as a sphere. From the weight of the outer spherical gas shells the pressure increases towards the center of the sphere. During the gravitational contraction of the pre-solar nebula towards its center, i.e. when forming a protostar, the gas has heated up by converting potential energy into thermal (kinetic) energy. This produces, together with a high gas density, a high pressure, which prevents the sphere from collapsing further inwards. Eventually, near the center, temperature and pressure are high enough to ignite nuclear reactions.
\subsubsection*{Structure}
At the core of the Sun the density and temperature (of the order of 13 million Kelvin, or $13 \cdot 10^{6}$ K) are high enough to fuse hydrogen and burn it into helium. This process also produces energy in the form of high-energy photons.  This continuous, long-lasting energy output from the nuclear reactions keeps the core of the Sun at high temperature to sustain the gravitational load from the outer gas shells. Due to the high density, the photons are continuously absorbed and re-emitted by nearby ions, and in this way the big energy output is slowly \emph{radiated} outwards,  while, towards the surface of the Sun, the density decreases exponentially, along with the temperature. Photons reaching today the Earth's surface were typically generated on the early times of \emph{Homo Sapiens}, as the typical travel time is $\sim170\,000$ years \citep{1992ApJ...401..759M}. 

At a distance from the center of approximately 70\% of the solar radius, the radiation process is not efficient enough to transport the huge amount of energy produced in the core. There the gas is heated up, and expands, it becomes buoyant and rises. This creates \emph{convection} cells in which hot material is driven up by buoyancy while cool gas sinks to the bottom of the cells, where it is heated again. These gas flows transport the energy to the outer part of the Sun, where the temperature is measured to be $\sim 5\,700\,$K and the density is low enough that the photons can escape without much further absorption. The outer region from where we receive most of the optical photons can be called the surface of the Sun, although it is not a layer in the solid state. It is called the \emph{photosphere} (sphere of light). Most of the photons we receive come from this layer are in the \emph{visible} part of the spectrum: light. This is why Nature favored in the late evolution process the development of vision instruments that are more sensitive %
in the spectral region in which most emission from the Sun occurs.
\begin{figure}[t]
\begin{center}
\includegraphics[width=\textwidth]{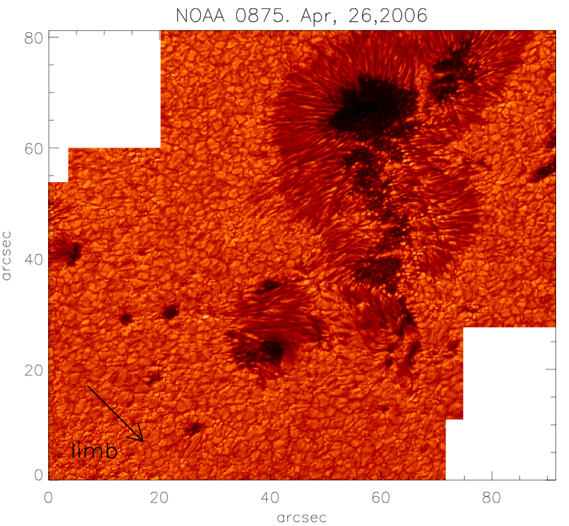}
\caption{High resolution image of the ``surface'' (photosphere) of the Sun with a resolution of $\sim140$ km. Granules are seen all around the photosphere outside the dark areas. They form the uppermost layers of the convection zone, in which the energy is transported from deep down outwards via gas motions. At the top, the gas cools down by radiating photons into space. Localized strong magnetic fields can also emerge and are seen as dark areas, the sunspots, which are a consequence of the less efficient energy transport.}
\label{fig:photosphere}
\end{center}
\end{figure}

Further out of this layer the atmosphere of the Sun extends radially, with decreasing density. In this outer part, with its low density, magnetic fields rooted inside the Sun cease to be pushed around by gas flows. This transition occurs together with a still not completely understood increase of temperature up to several million degrees. Therefore,  there must be a layer with a minimum temperature. Standard average models place it at a height of about 500 km with a temperature of about 4000 K, which is low enough to allow the formation of molecules like CO or water vapor. Beyond this layer the temperature rises outward. Again in standard models, the layer following the temperature minimum has an extent of about 1\,500 km and its temperature rises to 8\,000 -- 10\,000 K. This layer is called the \emph{chromosphere}. The present work deals with some of its properties. Outside the chromosphere, the temperature rises abruptly within the \emph{transition region}. The outermost part of the atmosphere, called \emph{corona}, drives a permanent outwards flow of particles moving along the magnetic field lines. This \emph{solar wind} extends to $100\,000$ times the solar radius, far beyond Pluto's orbit, to the outer border of our Solar System, the \emph{heliopause}. There the interaction with the interstellar medium creates a shock front, which is being measured these years by the Voyager 1 and Voyager 2 probes.

Beyond this layered structure, the Sun is far more complex. Some other properties, which we describe shortly, are: 

- The Sun vibrates. As a self gravitating compressible sphere, it vibrates. Pressure and density fluctuations mainly generated by the turbulent convection, are propagated through the Sun. Waves with frequencies and wavelengths close to those of the many normal modes of vibration of the Sun add up to a characteristic pattern of constructive 
interference. This vibration, although of low amplitude with few 100 m/s in the photosphere, can be measured and decomposed into eigenmodes by means of Doppler shifts and observations of long duration. The propagation of the waves depends on the properties of the medium. It is possible then to infer these properties from the measured vibration patterns. Some waves propagate only  close to  the surface, but others can propagate through the entire Sun. These latter waves provide means to infer some structural properties, such as temperature, of the solar interior and test models of the Sun.  
\emph{Global Helioseismology} provides means to infer the global properties of the interior of the Sun studying the vibration pattern, while \emph{local helioseismology} can depict the surroundings of the local perturbations.

- The Sun rotates. The conservation of angular momentum of a slowly rotating cloud that will form a star result, upon contraction, a rapid rotation. It is commonly accepted that most of the Sun's angular momentum was removed during the first phases of the life of  the Sun by braking via magnetic fields anchored in the surrounding interstellar medium and by a strong wind. The remaining angular momentum leads to today's solar rotation period. But being the Sun not a rigid body this rotation varies  from layer to layer and with latitude. Gas at the equator rotates at the surface with a period of 27 days, faster than at the poles where the rotation period is approximately 32 days. Using helioseismology observations we know that this differential rotation continues inside the Sun, until a certain depth, from which on the inner part rotates like a rigid sphere with a period of that at middle latitudes on the surface. This region corresponds to the layer where the convection starts, at around $0.7$ solar radii, and is called the \emph{tachocline}. The differential rotation creates meridional flows of gas directed towards the poles near the surface and towards the equator near the bottom of the convection zone.

- The Sun shows (complex) magnetic activity. The Sun possesses a very weak overall magnetic dipole field. However, the solar surface can host very strong and tremendously complicated magnetic structures, which can be seen through their effects on the solar plasma, e.g. less efficient energy transport (that leads to dark sunspots). All matter in the Sun is in the form of plasma, due to the high temperature.  The high mobility of charges that characterizes the plasma state, makes it highly conductive, causing magnetic field lines to be "frozen" into it. Provided that the gas pressure  is much higher than the magnetic pressure, the magnetic field lines follow generally the dynamics of the plasma. The source of these localized strong magnetic fields is still to be understood. The dynamo theory addresses this problem suggesting that the weak dipolar magnetic field is amplified at the bottom of the convection zone by the stochastic mass motion and shear produced by the convection and the differential rotation.

- The Sun has cycles. The Sun suffers fluctuations in time. Changes occur in the total irradiance, in solar wind and in magnetic fields. They happen in approximately regular cycles, like the 11 years sunspot cycle, and aperiodically over extended times, like the Maunder Minimum (a period of 75 years in the XVII century when sunspots were rare, and which coincided with the coldest part of the \emph{Little Ice Age}). These fluctuations modulate the structure of the Sun's atmosphere, corona and solar wind, the total irradiance, occurrence of flares and coronal mass ejections and also indirectly the flux of incoming high-energy cosmic rays. None of these variations are fully understood and their effect on the Sun itself or Earth is still under debate.  The generally accepted idea about the cyclic and more aperiodic fluctuations is that they are caused by variable magnetic fields. These are generated by dynamo mechanisms.

- The Sun evolves. The Sun is now in its main-sequence phase, where the main source of energy is the nuclear fusion of hydrogen to helium. After the initial phase of accretion of mass, a self gravitating star enters this phase, which lasts for most of its life. In the case of the Sun this phase will continue for approximately another five million years, after which the later evolution stages include a complex variation of the radius, with burning of helium as the source of energy in a later red giant phase. After this stage, the mass of the Sun is believed to be not large enough to undergo further fusion stages, and the Sun will slowly faint as a white dwarf star.

Readers can find further general information about the Sun in e.g. \cite{wikisun,Stix:2002lr} and many references therein. 

\section{The chromosphere\label{intro:chromo}}

In our short description of the Sun's structure we stated that the atmosphere of the Sun comprises a layer above the photosphere in which the temperature begins to rise again until the transition region where an abrupt increase of temperature, from approximately 10\,000~K to 1 million K, occurs. This first layer above the photosphere is called \emph{chromosphere}. The name comes from the greek of ``color sphere'', as it can be seen as a ring of vivid red color around the Sun during total solar eclipses\footnote{The apparent size of the Sun on the sky happens to be very similar to the apparent size of the Moon, leading to annular or total solar eclipses, during which the red ring can be seen.}.

The boundaries of the chromospheric layer are very rugged, resembling more cloud structures than a spheric surface. Above quiet Sun regions the chromosphere can be about 2\,000 km thick, but some structures seen in typical chromospheric lines can reach to much higher altitudes, like filaments (that can reach heights of $350\,000$ km). 

The solar chromosphere is a highly dynamic atmospheric layer. At most wavelengths in the optical range, it is transparent due to the fact that its density is low, much lower than in the photosphere below it.  Nevertheless, in strong lines like H$\alpha$ (at 6563 \AA) or \ion{Ca}{II} K and H (at 3934 \AA\, and 3969 \AA, respectively) we have strong absorption (and re-emission) which allows direct studies about its peculiar characteristic, like bright plages around sunspots, dark filaments across the disk, as well as spicules and prominences above the limb. Indeed, recent works, e.g. \cite{2003A&A...402..361T}, suggest that many of these chromospheric features could all have the same physical properties but within different scenarios. 
\begin{figure}[t]
\begin{center}
\includegraphics[width=\textwidth]{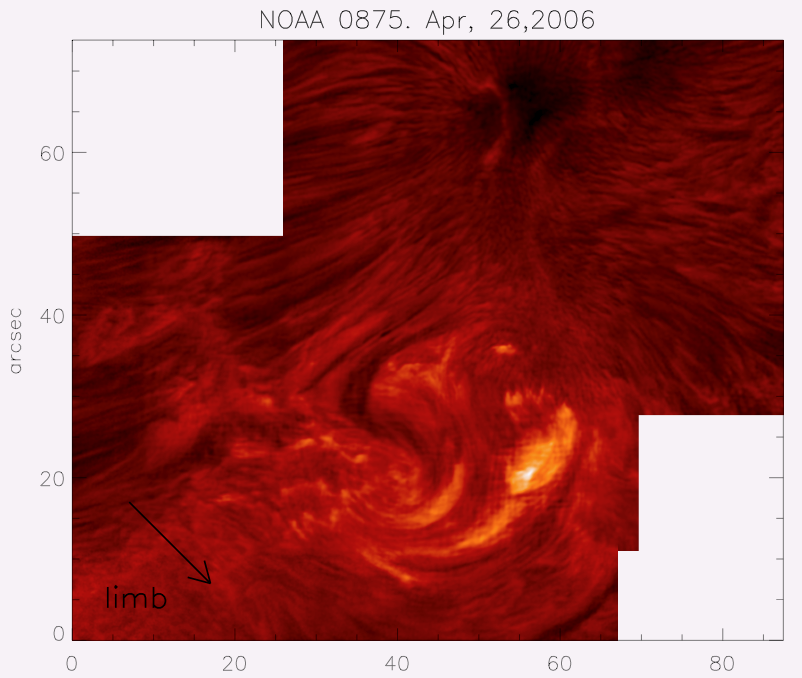}
\caption{High resolution filtergram taken in the center of the H$\alpha$ spectral line, showing the chromosphere of the Sun with an image resolution of $\sim150$ km. The same field of view as image \ref{fig:photosphere}. The localized strong magnetic fields causing sunspots in the photosphere are seen now as fibrils around the sunspots. Given the low $\beta$ parameter, the plasma is forced to follow the magnetic lines, providing visible tracers and the variety of structures seen in the chromosphere. In the image we can see a  carpet of spicules, plage region and a top view of a rising twisted magnetic flux tube above the active region. This image corresponds to the dataset ``sigmoid'' studied in Chapter \ref{chapter:hr}.}
\label{fig:chromosphere}
\end{center}
\end{figure}

The temporal evolution of the chromospheric structures is complex. The dynamics of a magnetised gas depends on  the ratio of the gas pressure  $P_\mathrm{gas}$ to the  magnetic pressure $P_\mathrm{mag}$, i.e. the plasma $\beta$ parameter, $\beta$\,=\,$P_\mathrm{gas}/P_\mathrm{mag}$, with
$P_\mathrm{mag}$\,=\,$B^2/(8\pi)$ and $B$ the magnetic field strength \footnote{It is very common in astrophysics, specially in solar physics, to use magnetic field strength synonymously with magnetic flux density. The reason is that in most astrophysical plasmas B=H in Gaussian units. We follow this use in this thesis.}. From the low chromosphere into the extended corona, this plasma parameter decreases from values $\beta>1$, where the magnetic lines follow the motion of the plasma (as in the photosphere and solar interior) to a
low-beta regime, $\beta\ll1$, where the plasma motions are magnetically driven, and the plasma follows the magnetic field lines, creating visible tracers of the magnetism. These effects give rise to a new variety of energy transport and phenomena, like magnetic reconnection, filaments standing high above the chromosphere or erupting prominences.

\section{Aim and outline of this work\label{ref:out}}
Since the discovery of the chromosphere and since the hand-drawings of \citet{secchi1877} we have been able to observe this solar atmospheric layer in much detail. Many theoretical models have been proposed to understand its peculiar characteristics. But, only in the last recent years we have been able to address the problem with fine spectropolarimetry and high spatial resolution. We can study the fine details and resolve small structures, following their dynamics in time. Within these recent advances it has been possible  both to test current theories and to observe new unexpected phenomena. This work thus aims at contributing to the understanding of the solar chromosphere.

This first Chapter provided a broad introduction to the context of this work. We have briefly presented some general properties of the Sun and the chromosphere. In the following pages, throughout Chapter 2, we summarize some theoretical concepts of radiative transfer and spectral line formation needed for this work. We also present general characteristics of the two spectral lines studied: H$\alpha$ and \ion{He}{i} 10830\, \AA. Chapter 3 presents in detail the observations. There we also summarize the characteristics of the used telescope and optical instruments, as well as the data reduction and post-processing methods applied to achieve spatial resolutions better than $0\farcs5$.
Next, in Chapter 4, we discuss results from data on the solar disc, dealing with the chromospheric dynamics and fast events observed in our data. We present the observations of magnetoacustic waves as well as other fast events.  Chapter 5 is devoted to the spicules above the solar limb. The analysis of the spectroscopic intensity profiles from spicules in the infrared spectral range can be used to compare current theoretical models with observations. Further, we present  high resolution images in H$\alpha$ of spicules. Finally, the concluding Chapter 6 of this thesis summarizes the main conclusions and gives an outlook for future work.



\chapter{Spectral lines}

Most of the information from the extraterrestrial cosmos, also from the Sun, arrives as radiation from the sky. It comes encoded in the dependence of the intensity on direction, time and wavelength. Also, the polarization state of the light contains information. These characteristics of the light we observe from any object have their origin in the interaction of atoms and photons under the local properties (temperature, density, magnetic field, radiation field itself, \dots).

To extract this encoded information from the recorded intensities it is important to understand how the radiation is created and transported in the cosmic plasmas and released into the almost empty space. 

This Chapter describes in the following sections the basis of radiative transfer and spectral line formation. We continue discussing the special properties of the spectral lines used in this work: the hydrogen Balmer-$\alpha$ line (named H$\alpha$ for short) at 6563~\AA, and the \mbox{\ion{He}{i} 10830\, \AA}\, multiplet. 

\section{Radiative transfer and spectral line formation}
Light, consisting of photons, interacts with the gas (of the solar atmosphere, in our case) via absorption and emission. Let  $I_{\lambda}(\vec{r},t,\vec{\Omega})$ be the specific intensity (irradiance) at the point $\vec{r}$ in the atmosphere, at time $t$, and into direction $\vec{\Omega}$, with   $|\vec{\Omega}|=1$. We further denote by $\kappa_{\lambda}$ and $\epsilon_{\lambda}$ as the absorption and emission coefficients, respectively. 

Along a distance $\mathrm{d}s$ in the direction $\vec{\Omega}$, the change of $I_{\lambda}$ is given by
\begin{equation}
\mathrm{d}\,I_{\lambda}= -\kappa_{\lambda}I_{\lambda} \mathrm{d}s + \epsilon_{\lambda} \mathrm{d}s\, ,
\end{equation}
or
\begin{equation}
\frac{\mathrm{d}\,I_{\lambda}}{\mathrm{d}s}= -\kappa_{\lambda}I_{\lambda}+ \epsilon_{\lambda}\, .
\label{radtran}
\end{equation}

 We define also the optical thickness between some points $1$ and $2$ in the atmosphere by
\begin{eqnarray}
\mathrm{d}\,\tau_{\lambda}= -\kappa_{\lambda} \mathrm{d}s &;& \tau_{\lambda,{1}}-\tau_{\lambda,{2}}=-\int_{2}^{1}\kappa_{\lambda}\mathrm{d}s \, ,
\end{eqnarray}
and the source function $S_{\lambda}$ of the radiation field as
\begin{equation}
S_{\lambda}= \frac{ \epsilon_{\lambda}}{\kappa_{\lambda}}\, .
\end{equation}
 
In the solar atmosphere, absorption and emission are usually effected by transitions between atomic or molecular energy levels, i.e. by bound-bound, bound-free and free-free transitions. If collisions among atoms and with electrons occur much more often than the radiative processes, the atmospheric gas attains statistical thermal properties such as Maxwellian velocity distributions and the population and ionization ratios according to the Boltzamnn and Saha formulae. These properties define locally a temperature $T$. It can be shown (e.g. \citealt{Chandrasekhar:1960lr})
that in these cases, called \emph{Local Thermodynamic Equilibrium} (LTE), the source function is given by the Planck function or black body radiation
\begin{equation}
S_{\lambda}=B_{\lambda}=\frac{2hc^{2}}{\lambda^{5}}\frac{1}{e^{hc / \lambda k T}-1}\, .
\end{equation}
$S_{\lambda}$ varies much more slowly with wavelength than the absorption/emission coefficients across a spectral line. Thus, within a spectral line, $S_{\lambda}$ can be considered independent of $\lambda$.

Generally, LTE does not hold, especially in regions with low densities (thus with only few collisions relative to radiation processes) and near the outer boundary of the atmosphere from where the radiation can escape into space. The solar chromosphere is a typical atmospheric layer where non-LTE applies. In this case, the population densities of the atomic levels for a specific transition depend on the detailed processes and routes leading to the involved levels.

Equation \ref{radtran} has the following formal solution
\begin{equation}
I(\tau_{2})=I(\tau_{1})e^{-(\tau_{1}-\tau_{2})}+ \int_{\tau_{2}}^{\tau_{1}}S(\tau')e^{-(\tau'-\tau_{2})}\,d\tau' \, ,
\end{equation}
or, for the case when $\tau_{1} \rightarrow \infty$ (optically very thick atmosphere) and $\tau_{2}=0$ \,($I(\tau_{2}=0)  \Rightarrow$ emergent intensity), then 
\begin{equation}
I_{\lambda}(\tau_{\lambda}=0)=\int_{0}^{\infty}S_{\lambda}(\tau'_{\lambda})e^{-\tau'_{\lambda}}\,d\tau_{\lambda}' \, .
\end{equation}
A second order expansion of $S(\tau_{\lambda})$ leads to the Eddington-Barbier relation
\begin{equation}
I_{\lambda}(\tau_{\lambda}=0) \approx S_{\lambda}(\tau_{\lambda}=1)\,.
\end{equation}
 This says that the observed intensity $I_{\lambda}$ at a wavelength $\lambda$ is approximately given by the source function at optical depth $\tau_{\lambda}=1$ at this same wavelength. In LTE, the intensity then follows the Planck function $B_{\lambda}(T(\tau_{\lambda}=1))$.
 
 In spectral lines, the opacity is much increased over the continuum opacity. Since the temperature decreases with height in the solar photosphere the intensity in spectral lines is decreased, and we probe higher and cooler layers. This explains the formation of absorption lines in LTE.
 
 In non-LTE, when collisional transitions between atomic levels occur seldom and near the outer atmospheric border, photons can escape and are thus lost for the build-up of a radiation field in the specific transition. Then, the upper level of the transition becomes underpopulated and the source function has decreased below the Planck function at the local temperature. It follows that, even for constant temperature atmospheres, a strong absorption line can be observed.
 
Outside the solar limb, in the visible spectral range, one observes spectral lines (and very weak continua) in emission. In spectral lines, high chromospheric structures are seen in front of a dark background.
 
\section{Hydrogen Balmer-$\alpha$ line (H$\alpha$)}

H$\alpha$ at $6563$ \AA\, is a strong absorption line in the solar spectrum for two reasons: 1) hydrogen is the most abundant element in the Sun, and in the Universe. 2) The Sun, as a G2 $\mathrm{V}$ star, has the appropriate effective temperature $T_{eff}\approx5\,800$ K to have the second level of hydrogen populated and thus to make absorptions in H$\alpha$ possible. 

As all strong lines, H$\alpha$ possesses a so-called Doppler core and damping wings. The Doppler core of H$\alpha$ and of other Balmer lines is much broader than of other strong lines from metals (atomic species with $Z>2$). The reason is the large thermal velocity of hydrogen compared to that of metals, thus leading to large Doppler widths
\begin{equation}
\Delta\lambda_{D}=\frac{\lambda_{0}}{c}\sqrt{\frac{2\,\mathcal{R}\,T}{\mu}}\, ,
\label{dopwi}
\end{equation}
where $\lambda_{0}$ is the rest wavelength, $c$ the speed of light, $\mathcal{R}$
 the universal gas constant, $T$ the temperature and $\mu$ the atomic weight (H has the minimum value among the chemical elements of $\mu = 1.008$). An eventual ``microturbulent'' broadening has been omitted in Eq. {\ref{dopwi}}.
 
 Another property of the Balmer transitions between the according hydrogen levels is the following: Chromospheric lines such as the \ion{Ca}{ii} H and K and the \ion{Mg}{i} h and k lines are weakly coupled to the local temperature through collisional transitions, effected by electrons, between the involved energy levels. Thus, these lines still contain information about the temperature of the electrons, although only in a ``hidden'' manner.
 
 However, for the Balmer lines of hydrogen and here especially for H$\alpha$, there exist also the routes for level populations through radiative ionization to the continuum and radiative recombination. These routes are taken much more often than the collisional transitions between the involved levels. The ionizing radiation fields, i.e. the Balmer and Paschen continua, originate in the lower to middle photosphere and are fairly constant, irrespective of the chromospheric dynamics. Only when many high-energy electrons, as during a flare, are injected into the chromosphere the H$\alpha$ line reacts to temperature and gets eventually into emission.
 
 Nonetheless, the chromosphere observed in H$\alpha$ exhibits rich structuring, due to absorption by gas ejecta, due to Doppler shifts of the H$\alpha$ profile in fast gas flows along magnetic fields, and due to channeling of photons around absorbing features.

\section{\ion{He}{i} 10830 \AA\, multiplet {\label{sec:limb:he}}}
Helium is the second most abundant element in the Universe, also in the Sun. It was first discovered in the Sun in 1868 (from where it was named after the greek word of Sun). 
At the typical chromospheric temperatures there is not enough energy to excite electrons to populate the upper levels from where these transitions occur. In coronal holes  the helium lines are substantially weaker compared with the quiet Sun outside the limb. More information about recent advances in measuring chromospheric magnetic fields in the He I 10830 \AA\,  line can be found in {\cite{2007AdSpR..39.1734L}}.

The energy levels that take part in the transitions of the \ion{He}{i} 10830 \AA\, multiplet are basically populated via an ionization-recombination process \citep{1994isp..book...35A}. The much hotter corona irradiates at high energies both outwards to space and inwards, to the chromosphere. The EUV coronal irradiation (CI)  at 
wavelengths lambda $\lambda<504$~\AA\ ionizes the neutral helium, and subsequent recombinations of singly ionized helium with free electrons lead to an overpopulation of the upper levels of the \ion{He}{i} 10830 multiplet.

Alternative theories suggest other mechanisms that may also contribute to the formation of the helium lines via the collisional excitation of the electrons in regions with higher temperature \citep[e.g.][]{1997ApJ...489..375A}.

\begin{wrapfigure}[18]{r}{0.5\textwidth}
\vspace{-1.1cm}
\begin{center}
\includegraphics[width=0.5\textwidth]{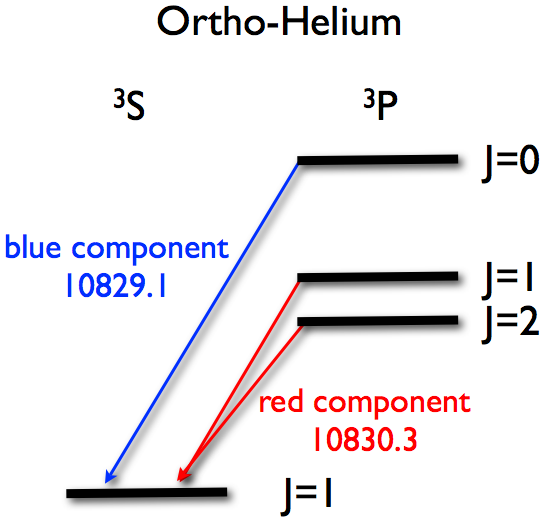}
\caption{Schematic Grotrian diagram for the \ion{He}{i} 10830~\AA\ multiplet emission lines.}
\label{fig:he:levels}
\end{center}
\end{wrapfigure}

The \ion{He}{i} 10830~\AA\ multiplet consists of the three transitions of the orthohelium (total spin of the electrons $S$=1) energy levels, from the upper term with angular momentum $L=1$ to the lower with $L=0$, in particular from  $^3$P$_{2,1,0}$, which has three sublevels ($J=2,1,0$),  to the lower metastable term $ ^3$S$_{1}$, which has one single level ($J=1$) (see Fig. \ref{fig:he:levels}). The two transitions from the J=2 and J=1 upper levels appear mutually blended, i.e. as merely one line, at typical chromospheric temperatures, and form the so-called red component, at 10830.3~\AA. The two red transitions are only 0.09~\AA\ apart.  The blue component, at 10829.1~\AA,  corresponds to the transition from the upper level with J=0  to the lower level with J=1.

The formation height of these lines is believed to be between 1\,500 and 2\,000 km,  (e.g. \citealt{Centeno06}) although, as we already mentioned, the chromosphere is strongly rugged. The Land\'e factors of the lines are not zero, meaning that they are sensitive to external magnetic fields.

A more detailed description about the properties of the \ion{He}{i} 10830 multiplet, in particular related to the emission profiles 
observed in  spicules above the limb is given in Chapter~\ref{ch:spicules}.

\chapter{Observations}

For the present work we used data from two different instruments, both mounted on the same telescope, the \emph{ Vacuum Tower Telescope} (VTT, Sec. \ref{sec:telescope}) in Tenerife. One of the instruments, the \emph{G\"ottingen Fabry-Perot Interferometer} (G-FPI, Sec. \ref{obs:fpi}) is able to achieve very high spatial resolution while the other, the \emph{Tenerife Infrared Polarimeter} (TIP, Sec. \ref{obs:tip}), is able to obtain full Stokes spectropolarimetric data with very high spectral resolution. Both instruments, in combination with the \emph{Kiepenheuer Adaptive Optics System} (KAOS, Sec. \ref{obs:kaos}), provided the data for this work.

In this Chapter we will describe the telescope, the instrumentation, the observations, and the reduction techniques. The latter are aimed at removing as many instrumental effects as possible. 

\section{Angular resolution and \emph{Seeing}\label{seeing}}
When using any kind of an optical imaging system, the angular resolution in the focal plane is limited by diffraction at the aperture of the instrument. For circular apertures the image of a point source (the PSF) is an Airy function with a certain Full Width at Half Maximum (FWHM). Two point sources closer than the FWHM of a certain instrumental PSF  are difficult to distinguish. If one considers diffraction of a telescope with a circular aperture of diameter $D$, the angular resolution limit is, in the usual Rayleigh definition,
\begin{equation}
\alpha_{min}=1.22 \, \frac{ \lambda}{D} \, .
\label{ec:res}
\end{equation}
The factor 1.22 is approximately the first zero divided by $\pi$ of the Bessel function involved in the Airy function. In the focal plane of such a telescope with a focal length $f$ the spatial resolution is therefore $d=1,22 \, \lambda\, f/D$. For good sampling this should correspond to, or even be larger than, the resolution element of the detector (2 pixels). In the case of the VTT, with a main mirror of $D=70$ cm, the diffraction limited resolution is $0\farcs24$ at $6563$ \AA\,(H$\alpha$) and $0\farcs39$ at $10830$ \AA\, (\ion{He}{i} triplet). In solar observation, it is common to use as the diffraction limit simply $\alpha_{min}=\lambda/D$. At this angular distance the modulation transfer function (MTF) has become zero.

Unfortunately all imaging systems on the ground are subject to aberrations that degrade the image quality, resulting in a much lower spatial resolution than the diffraction limit. The light we observe from the Sun travels unperturbed along approximately 150 million km, but during the last few microseconds before detection it becomes distorted due to its interaction with the Earth's atmosphere and our optical instrument.

The refraction index of the air is very close to 1 at optical wavelengths, but depends on the local pressure and temperature. Their fluctuations in space and time produce aberrations of the wavefronts from the object to be observed\footnote{The local values of the temperature and pressure depend on the complicated turbulent dynamics of the atmosphere. This includes friction and heating of the Earth's irregular surfaces, condensations and formation of clouds, shears produced by strong winds, \dots For more information we refer to e.g.  \citealt{2002RvMP...74..551S}.}. Since the time scale of the variation of the aberrations of $\approx 10$ ms is usually smaller than the integration time, it also produces smoothing of the image details. Thus, the information at small scales is lost.

Further, the turbulent state of the air masses through which the light is passing varies on small angular scales. This produces an anisoplanatism of the wavefronts arriving at the telescope, with angular sizes of the isoplanatic patches not larger than approximately $ 10\arcsec$.

Beside the atmospheric factors, the final quality of the image is influenced by local factors like the aerodynamical shape of the telescope building or convection around the building and the dome. 

Finally, the internal \emph{seeing} of the telescope plays an important role for the image quality. Convection along the light path in the telescope triggered by heated optical surfaces can be avoided by allowing air flowing freely through the structure or, quite the contrary, by evacuating the telescope.

In solar physics we usually measure the average image quality of the observations estimating the diameter of a telescope that would produce, from a point source, an image with the same diffraction-limited FWHM as the atmospheric turbulence or internal \emph{seeing} would allow even with a much larger telescope aperture. This is called the Fried parameter ($r_{0}$).  Typically, upper limits for the ``Observatorio del Teide'' are $r_{0} \approx 15$\, cm during night-time  and $r_{0} \approx 7$\, cm during day.

Besides these structural requirements for best \emph{seeing} conditions, there are nowadays methods for correcting the images for seeing distortions to obtain near diffraction limited resolution. In this thesis we have used various methods: We correct partially the aberrations in real time using adaptive optics (Sec. \ref{obs:kaos}) which can increase the $r_{0}$ around the center of the field of view up to $r_{0}\sim 25$\,cm and we also apply post-processing methods of image reconstruction (Sec. \ref{datared}) to approach the upper limit of $r_{0} \lesssim 70$ cm.

\newpage
\section{Telescope\label{sec:telescope}}
\begin{wrapfigure}[19]{r}{0.4\textwidth}
\vspace{-0.4cm}
\begin{center}
\includegraphics[height=7cm]{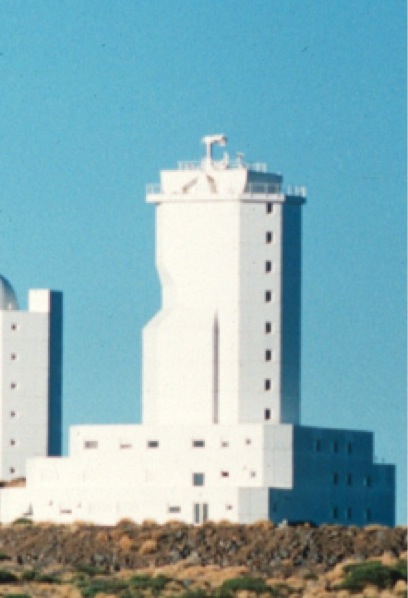}
\caption{Building which houses the solar \emph{Vacuum Tower Telescope}.}
\label{fig:foto:vtt}
\end{center}
\end{wrapfigure}

The \emph{Vacuum Tower Telescope} \citep[VTT,][ Fig. \ref{fig:foto:vtt}]{1985spit.conf.1191S} is located at the Spanish ``Observatorio del Teide'' (2400 m above sea level, 16\fdg  30' W, 28\fdg 18' N) in Tenerife, Canary Islands. It is operated by the Kiepenheuer-Institut f\"ur Sonnenphysik, Freiburg, with contributions from the Institut f\"ur Astrophysik in G\"ottingen, the Max-Planck-Institut for Sonnensystemforschung, Katlenburg-Lindau,  and the Astrophysikalisches Institut Potsdam.

\begin{figure}[t]
\begin{center}
\includegraphics[width=0.5\textwidth]{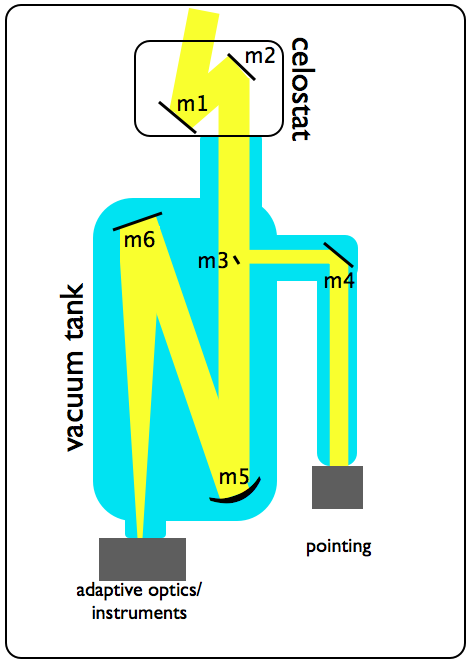}
\caption{Optical setup of the VTT. The coelostat (mirrors \emph{m1,m2}) follows the path of the Sun on the sky and directs the light to the entrance window of the vacuum tank (blue shaded). Mirror \emph{m3} takes out a small amount of the light and feeds the guiding telescope mounted outside the vacuum tank. The collimating mirror \emph{m5} produces, together with the flat mirror \emph{m6}, the solar image in the primary focal plane behind the exit window of the vacuum tank. There, a flat mirror can be mounted under $45^{\circ}$ to the vertical (not shown) to feed post-focus instruments in optical laboratories. The adaptive optics system is located below the exit window, and it is used optionally. }
\label{fig:vtt:optical}
\end{center}
\end{figure}

The VTT optical setup is depicted in Fig. \ref{fig:vtt:optical}. At the top platform of the building, a coelostat achieves to follow the path of the Sun on the sky, by means of two flat mirrors of very high optical quality. The primary coelostat mirror rotates clockwise (seen pole-on) about an axis which is contained in the mirror surface and is parallel to the Earth's rotation axis. It reflects the sunlight towards the secondary mirror. The latter redirects the beam towards the fixed telescope in the tower. The telescope is an off-axis system. It consist of a slightly aspherical main mirror of 70 cm diameter and a focal length of 46 m, and of a folding flat mirror. The free aperture of the circular entrance pupil with D=70 cm gives the telescopic diffraction limit for the  angular resolution of $\alpha_{min}=\lambda/D \approx0\farcs16$ for $\lambda$ in the visible spectral range. 

To avoid turbulent air flows inside the telescope caused by heated surfaces, the telescope is mounted in a tank that is evacuated to 1 mbar. The vacuum tank has high quality transparent entrance and exit windows located below the coelostat and close to the primary focus, respectively.

Shortly after the entrance window, a small part of the sunlight is reflected out to a second imaging device. This uses a quadrant cell to track the image of the solar disc and to correct slow image motions, e.g. due to a non-perfect hour drive of the coelostat. Telescope pointing to a target inside and near the solar disc is achieved by moving this tracking device as a whole in the image plane. The imbalanced illumination of the quadrant cell is transformed to a tip-tilt motion of the secondary coelostat mirror.

After the main vacuum tank, the adaptive optics (Sec.\ref{obs:kaos}) device is located. This optical system is able to correct in real time the low order aberrations of the incoming wavefronts of the light beam caused by the turbulence in the Earth's atmosphere. After the adaptive optics system, which can optionally be moved in or out of the path, the light path continues to the vertical slit spectrograph or to a folding mirror that can be used to direct the light to different other available science instruments.

\subsection{Kiepenheuer Adaptive Optics System\label{obs:kaos}}
As mentioned in the beginning of this Chapter (Section \ref{seeing}) the atmosphere of the Earth degrades the quality of the images during observations. KAOS \citep[Kiepenheuer Adaptive Optics System, ][]{2003SPIE.4853..187V,2007msfa.conf..107B} is a realtime correction device that calculates and corrects the instantaneous aberrations of the wavefront using special deformable mirrors.

\begin{figure}[t]
\begin{center}
\includegraphics[width=0.7\textwidth]{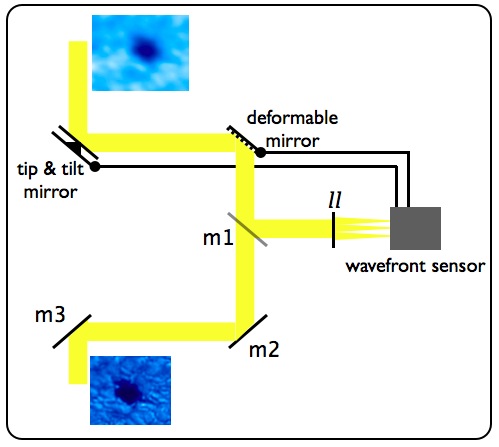}
\caption{Scheme of typical AO. Inside the closed loop, a fraction of the incoming light is directed to the KAOS camera (semitransparent mirror \emph{m1}), where a lenslet array (\emph{ll}) produces many subfield images with light from different parts of the pupil. The calculated instantaneous aberration is compensated using the two (tip\&tilt  and deformable) mirrors, every 0.4 ms.}
\label{fig:kaos:optical}
\end{center}
\end{figure}

The optical scheme of a typical adaptive optics  (AO) system is shown in Fig. \ref{fig:kaos:optical}. By means of a dichroic semitransparent beam splitter, part of the light entering the system is directed to the wavefront sensor. The latter, a Shack-Hartmann sensor, consists of a lenslet array positioned in an image of the entrance pupil and  a fast CCD detector. Each lenslet, cutting out a subaperture of the pupil image, produces an image of a small area on the Sun on a subarea of the CCD. Using a good, i.e. as sharp as possible, subimage of the present scenery on the Sun and with a correlation algorithm, it is possible to compute the displacement of each subimage and to estimate from this the aberrations of the wavefront.
Every aberration can be expressed by a sum of adequate polynomials (for example Zernike polynomials) with appropriate coefficients. Each polynomial represents a specific wavefront aberration, e.g.  tilt, defocus, astigmatismus \dots The AO is able to correct the low orders of the aberration, that is those with the largest scales. For this purpose  it has two active optical surfaces (both of them in the main lightbeam, so the correction is done in a closed loop). In the case of KAOS the first element is the tip-tilt mirror that is able to displace the whole image in two perpendicular directions, thus tracking on the reference image. The second optical element is a bymorphous deformable mirror with 35 actuators. With appropriate voltages, the surface of this mirror obtains a shape that corrects the aberrations of the incoming wavefront up to the $27^{th}$ Zernike polynomial. This correction is done in a fast closed loop at 2100 Hz. The bandwidth of KAOS is 100 Hz. It thus operates at timescales comparable to that of the variation of the turbulence in the atmosphere.

As already mentioned, the aberration of the wavefront is not constant, i.e. not isoplanatic across the whole field of view (FoV). The wavefront camera has a restricted FoV of $12\arcsec \times 12\arcsec$ where the assumption of isoplanatism is approximately valid. The center of this subfield of AO correction is called \emph{lockpoint}. The restricted area of isoplanatism is one of the main limitations of current AO systems. The corrections are calculated for the lockpoint feature we are tracking on and applied to the whole FoV of the telescope. Therefore the correction becomes increasingly inaccurate with increasing distance from the lockpoint. The quality of the image is degraded outwards from the center of the FoV, where the \emph{lockpoint} is usually located.  Fortunately this can be taken into account using {\em post factum} image reconstruction like speckle interferometry and blind deconvolution.

In night-time astronomy, AO systems lock on a star image, so the displacements of the subfields imaged by the lenslet are easily calculated. In solar observations, the image used by the AO comes always from an extended source, making the calculations of the displacements much more demanding. In solar AOs, a reference image is taken and updated regularly during operation, and correlations between this image and the subfield images are used. For well defined maxima of the correlation functions we need features with sufficient contrast inside the FoV to lock on with the algorithm, e.g. a pore or the granulation pattern. Moreover, the wavefront sensor can only work with a high light level, e.g. integrated over some wavelength. So it is not possible to lock for example on  features within the H$\alpha$ line with low intensity. Also, as we will explain in Sec. \ref{obs:tip}, near or off-limb observations are difficult as the AO algorithm is not able to track on that kind of references, as the one-dimensional limb image.

\section{High spatial resolution}
For our study of the dynamics of chromospheric structures, we are interested in observations with the highest possible spatial resolution\footnote{It has become a widespread custom in solar observations  to use  ``spatial resolution'' synonymously  with ``angular resolution''.}, with the highest achievable temporal cadence, and with as much spectral information as possible. For that purpose we used the ``G\"ottingen'' Fabry-Perot Interferometer (G-FPI). Here, the designation FPI stands as \emph{pars pro toto}, for the whole post-focus instrument, a two-dimensional spectrometer based on wavelength scanning Fabry-Perot etalons. It was developed at the Universit\"ats-Sternwarte G\"ottingen \citep{1992A&A...257..817B,1993PhDT.......243B,1995A&AS..112..371B}. Subsequently, it had undergone several upgrades \citep{2001A&A...365..588K, 2006A&A...451.1151P,Gonzalez:2007fk}. For the present work, the G-FPI with the high-efficiency performance described by \cite{2006A&A...451.1151P} was employed.

Basically, this instrument was able, at the time the data for this study were taken, to produce an image from a selected wavelength range with a narrow passband of 45 m\AA\, 
FWHM at 6563 \AA\, (H$\alpha$). A recent upgrade has reduced the FWHM. The spectrometer also can be tuned to almost any desired wavelength, being able to scan a spectral line, producing 2D filtergrams (images) at, e.g., 20 spectral position along a line. If we scan iteratively one spectral line we obtain a time sequence of very high spatial resolution, at several spectral positions and with a cadence which would be the time required to scan the full line, which is typically in the order of 20 seconds for our data.

The main limitation of this kind of observational procedure is that the images corresponding to a single scan are not obtained simultaneously, as they are taken consecutively. This is of special importance when we compare the images in the two wings of a spectral line, as the small-scale solar structure under study may have changed during the time needed to scan between these positions. This should be taken into account when studying features whose typical timescale of variation is comparable to the scanning time. In Sec. \ref{obser}, we will see that this limitation can partly be  compensated when we have a long temporal series.

\subsection{Instrument\label{obs:fpi}}
The G\"ottingen Fabry-Perot Interferometer \citep{1995A&AS..112..371B,volkmer95,2001A&A...365..588K,2006A&A...451.1151P} is a speckle-ready two-dimensional (2D) spectrometer. It is able to scan a spectral line producing a set of speckle images at several spectral position with a narrow spectral FWHM, while taking simultaneous broadband images, needed for the {\em post factum} image reconstruction.

\subsubsection*{Fabry-Perot interferometer (FPI)}

A Fabry-Perot interferometer, or etalon, is an interference filter possessing two plane-parallel  high-reflectance layers of high quality ($ \sim\lambda/100$). Light entering the filter is many times reflected between the plane-parallel reflecting surfaces. These reflections will produce  destructive interference for transmitted light at all wavelengths but the ones for which two times the spacing $d$ of the plates is very close to a multiple of the wavelength. This effect gives rise to a final Airy intensity function \citep{Born:1999lr}:
\begin{equation}
I=I_{max}\frac{1}{1+\frac{4R}{(1-R)^{2}}\sin^{2}\frac{\delta}{2}} \, ,
\end{equation}
where  the maximum intensity $I_{max}=\frac{T^{2}}{(1-R)^{2}}$ , $T$ is the transmittance, $R$ is the reflectance ($R=1-T$ if absorption is negligible), and the dependence on wavelength $\lambda$, angle of incidence $\Theta$, and refractive index $n$ of the material between the surfaces is
\begin{equation}
\label{eq:fpid}
\delta=\frac{4\pi}{\lambda}nd\,\cos\Theta \, .
\end{equation}

\begin{figure}[t]
\centering
  \subfloat{\vspace{-0.2cm}
    \includegraphics[width=0.5\textwidth]{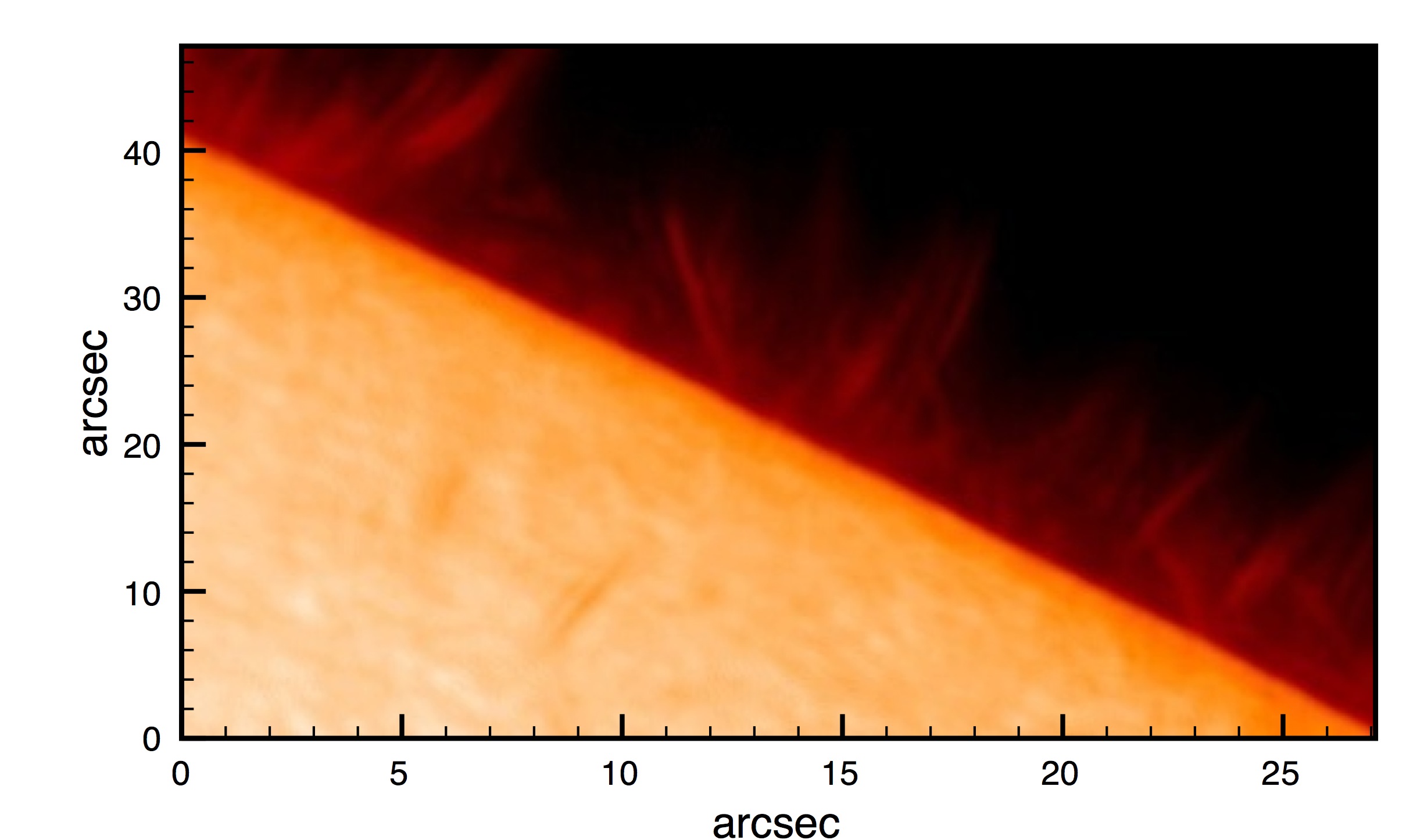}}%
  \quad%
  \subfloat{
    \includegraphics[width=0.43\textwidth]{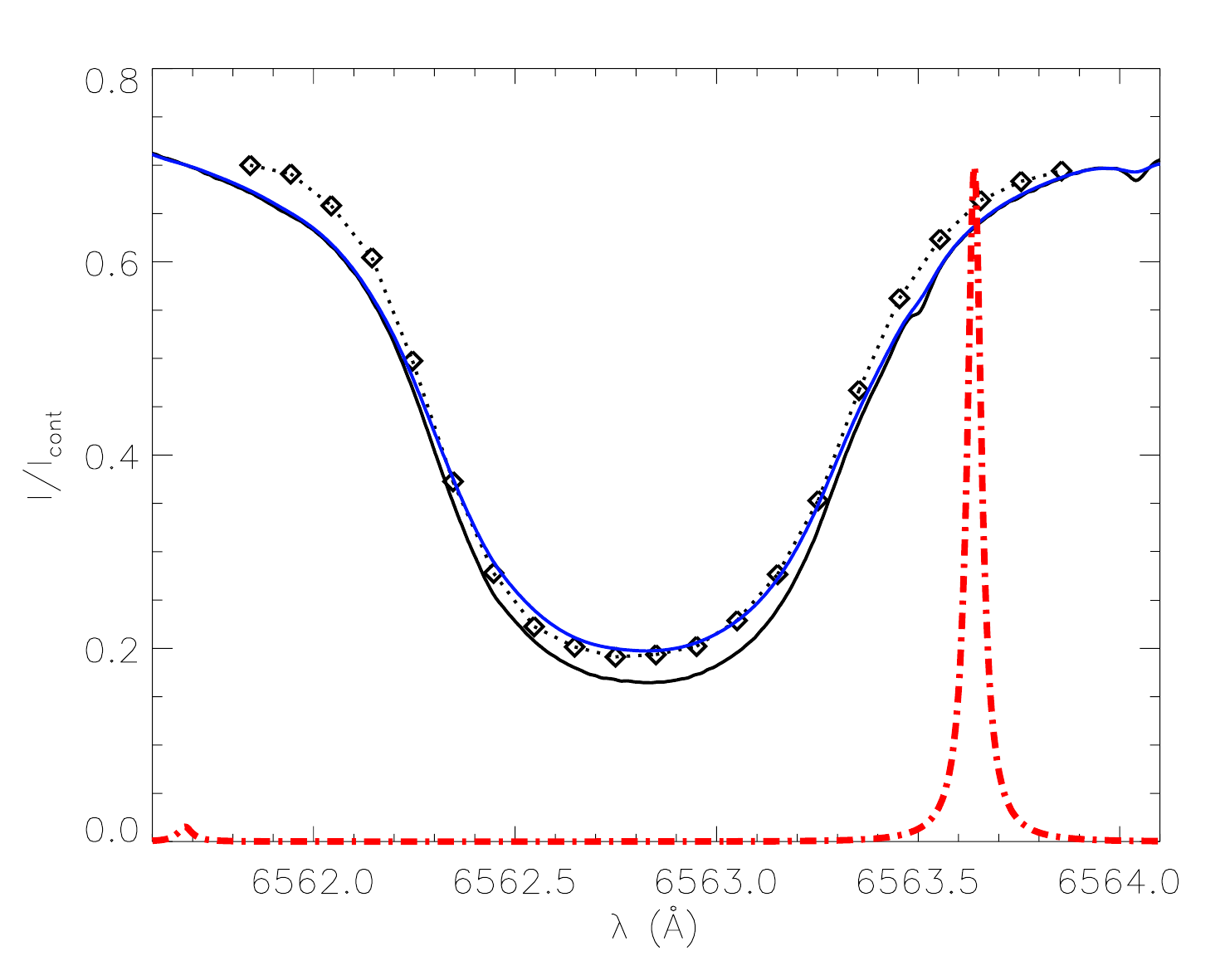}}
\caption{Example of the narrow-band scanning with the G-FPI. {\bf Left}: One narrow-band frame from a two-dimensional spectrometric scan through the hydrogen Balmer-$\alpha$ line (H$\alpha$).{ \bf Right}: H$\alpha$ line; \emph{solid black} from the Fourier Transform Spectrometer (FTS) atlas  (Brault \& Neckel, quoted by \citealt{Neckel:1999lr}); \emph{blue}: FTS profile convolved with the Airy transmission function of the FPIs; \emph{dashed} average $H\alpha$ profile observed with the spectrometer at 21 wavelength position (\emph{rhombi}) with steps of 100 m\AA. The \emph{red} line is the Airy transmission function, positioned at the wavelength in which the image in the left panel was taken, and re-normalized to fit on the plot..}
\label{fpi:scan}
\end{figure}

\begin{figure}[t]
\begin{center}
\includegraphics[width=\textwidth]{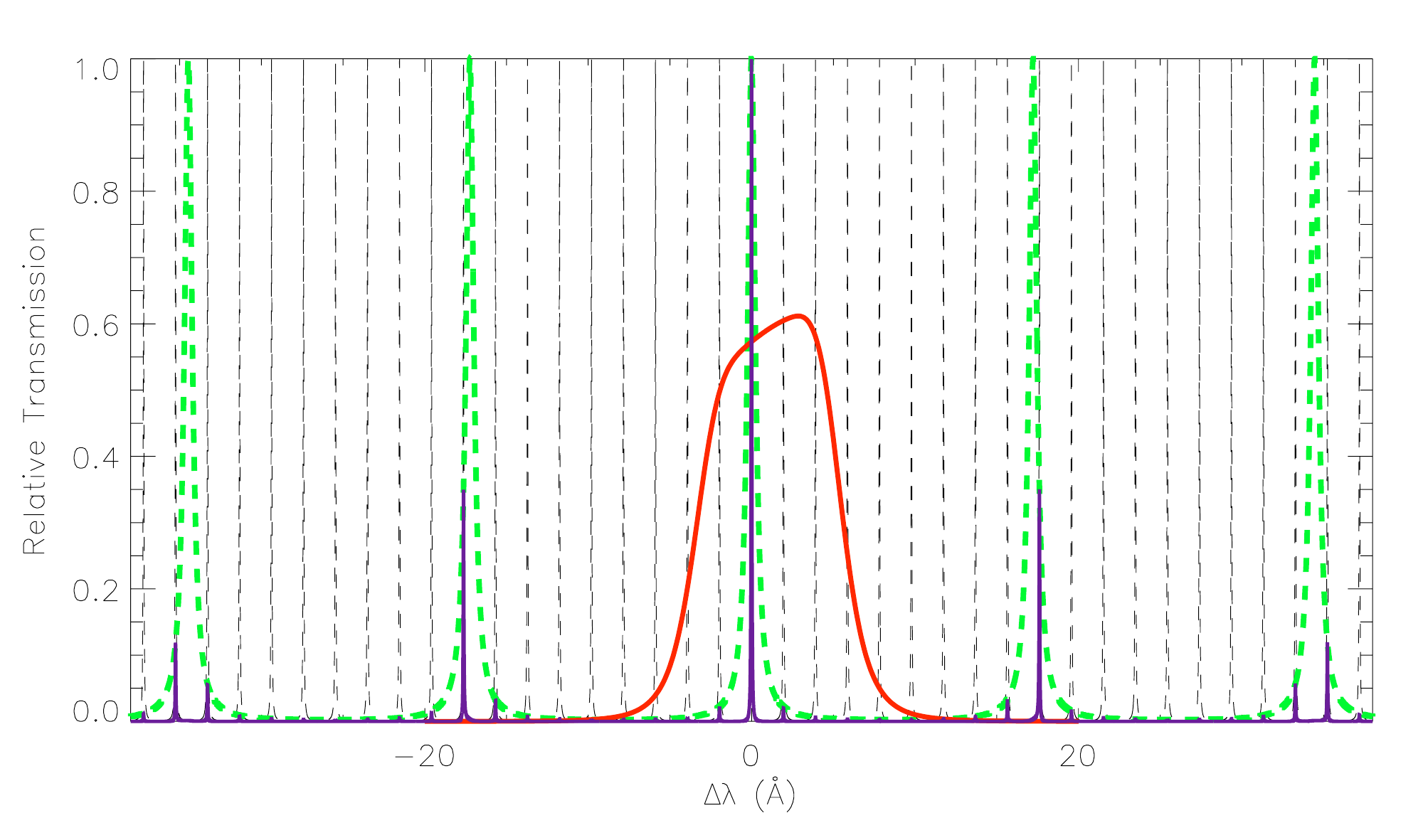}
\caption{Transmission functions for the narrow-band channel of the G-FPI with the H$\alpha$ setup. The periodic Airy function of the narrow-band FPI (dashed line) coincides in the central wavelength with that of the broadband FPI (strong dashed green line). The global transmission of both FPIs has one single strong and narrow peak at the central wavelength (purple strong line). An additional interference filter (red line) is mounted to restrict the light to the scanned spectral line.}
\label{fig:gfpi:transimance}
\end{center}
\end{figure}

The narrow transmittance of the filter can be tuned to any desired wavelength by changing the spacing $d$ (or the refractive index $n$, for pressure controlled FPIs). 
One single FPI produces a channel spectrum according to the interference condition, i.e. for normal incidence ($\Theta = 0^{\circ}$) and assuming $n$=1,
\begin{equation}
m\lambda = 2 d
\end{equation}
with $m$ being the order. From here, the distance to the next transmission peak, or \emph{free spectral range (FSR)}, follows as
\begin{equation}
\emph{FSR}=\frac{\lambda^{2}}{2d} \, .
\end{equation}
To suppress all but the desired transmission, the G-FPI has a second Fabry-Perot etalon with different spacing, i.e. different \emph{FSR}. Both Fabry-Perot etalons need to be synchronized when scanning in order to keep the desired central transmittance peaks coinciding. The combination of two FPI with different \emph{FSR} removes effectively the undesired  transmission peaks from other orders. An additional interference filter  ($FWHM \approx 8\, \AA 
$) is used to reduce the incoming spectral range to the spectral line under observation. The combination of these three  elements produces a single narrow central peak, as depicted in Fig. \ref{fig:gfpi:transimance}. 

The FP etalons are mounted close to an image of the telescope's entrance pupil in the collimated, i.e. parallel, beam. On the one hand, this avoids the ``orange peel'' pattern in the images, which one obtains with the telecentric mounting near the focus and which arises from tiny imperfections of the etalon surfaces. On the other hand, in the collimated mounting one has to deal with the fact that the wavelength position of the maximum transmission depends on the position in the FoV. This can be seen from Eq. \ref{eq:fpid} where the angle of incidence $\Theta$ changes with position in the FoV.

For the {\em post factum} image reconstruction (Sec. \ref{datared}) we have to acquire simultaneously short-exposure images from the narrow-band FPI spectrometer and broadband images. The latter are taken through a broadband interference filter ($FWHM \approx 50\, \AA $) at wavelength close to the one observed with the spectrometer. Two CCD detectors, one for each channel, with high sensitivity and high frame rates were used which allow a high cadence of short exposures. All processes (simultaneous exposures, synchronous FPI scanning and observation parameters) are controled by a central computer. The imaging on the two CCDs is aligned with special mountings and adjusted to have the same image scale on the two detectors.

\begin{figure}[t]
\begin{center}
\includegraphics[height=7cm]{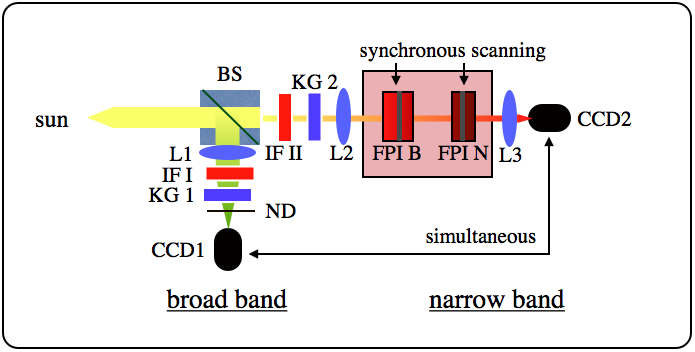}
\caption{Schema of the ``Gottingen'' Fabry-Perot interferometer optical setup. After KAOS, the light is transferred from the telescope's primary focus to the spectrometer. A beam splitter BS directs 5\% of the light into the broadband channel consisting of a focusing lens L1, a broadband interference filter IF1 ($FWHM \approx 50\,\AA$), an infrared blocking filter KG1 (``Kaltglas''), a neutral density filter ND, and the CCD1 detector. 95\% of the light enter the spectrometer through a field stop at the entrance focus. Then follow: infrared blocking filter KG2, interference (pre-) filter IFII ($FWHM \approx 6 \AA\dots10\AA$, depending on the spectral line and wavelength range), collimating lens L2, the two FPI etalons FPI B and FPI N ($FWHM \approx 45 m\AA$ at H$\alpha$), the focusing camera lens L3 and the CCD2 detector. CCD1 and CCD2 take short-exposure (3-20 ms) images strictly simultaneously.}
\label{fig:gfpi:optical}
\end{center}
\end{figure}

The optical setup is shown schematically in Fig. \ref{fig:gfpi:optical}. From the focal plane following KAOS the image from the region of interest on the Sun is transferred via a $1:1$ re-imaging system into the optical laboratory housing the FPI spectrometer. In front of the focus at the spectrometer entrance, a beam splitter directs 5\% of the light into the broadband channel. The latter contains a focusing lens, the broadband interference filter (IF1), a filter blocking the infrared light (KG1, from \emph{Kaltglas} = ``cold glass'', notation by Schott AG), a neutral density filter to reduce the broadband light level, and a detector CCD1.

Most of the light (95 \%), enters the narrow-band channel of the spectrometer through a field stop at the entrance focus. After the field stop follow: an infrared  blocking filter (KG2), the narrow interference filter (IF2), a collimating lens giving parallel light, the two Fabry-Perot etalons (FPI-B and FPI-N), a camera lens focusing the light on the detector CCD2. Figure \ref{fpi:scan} gives an example of the type of observation one can obtain with this narrow-band spectrometer.

The instrument has additional devices for calibration and adjustment: a feed of laser light, facilities to measure with a photomultiplier and to aid identifying the spectral line to be observed, and a feed of continuum light for various purposes, e.g. co-aligning the transmission maxima of the etalons or measuring the transmission curve of the pre-filter IF2.

\subsection{Observations\label{oberv}}
For the study of the chromospheric dynamics on the basis of high resolution observations we have used three  data sets. Table \ref{table:obs:HRb} lists the details for each data set:
\begin{itemize}
\item Dataset \textit{mosaic} focuses on the study of a large active solar region, where we find fast moving dark clouds, as we will discuss in Sec. \ref{hr:darkclouds}. These data were obtained before the instrument upgrading in 2005 \citep{2006A&A...451.1151P} with the old cameras. The exposure time was six times longer than with the new CCDs and the FoV of a single frame is one fourth of that of the new version of the G-FPI. The observers of these data were M\'onica S\'anchez Cuberes, Klaus Puschmann and Franz Kneer.

\item Dataset \textit{sigmoid} uses the improvements of the instrument from 2005 and was obtained during excellent seeing conditions from a very active region. During the time span of our observations at least one flare was recorded from this region in our FoV. Our focus with these data is the study of fast events and magnetoacustic waves (Sec. \ref{waves1}) with the original intention to detect Alfv\'en waves. Examples of these data were also used to compare the results from different methods of \emph{post factum} image reconstruction, as we will show in Sec. \ref{sec:comp}.

\item With dataset \textit{limb} and in Sec. \ref{sec:comp} we apply blind deconvolution methods for image reconstruction (see Sec. \ref{momfbd}). The observations were taken with the G-FPI, renewed in 2005, to study with very high spatial resolution the evolution of spicules as seen in the H$\alpha$ line.
\end{itemize}

\begin{table}[b]
\begin{center}\begin{tabular}{|r|c|c|c|}\hline
\textbf{ Data set name}  & \textit{``mosaic''} & \textit{``sigmoid''} & \textit{``limb''} \\\hline\hline
  Date			 & May,31$^{st} $,2004  &  April,26$^{th} $,2006 & May,4$^{th}$, 2005  \\\hline 
  Object			  & AR0621  & AR10875  & limb  \\\hline 
  Heliocentric angle	 & $\mu=0.68$  &$\mu=0.59$  & $\mu=0$  \\\hline 
  Scans \#			 & 5  & 157  & 5  \\\hline 
  Cadence			 & 45 s  & $\sim22$ s (see Sec. \ref{obser})  & $\sim19$ s  \\\hline 
  Time span  		 & 4 min  & 55 min  & 2 min  \\\hline 
  Line positions	\#	&  18 & 21 & 22  \\\hline
  FWHM 			& \multicolumn{3}{|c|}{50 \AA\, broadband / 45 m\AA\, narrow-band}  \\\hline   
  Broadband filter  & \multicolumn{3}{|c|}{6300 \AA}   \\\hline  
  Stepwidth		& 125 m\AA  & 100\,m\AA & 93 m\AA  \\\hline   
  Exposure time	& 30 ms  &  \multicolumn{2}{|c|}{5 ms} \\\hline   
  Seeing condition	& good  & $r_{0} \approx 32$ cm  & $r_{0} \approx 20$ cm  \\\hline   
 KAOS support & \multicolumn{3}{|c|}{yes}  \\\hline 
  Image reconstruction& speckle & AO ready speckle  &  MFMOBD  \\\hline 
   Field of view		 & 33\arcsec $\times$ 23\arcsec (total 103\arcsec $\times$94\arcsec  )  &  \multicolumn{2}{|c|}{77\arcsec $\times$ 58\arcsec}  \\\hline 
   
  \end{tabular} \caption{Characteristics of the data sets taken with the G-FPI used in this work.\label{table:obs:HRb}}
\end{center}
\end{table}

\subsection{Data reduction\label{datared}}
After the recording of the data, several processing steps have to be carried out in order to minimize the instrumental effects. These are mainly to take into account the differential sensitivity of the CCDs from one pixel to another or the fixed imperfections on the optical surfaces positioned close to one of the focal planes. This concerns for example dust on the beam splitter, on the infrared blocking filters and interference filters and the CCDs. In this step we also remove an imposed bias signal applied electronically to every frame. This is the usual treatment of any CCD data.

\begin{figure}[t]
  \centering
  \subfloat[Broad band raw frame]{
    \includegraphics[width=0.45\textwidth]{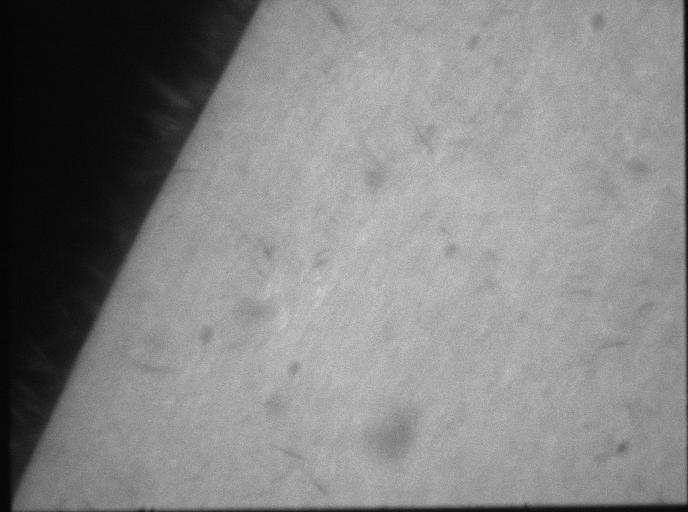}}%
  \quad%
  \subfloat[Flat field frame ]{
    \includegraphics[width=0.45\textwidth]{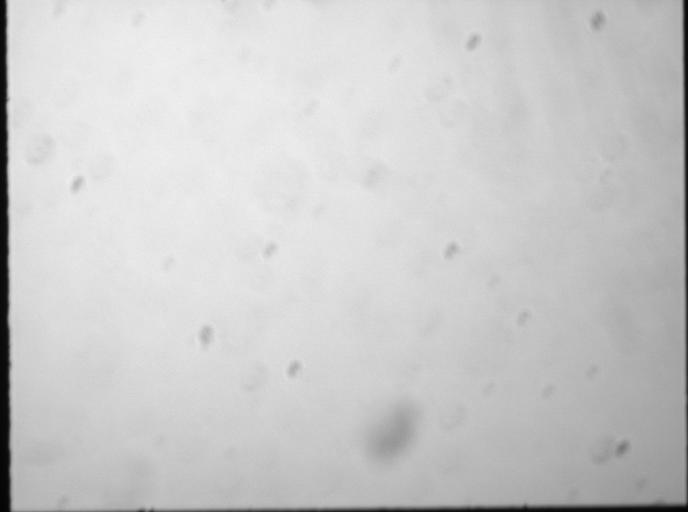}}
    \\
  \quad%
    \subfloat[Dark frame ]{
    \includegraphics[width=0.45\textwidth]{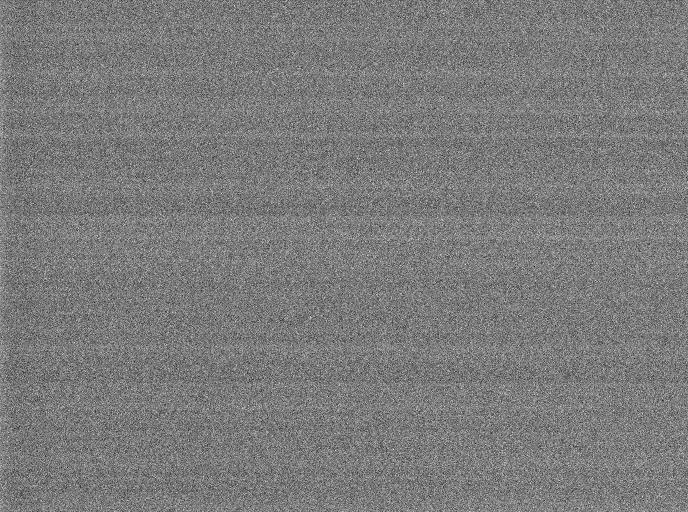}}
  \quad%
  \subfloat[Reduced frame]{
    \includegraphics[width=0.45\textwidth]{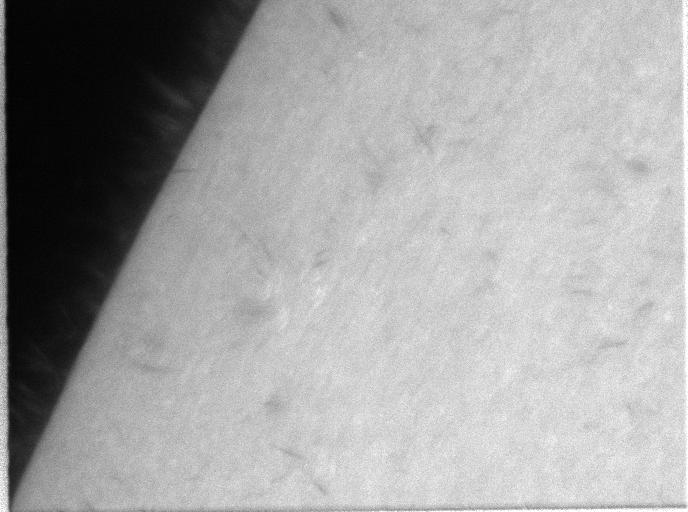}}
    \caption{Example of the standard data reduction process. Every frame taken with the CCD   (a) includes instrumental artifacts like shadows from dust particles on the CCD chips or the filters near the focus (Fig. b) and the intrinsic differential response of each pixel (c). Subtracting the dark frame and dividing by the flat response provides a clean frame (d).}
\label{fig:obs:red2}

\end{figure}

For this purpose we take flat fields, dark, continuum and target images (see Fig. \ref {fig:obs:red2}).

\emph{Target}. A target grid is located in front of the instrument, in the primary focal plane. Target frames therefore display in both channels a grid of lines that are used to focus and align the cameras in both channels. This is crucial for the image reconstruction. 

\emph{Continuum} data are taken with the same scanning parameters as with sunlight but using a continuum source, so we can test the transmission of the scanning narrow-band channel. 

\emph{Dark} frames are taken with the same integration time but blocking the incident light. These frames have information of the differential and total response of the CCD array without light, in order to remove this effect from the scientific data. 

\emph{Flat fields}  are frames with the same scanning parameters and with sunlight, but without solar structures. In this way we can see the imperfections and dust on the optical surfaces fixed on every frame taken with the instrument, and remove them dividing our science data by these flat frames. To avoid signatures from solar structures in the flat frames, the telescope pointing is driven to make a random path around the center of the solar disc far from active regions.

Thus, to reduce the instrumental effects we use the following formula, for each channel and for each spectral position independently:
\begin{equation}
 reduced frame=\frac{raw\,frame  - mean\,dark}{mean\,flatfield - mean\,dark} \, .
\end{equation}

Our instruments produce data sets that can be subject to \emph{post factum} image reconstruction. We have applied speckle and blind deconvolution methods to minimize the wavefront aberrations and to achieve spatial resolution close to the diffraction limit imposed by the aperture of the telescope. 

The aberrations are changing in time and space. In a long exposure image, the temporal dependence will produce the summation of different aberrations, blurring the small details of the image.  Therefore, for post-processing, all image reconstruction methods need input \emph{speckle} frames with integration times shorter than the typical timescale of the atmospheric turbulence. With this condition fulfilled, the images appear distorted and speckled but not blurred, and still contain the information on small-scale structures. Another common characteristic of speckle methods is the way to address the field dependence of the aberrations. In a wide FoV each part of the frame is affected by different turbulences. That is, inside the atmospheric column affecting the image, there are spatial changes of the wavefront aberration. Therefore, the FoV is divided into a set of overlapping subfields smaller than the typical angular scale of change of the aberrations (5\arcsec -- 8\arcsec), the isoplanatic patch.

Speckle interferometry denotes the interference of parts of a wavefront from different sub-apertures of a telescope. This results in a speckled image of a point source, e.g. of a  star. The effect is used for ``speckle interferometric'' techniques of postproccesing. They are able to remove the atmospheric aberrations of the wavefronts that degrade the quality of the images. In the following Sections we introduce the basic background of the methods used and provide some examples and further reference.

\subsubsection{Speckle interferometry of the broadband images\label{SIb}}
This method is based on a statistical approach to deduce the influence of the atmosphere. It was developed following the ideas of \cite{1965JOSA...55.1427F,1970A&A.....6...85L,1973JOSA...63..971K,1977OptCo..21...55W,von-der-Luehe:1984fk} . The code used for our data was developed at the  Universit\"ats-Sternwarte G\"ottingen \citep{1996A&AS..120..195D} . The \emph{sigmoid} dataset uses the latest improvements to take into account the field dependence of the correction from the AO systems \citep{2006A&A...454.1011P}.

In what follows we present a brief overview of the method:
The observed image (\emph{i}) is the convolution ($\star$) of the true object (\emph{o}) with the \emph{Point Spread function ($PSF$)}. The $PSF$ is the intensity distribution in the image plane from a point source with intensity normalized to one, i.e. 
\begin{equation}
\int\int PSF (x,y) dx dy = 1 \, ,
\end{equation}
where the integration is carried out in the image plane. The $PSF$ depends on space, time and wavelength. Its Fourier transform ($\mathscr{F}$) is the \emph{OTF, Optical Transfer Function}
\begin{equation}
\mathscr{F} ( i ) = \mathscr{F} (o \star PSF ) \hspace{0.5cm} \rightarrow \hspace{0.5cm} I=O \cdot OTF\, .
\label{ec:obs:obser}
\end{equation}
A normal long exposure image would be just the summation of N speckle images:
\begin{equation}
\sum^{N}_{i=1} I_{i} = O \cdot  \sum^{N}_{i=1} OTF_{i} \, .
\label{ec:obs:long}
\end{equation}
The $OTF_{i}$ are continuously changing in time, which leads to a loss of information. The temporal phase change of the $OTF_{i}$ will, upon this summation, reduce strongly or even cancel the complex amplitudes at high wavenumbers. \cite{1970A&A.....6...85L} proposed to use the square modulus, to avoid cancellations:  
\begin{equation}
\frac{1}{N}\sum^{N}_{i=1} |I_{i}|^2 = |O|^{2} \cdot \frac{1}{N} \sum^{N}_{i=1} |OTF_{i}|^2 =  |O|^2 \cdot STF \, .
\label{ec:obs:stf}
\end{equation}
\noindent
Yet this procedure also removes the phase information on $o$. Thus, the phases have to be retrieved afterwards. \emph{STF} is the \emph{Speckle Transfer Function}, it contains the information on the wavefront aberrations during N speckle images. To deduce this STF is therefore one of the aims of the speckle method. On the Sun, point sources do not exist. It is thus not a trivial task to determine the $STF$. There are, however, models of $STF$ for extended sources from the notion that they  depend only on the seeing conditions, through the \emph{Fried} parameter $r_{0}$ \citep{1973JOSA...63..971K}. This parameter can be calculated \emph{statistically} using the spectral ratio method \citep{von-der-Luehe:1984fk}. As this is a statistical approach, a minimum number of speckle frames must be used, more than 100.

To recover the phases of the original object the code uses the speckle masking method \citep{1977OptCo..21...55W,1983OptL....8..389W}. It recursively recovers the phases from low to high wavenumbers.

Finally a noise filter is applied, zeroing all the amplitudes at wavenumbers higher than a certain value, which depends on the quality of the data.
\begin{figure}[t]
  \centering
    \subfloat[Average of 330 speckle images (total exposure time $\sim1,6$ s). ]{
    \includegraphics[width=0.47\textwidth]{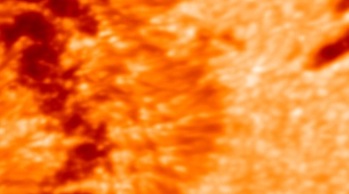}}
  \quad%
  \subfloat[Single speckle frame, 5 ms exposure time.]{
    \includegraphics[width=0.47\textwidth]{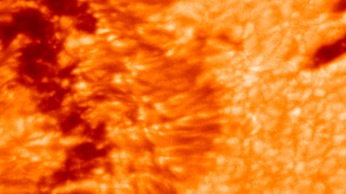}}%
    \\
  \quad%
    \subfloat[Reconstructed broadband image, using 330 speckle frames. ]{
    \includegraphics[width=\textwidth]{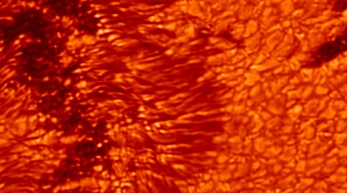}}
    \caption{Example of improvement of broadband images with the speckle reconstruction. The size of the image is $\sim$ 34\arcsec $ \times $ 19\arcsec. The achieved spatial resolution is close to the diffraction limit, $ 0\farcs22$, with the diffraction limit $\alpha_{min}=\lambda/D  \, \hat{=}\, 0\farcs19$ at $\lambda=6563$ \AA\,(H$\alpha$) and telescope aperture $D=70$~cm. }
\label{fig:obs:red}

\end{figure}

\begin{figure}[t]
\begin{center}
\includegraphics[width=\textwidth]{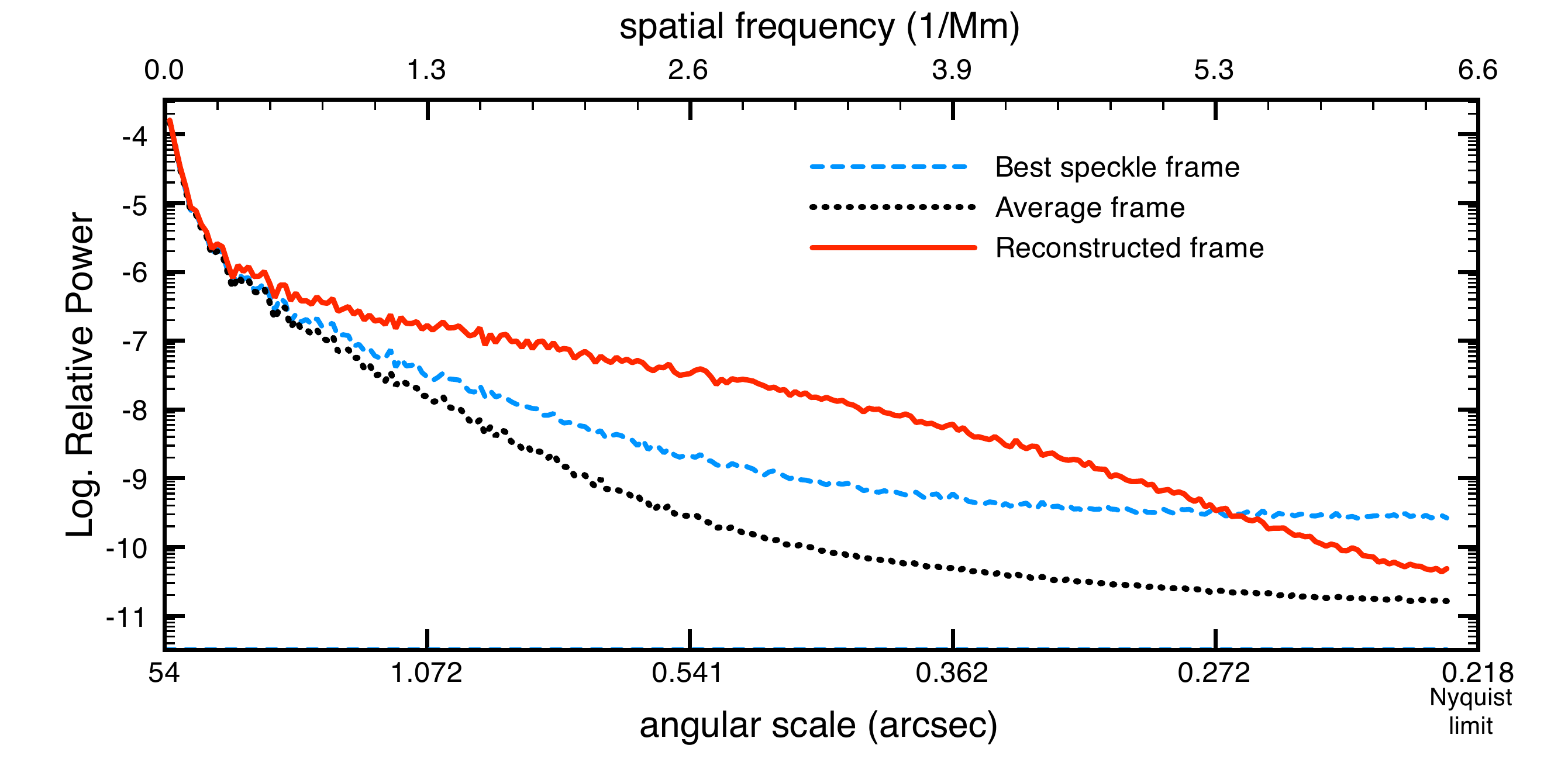}
\caption{Power spectra showing the influence of the \emph{post factum} reconstruction. Ordinate is the relative power on logarithmic scale, and abscissa is the spatial frequency, from the largest scales near the origin to the smallest scales at the Nyquist limit, corresponding to two pixels. A long exposure image (\emph{black dotted line}), taking the average of all speckle images, has very low noise, but the power is also low at all frequencies $\geqslant 0.8$ Mm$^{-1}$ (blurring effect). A single speckle frame (\emph{dashed blue line}) has more power at all frequencies, but also much more noise (more than two order of magnitude). The speckle reconstructed frame (\emph{red solid line}) keeps the noise low while it possesses higher power at all frequencies, where the spatial information on small-scale structures is stored.}
\label{fig:obs:speckle:power}
\end{center}
\end{figure}
\subsubsection*{Influence of the AO on the speckle interferometry\label{SIbao}}
As explained in Sec. \ref{obs:kaos} the AO systems provide a realtime correction of the low order aberrations (up to a certain order of Zernike polynomials). Nonetheless, given the anisoplanatism of the large field of view, the corrections are calculated for the lock point and applied to the whole frame, resulting in a degradation of the image correction from the lock point outwards. The problem arises from the different atmospheric columns traversed by the light from different parts in the FoV. This creates, after the AO correction, an annular dependence of the correction about the lock point and therefore an annular dependence of the $STF$s when processing the data. \cite{2006A&A...454.1011P} provided a modified version of the reconstruction code that computes different $STF$s for annular regions around the lock point, providing a more accurate treatment over the field of view.

The \emph{sigmoid} dataset was reduced using this last version of the code, improving substantially the quality of the results. Both AO and speckle interferometry work best with good seeing, and this data set was recorded under very good seeing conditions.

\subsubsection{Speckle reconstruction of the narrow-band images\label{SIn}}
The narrow-band channel scans the selected spectral line, taking several ($\sim 20$) images per spectral position. The statistical approach as for the broadband data can not be applied given the low number of frames per spectral position. 
To reconstruct these images from this channel we use a method proposed by \cite{1992A&A...261..321K} and implemented in the code by \cite{2003PhDT.........2J}. For each narrow-band frame, there is a frame taken simultaneously in the broadband channel, which is degraded by the same wave aberrations. The images in the broadband channel were taken at 6300 \AA, i.e. at a wavelength 260 \AA\, shorter than that of H$\alpha$. We neglect the wavelength dependence of the aberration.
For each position in the spectral line, for each subfield, we have a set of pairs of simultaneous speckle images from the narrow- and broadband channel, with a common $OTF_{i}$ for each realization in both channels:
\begin{equation}
  I_{Broad_{i}} = O_{Broad} \cdot OTF_{i}
  \label{ec:obs:narrow1}
\end{equation}
\begin{equation}
   I_{Narrow_{i}} = O_{Narrow} \cdot OTF_{i}
  \label{ec:obs:narrow1b}
\end{equation}
Using Equation \ref{ec:obs:narrow1} in  \ref{ec:obs:narrow1b}, the reconstructed narrow-band image $O_{Narrow}$ is obtained from the minimization of the error metric
\begin{equation}
E= \sum_{i=1}^{N} \Big | O_{Narrow} \cdot \frac{I_{Broad_{i}}}{O_{Broad_{i}}}-I_{Narrow_{i}} \Big |^{2} \, ,
\label{ec:obs:narrow2}
\end{equation}
where $N$ is the number of images taken at one wavelength position. Minimization of $E$ with respect to $O_{Narrow}$ yields
\begin{equation}
  O_{Narrow} = H\cdot \frac{\sum_{i=1}^{N}I_{Narrow_{i}} \cdot I_{Broad_{i}}^{*}}{\sum_{i=1}^{N}|I_{Broad_{i}}|^{2}} \cdot O_{Broad_{i}} \, .
  \label{ec:obs:narrow3}
\end{equation}
Here we have included a noise noise filter ($H$) to remove the power at spatial frequencies higher than a certain threshold above which the noise dominates. The noise power is obtained from the flat field data.

\subsubsection[Multi object multi frame blind deconvolution]{Multi object multi frame blind deconvolution (MOMFBD)\label{momfbd}}
The speckle interferometry method presented above relies on a statistically average influence of the wavefront aberration. In this Section we shortly present another approach that we have also used in this work. It is based on the simultaneous estimation of the object and the aberrations in a maximum likelihood sense using different simultaneous channels and several speckle frames. For more information see e.g. \citep{Lofdahl:2002qy,2005SoPh..228..191V,2007msfa.conf..119L}.

The method used is called \emph{Multi Object Multi Frame Blind Deconvolution} (MOMFBD), which historically is a modification of the ``Joint Phase Diverse Speckle'' image restoration. The original method is based on the possibility of separating the aberrations from the object if we observe simultaneously in two channels introducing a known aberration, like defocussing the image, in one of them. Mathematically, both phase diversity and multi-object methods are particularizations from the ``Multi Frame Blind Deconvolution''. Using a model of the optics, including its unknown pupil image, it is possible to jointly calculate the unaberrated object and the aberration, in a maximum likelihood sense.

Coming back to Eq. \ref{ec:obs:obser} for a single isoplanatic speckle subfield,  the Optical Transfer Function (OTF) is the Fourier transform of the Point Spread Function (PSF), which is the square modulus of the Fourier transform of the pupil function (P), that can be generalized with an expression like
\begin{equation}
P= A\cdot exp(i\phi) \, ,
\label{ec:momfbd:pupil}
\end{equation}
where $A$ stands for the geometrical extent of the pupil (A$=1$ inside pupil, A$=0$ outside). This unknown phase  $\phi$ can be then parametrized using a polynomial expansion:
\begin{equation}
\phi = \sum_{m\in M} \alpha_{m} \psi_{m} \, ,
\label{eq:momfdb:expan}
\end{equation}
where $\psi_{m},m \in M$, is a subset of a certain basis functions. The MOMFBD uses a combination of Zernike polynomials \citep{1976JOSA...66..207N} for tilt aberrations and Karhunen-Lo\`eve for blurring effects, as they are optimal for atmospheric blurring effects \citep{1990SPIE.1237..668R} . The $\{\alpha_{m} \}$ coefficients have therefore the information of the instantaneous wavefront aberration, whether it comes from seeing conditions, telescope aberrations or AO influence. It is interesting to note that the expansion of the phase aberration is therefore finite ($m \in M$) in our calculation, that leads to a systematic underestimation of the wings of the PSF 
\citep{2005SoPh..228..191V}

For the calculation of the solution, the MOMFBD code uses a metric quantity that depends only on the $\{ \alpha_{m} \} $ parameters and is expressed as the least square difference between the $j$ speckle data frames, $D_{j} $, and the present estimated synthesized data frame, obtained by convolving the present estimation of $PSF$  and object. 
\begin{equation}
L(\{\alpha_{m}\})= \sum_{u} \Big[ \sum_{j}^{J} |D_{j}|^2 - \frac{|\sum_{j}^{J}O^{*}_{mj}\widehat{OTF}_{mj}|^2}{\sum_{j}^{J}|\widehat{OTF}_{mj}|^{2}+\gamma}\Big]
\end{equation} 
where the $\gamma$ term accounts for the noise and corresponds to an optimum low pass filter \citep{Lofdahl:2002qy} and the $u$ index for the spatial index in the Fourier domain.

This mathematical expression, from \cite{1996ApJ...466.1087P}, to solve the blind deconvolution problem depends on the noise model used. In our case the MOMFBD assumes additive Gaussian statistics, which gives the simplest form of $L$ and the fastest code, and turns to be appropriate for low contrast objects.

The solution of the problem of image reconstruction is to find the set of $\{\alpha_{m} \}$ that minimizes the metric $L(\{\alpha_{m}\})$, providing an estimation of the OTF, and from there the new estimation of the objects. Details on the process and optimization used can be found in \cite{Lofdahl:2002qy}. The final converging solution provides thus the real object and instantaneous aberration simultaneously.

With only one channel the $\{\alpha_{m} \}$ are independent, but if we can specify linear equality constraints (LEC) to these parameters we can reduce the number of unknown coefficients for multiple channels.

The Phase Diversity method is one example of LEC. By defocussing one of the cameras on a simultaneous channel we introduce a known phase contribution in the expansion of Eq. \ref{eq:momfdb:expan}. This creates a set of related pairs of $\{\alpha_{m} \}$.  Typically, 10 or even less realizations of such pairs of images are enough for a good restoration.

Different channels observing simultaneously  in different, yet close, wavelengths can be used also to constrain the $\{\alpha_{m} \}$, as the instantaneous aberration can be considered the same for all channels. In our case we have several speckle images per position and two simultaneous channels. The broadband channel and the narrow-band channel scanning the spectral line at  21 positions with 20 frames per position. We have therefore a set of 21 pairs of 2 simultaneous objects, with 20 frames for each object and channel.

One interesting outcome of this multi object approach is that, if the observed data frames are previously aligned using a grid pattern, the resulting images are then perfectly aligned between simultaneous channels, which greatly reduces possible artifacts on derived quantities as Dopplergrams or magnetograms.

In this work we have used this MOMFBD approach to process the data where our usual speckle interferometry method was not applicable. This mainly applies for on-limb observations, as the limb darkening gradient on the field of view influences the statistics. Also, with the actual presence of the off-limb sky, the data are not suitable for the narrow-band speckle reconstruction, as we don't have a broadband counterpart for the emission features present off the limb.

The \emph{limb} data set was reduced using this code (see Sec. \ref{sec:limb:ha}), as well as some other data frames for comparison purposes with the speckle interferometry  (Sec. \ref{sec:comp}).

The MOMFBD code was implemented by \cite{2005SoPh..228..191V} and was made freely available at \verb"www.momfbd.org". Given the high processing power needed it is written and greatly optimized in \verb"C++". It is developed to run in a multithreaded clustering environment, where the work is split in workunits and sent back from the slave machines to the master once the processing is done. A typical run with one of our H$\alpha$ scans in broad and narrow-band channel, reconstructing the first 50 Karhunen-Lo\`eve modes, takes $\sim7$ hours to process with 20 CPU cores of $3.2$ GHz.

\section{Infrared spectrometry}

For this work we have also used spectroscopic data in the infrared region, to study the spicular emission in the \ion{He}{i} 10830 \AA\, multiplet. 
For this purpose we used the echelle spectrograph of the VTT and the Tenerife Infrared Polarimeter (TIP).

In this Section we summarize the instrument characteristics, the optical setup
 and the observations performed for the study of the emission profiles of spicules, which will be presented in Chapter \ref{ch:spicules}.

\subsection{Instrument\label{inst:tip}}
TIP was developed at the Instituto de Astrof\'isica de Canarias  \citep{Martinez-Pillet:1999lr} and recently upgraded with a new, larger infrared CCD detector \citep{Collados:2007fk}. It is able to record simultaneously all four Stokes components with very high spectral resolution in the infrared region from $1 \mu m $ to $2.3 \mu m$, with a fast cadence and very high spatial resolution along the slit.

\begin{figure}[t]
\begin{center}
\includegraphics[width=\textwidth]{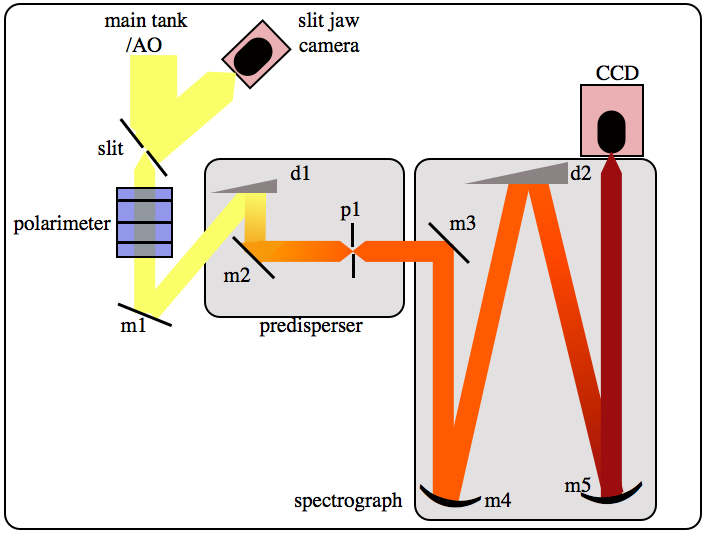}
\caption{Optical schema of the Tenerife Infrared Polarimeter (TIP) with slit jaw camera, predisperser and spectrograph of the VTT. After the AO correction, the light from the prime focus of the telescope enters the instrument through the slit. The light reflected from the slit jaws is recorded with video cameras to create context frames.  After the slit, the polarimeter with the ferroelectric liquid crystals modulates the polarization of the light beam. The predisperser selects, with mask (p1), the spectral region to observe, and the spectrograph disperses the light into its spectral components. The nitrogen-cooled CCD detector records the modulated polarization of the spectra. d1 and d2 are the diffraction gratings.}
\label{fig:tip:optical}
\end{center}
\end{figure}

The optical setup of the instrument is shown in Fig. \ref{fig:tip:optical}. After the main tank and the AO system, a narrow ($\sim100 \,\mu$m wide) slit is mounted in the plane of the prime focus of the telescope. The light reflected from the slit jaws enters a camera system to provide images, to point the telescope and to have the region of interest imaged onto the slit. The small fraction of light entering the slit goes through  the polarimeter, where the Stokes components are modulated. Then, the predisperser and spectrograph  decompose the light into its spectral components. At the end of the optical path the detector is mounted, a CCD cooled below 100 K
 to reduce the thermal excitation of electrons in the CCD pixels.
 
\subsubsection*{The polarimeter\label{polarimeter}}
TIP is able to obtain simultaneously the full set of the four Stokes parameters that determines the polarization of the light, from each point in the slit. However, this work concentrates only on the intensity measurements. The polarization measurement is performed by means of two ferroelectric liquid crystals (FLC). These are electro-optic materials with fixed optical retardation, whose axis can be switched between two orientations by applying voltages of approximately $\pm$ 10V. This amplitude of the rotation of the retardation axis is somewhat dependent on the temperature, and is $\sim 45^{\circ}$ at $20-25$C. With two FLCs, with two possible states each, we can create four different combinations of modulation of the incident light. The four modulated intensities are four different linear combinations of \{I,Q,U,V\} with different weights on each parameter. With four consecutive measurements we can therefore retrieve the four components of the Stokes vector. Thus, TIP is able to obtain simultaneously the four components of the polarization for each full cycle of the polarimeter. Although TIP makes a full cycle of the FLCs in less than one second, we have to accumulate several spectrograms in order to increase the signal to noise ratio, especially when measuring weak signals like the polarization of spicules outside the solar limb.

In the sequence following the light path, the physical setup of the polarimeter consists of a UV-blocking filter to protect the FLCs from intense high energy radiation at short wavelength. Then, the first FLC with a retardation of $\lambda/2$ and the second FLC with $\lambda/4$ follow. The retardances of $\lambda/2$ and $\lambda/4$ are nominal values. The actual retardances differ from these values and depend on wavelength. Finally a Savart plate splits the light into two orthogonal linearly polarized beams.

As part of the instruments we need a calibration optic subsystem (see explanation in Sec.  \ref{tip:reduc}) to account for the influence of the mirrors following the telescope. For this reason, in front of the AO system, there is a polarization calibration unit (PCU) that can be moved into the light path. It is composed of a retarder with nominal retardance of $\lambda/4$ in the optical spectral range, and a fixed linear polarizer. The retarder rotates a full cycle with measurements taken every 5 degrees, creating a set of 73 modulations of the light beam that are used to model the influence of the optics behind the telescope, but including AO, till the detector. The influence of the coelostat mirrors and the telescope proper on the polarization state are taken into account with a polarization model of these parts by \cite{2005A&A...443.1047B}.

\subsection{Observations\label{obs:tip}}

Table \ref{table:obs:tip} summarizes the details of the observing campaign for the course of this work. It focuses on studying the emission profiles observed in spicules in the \ion{He}{i} 10830 \AA\  multiplet. 

The strong darkening close to the solar limb and the presence of the
limb make it difficult to use KAOS for off-limb observations, since the
correlation algorithm of KAOS was not developed for this kind of observations. 

We scanned the full height of the spicule extension, starting inside the disc. We made a single spatial scan with long integration time per position. As the \emph{lock point} of the AO was placed on a nearby facula inside the disc was chosen. Apart from the facula used for AO tracking, it was a quiet Sun region. In the present work we study only the intensity component of the Stokes vector \citep[see definition in e.g.][]{Chandrasekhar:1960lr,wikistokes}.

\begin{table}[t]
\begin{center}\begin{tabular}{|r|c|c|c|}\hline
\hline
  Date			 & Dec,4$^{th} $,2005  
  \\\hline 
 Location			  & NE limb 
  \\\hline 
   Spectral sampling \#			 & 10.9 m\AA/px 
    \\\hline 
   Time span  		 & 1 scan in 66 min.  
   \\\hline 
  Slit 	&  40\arcsec $ \times $ 0\farcs5 
   \\\hline
  Integration time 			& 5$ \times $2.5 s 
  \\\hline   
  Step size  & 0\farcs35 
  \\\hline  
  Max. height off-limb		& 7\arcsec 
  \\\hline      
  Seeing condition ($r_{0}$)	& $\sim7$cm (max 12 cm) 
  \\\hline   
 KAOS support &  yes 
 \\\hline 

  \end{tabular} \caption{Characteristics of the data taken with TIP used in this work. $r_{0}$ is the Fried parameter.\label{table:obs:tip}}
\end{center}

\end{table}

\subsection{Data reduction\label{tip:reduc}}
As for the G-FPI case, the data reduction process aims to remove the instrumental effects as well as the atmospheric influence. For TIP data this involves three  steps. The first is common to all CCD observations and consists in removing instrumental effects, the second is the polarimetric calibration of the signal, and the third is the spectrosposcopic calibration. 

\subsubsection*{Reduction of CCD effects}
\begin{figure}[t]
  \centering
  \subfloat[Frame of raw data]{
    \includegraphics[width=0.45\textwidth]{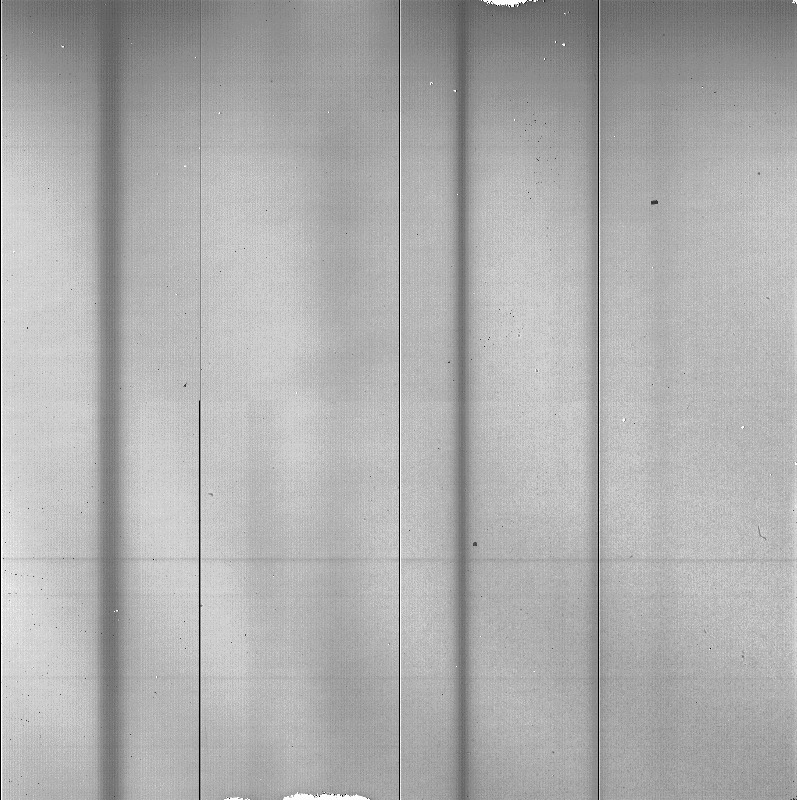}}%
  \quad%
  \subfloat[Flat field ]{
    \includegraphics[width=0.45\textwidth]{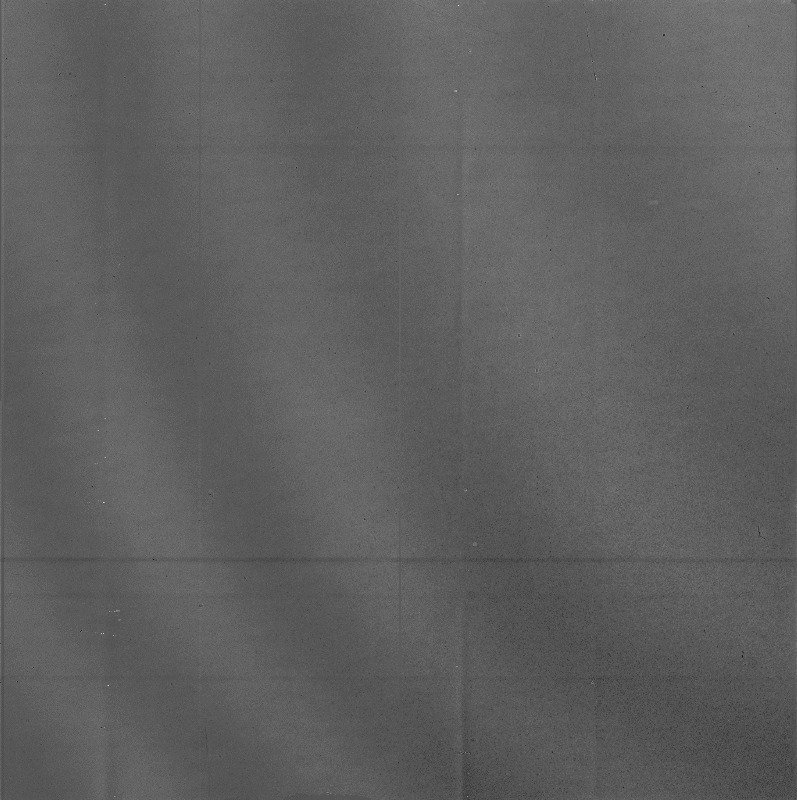}}
    \\
  \quad%
    \subfloat[Dark frame ]{
    \includegraphics[width=0.45\textwidth]{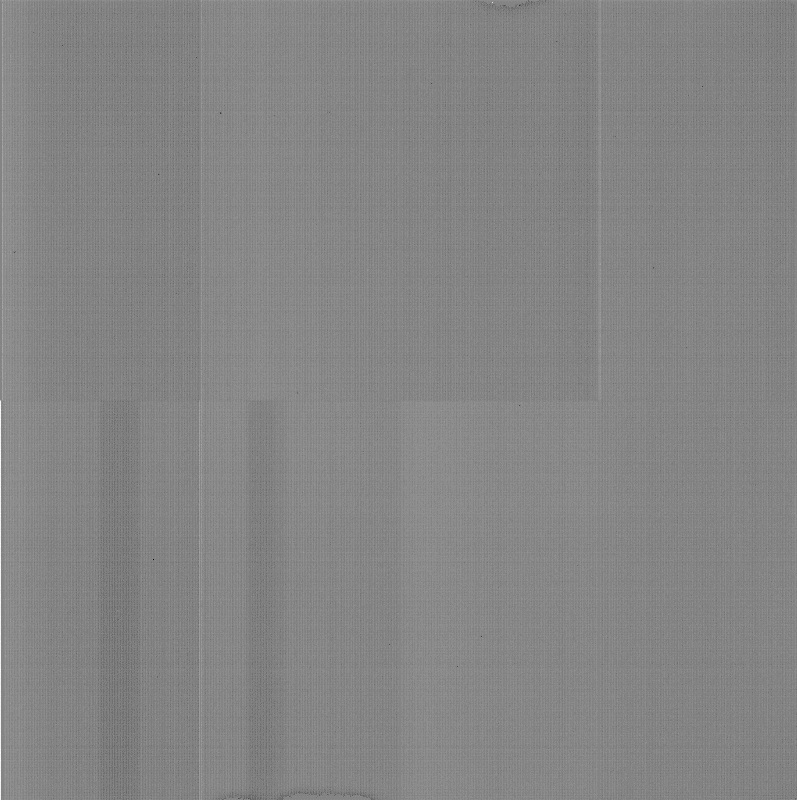}}
  \quad%
  \subfloat[Reduced frame]{
    \includegraphics[width=0.45\textwidth]{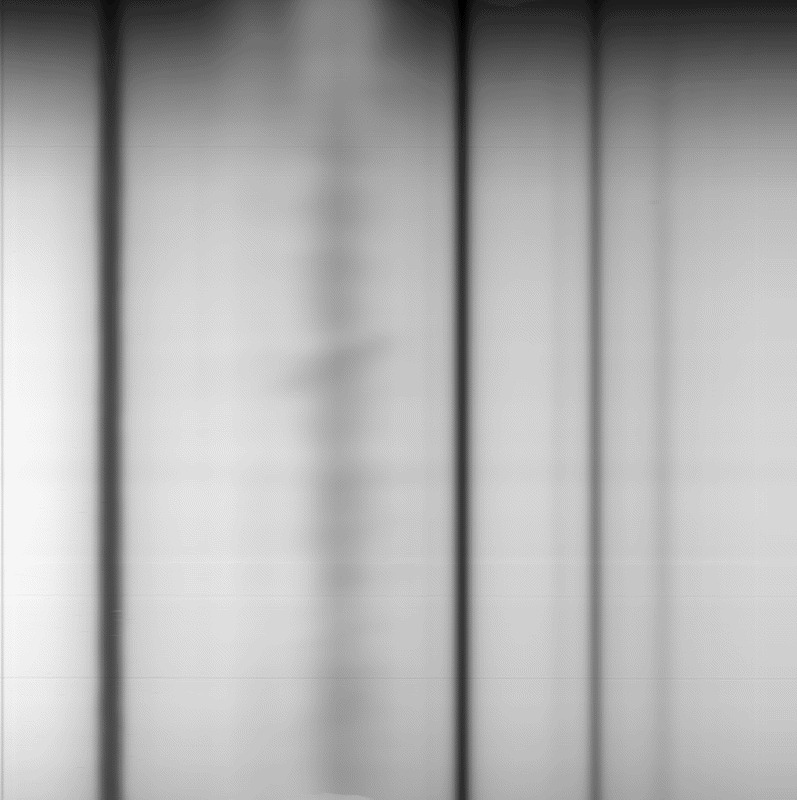}}
    \caption{Examples of the standard data reduction process for spectral data. The Flat field frame (b) is calculated dividing average flat field data by the mean spectra of the average. 
    }
\label{tip:flat}
\end{figure}

This processing is basically the same for all CCD observations: removal of dark counts and correction for differential sensitivity of the pixel matrix with the gain table (using the flat fields). The only difference to G-FPI data reduction is when creating the flat fields. The mean flat field frame is not \emph{flat}. Although being a spatial average, it still contains spectral information. To retain only the gain table information we divide the flat field by the mean spectrogram, so that only the differential response of the pixels is left (see Fig. \ref{tip:flat}). The mean spectrogram is obtained by averaging the flat field spectrograms over the spatial coordinate.

\subsubsection*{Polarimetric calibration}
The signals recorded with the CCD are not directly the Stokes parameters \cite[see description in e.g. ][]{Chandrasekhar:1960lr} . With two FLCs we have four different combinations in one full cycle. For each configuration in the cycle, we measure intensities as a particular linear combination of \{I,Q,U,V\} with different weights, so we can solve the ensuing system of equations. Also, in each CCD frame, we measure light of two orthogonal linearly polarized beams (see Sec. \ref{inst:tip}).

An important problem in polarimetric observations is that each reflecting surface of the telescope changes the polarization state of the incoming light. So the optical path, with all the reflecting surfaces from the coelostat to the CCD, introduces a complex modulation of the incoming polarization. At the VTT there is a polarization calibration unit (PCU) mounted in front of the AO system. This device feeds the subsequent optical components with light of well defined polarization states.  So, once we have a set of Stokes parameters from different configurations of the PCU, we can obtain the modulation induced by the optical path, the Mueller matrix $ \mathbb{M}$,  from the PCU to the polarimeter:
\begin{equation}
\left(\begin{array}{c}I \\Q \\U\\V\end{array}\right)_{polarimeter} = \mathbb{M} \cdot \left(\begin{array}{c}I \\Q \\U\\V\end{array}\right)_{input}
\end{equation}
The inverse  matrix of $\mathbb{M}$ will therefore relate the polarization state of the light that reaches the polarimeter with the light arriving at the PCU position. However, the light path from the coelostat to the PCU (in front of the AO) cannot be calibrated with this system, so the reduction routines use a theoretical model of this part of the telescope.

This process is already implemented with available reduction pipelines. Further investigation of \emph{crosstalk} or other additional polarimetric reduction are needed to reduce the instrumental effect in our data. However, this is not necessary for our case, since this work concentrates only on the intensity component.
\subsubsection*{Spectroscopic reduction}

The last type of reduction procedure is related to the nature of spectroscopic data and consists of the calibration in wavelength, the continuum correction and a low pass filtering to remove noise.

\begin{figure}[t]
\center
\includegraphics[width=\textwidth]{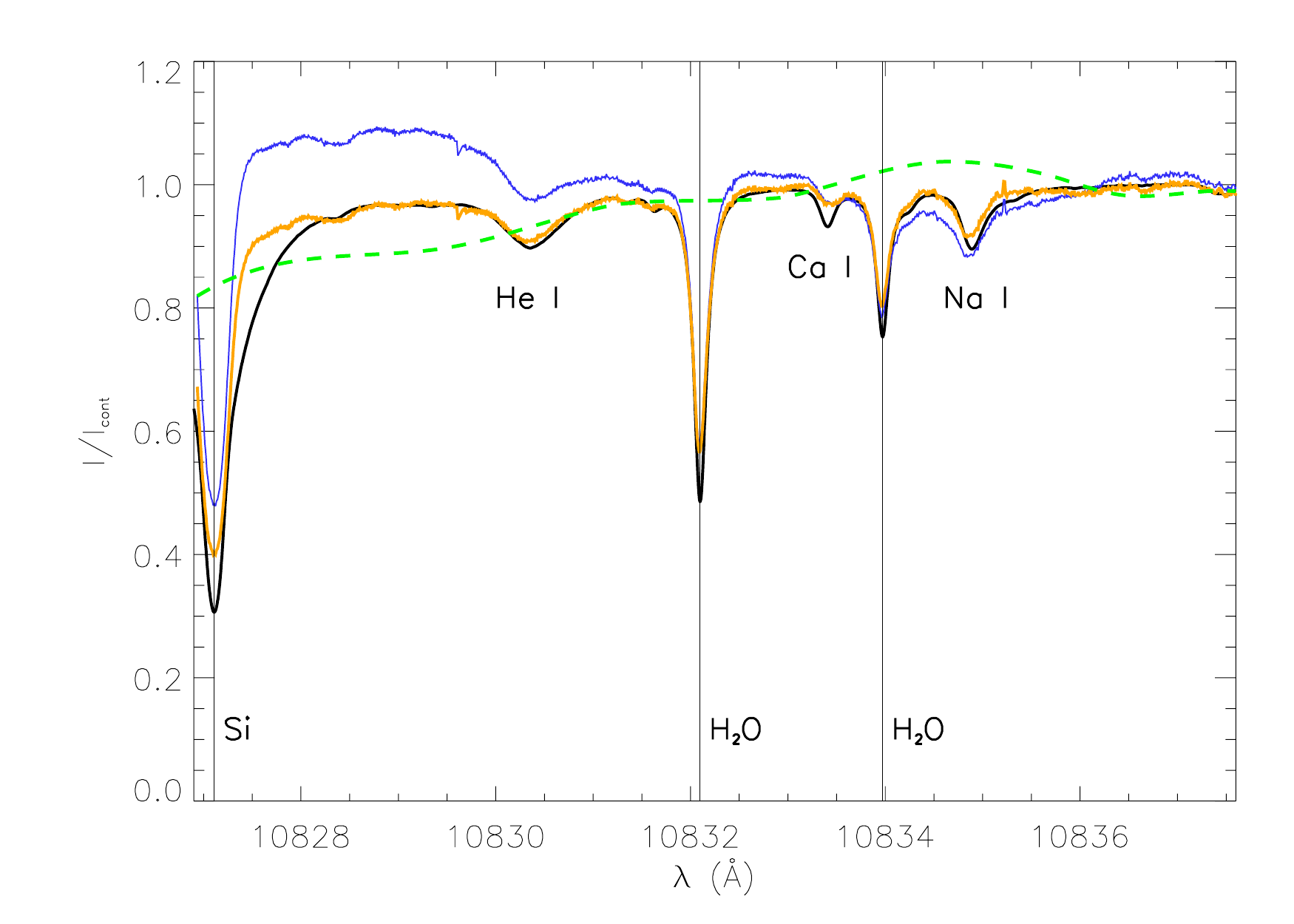}
\caption{Example of intensity calibrated spectra on the disc near the limb. Raw spectrogram (blue line) has to be corrected for the continuum level to agree with the values in the FTS atlas \citep[][ black line]{Neckel:1999lr}. Using the continuum at several positions we can estimate the continuum correction (green dashed line). The corrected data (not filtered) are shown in orange. For the wavelength calibration we use the two telluric  H$_{2}$O lines (labeled in the figure). The region of the \ion{He}{i} 10830 \AA\  multiplet is also labeled, as well as some other lines in the range (Si, \ion{Ca}{i} , \ion{Na}{i}).}
\label{fig:tip:cont}
\end{figure}

To calibrate our spectrograms in wavelength we make use of the two telluric lines present in our spectral range of the TIP data. Solar lines are subject to Doppler shifts from local flows and solar rotation. Yet, telluric absorption lines are formed in the atmosphere of the Earth. Therefore, they are always narrow due to only small Doppler broadening and are located at fixed wavelength. This provides a fixed reference coordinate that we use with the FTS atlas \citep{Neckel:1999lr}. Comparing both spectra we can accurately measure the spectral sampling which is for all data sets $10.9 $m\AA/pixel\,. See wavelength scale abscissa of Fig. \ref{fig:tip:cont}.

The transmission of the filters is not a constant in the transmitted wavelength range, so this creates an intensity variation curve in all our spectrograms. For normalization, we have to find the correct level of the continuum intensities of the spectrograms observed on the disc. For this, we use several spectral positions between spectral lines and calculate the ratio between the observed data and the values from the  FTS atlas. We interpolate to create the continuum correction (see green dashed line on Fig. \ref{fig:tip:cont}).

An electronic signal was also found in some observed spectrograms with a frequency higher than those containing information on the solar spectrogram. We used for all data a low-pass filter which removes the power at all frequencies higher than a certain threshold, preserving the spectral line information.

Once we have filtered and corrected the signal for all instrumental effects we have to remove finally the scattered light.  We define the position of the solar limb as the height of the first scanning position (counting from inside the limb outwards), where the helium line appears in emission. For increasing distances to the solar limb a decreasing amount of sunlight is added to the signal by scattering in the Earth's atmosphere and by the telescope's optical surfaces. Since the true off-limb continuum must be close to zero, i.e. below our detection limit, the observed continuum signal measures the spurious light.  Therefore, we removed the spurious continuum intensity level by using the information given by a nearby average disc spectrogram. This first subtraction estimates the continuum level on a region 6 \AA \, away from the  \ion{He}{i} 10830~\AA \, emission lines.  After this correction with a coarse estimate of the spurious light, a second correction was applied to remove the residual continuum level 
seen around the emission lines. This was needed since the transmission 
curve of the used prefilter is not flat but variable with wavelength.

\chapter[High resolution imaging of the chromosphere]{High resolution imaging of the chromosphere\label{chapter:hr}\Large{\protect\footnote{Contents from this Chapter have been partially published as \citet{2005ESASP.600E..70S,sanchez07}}}}

Since the discovery of the chromosphere 150 years ago, it has remained a lively and exciting field of research.  Especially the chromosphere of active regions exhibits a wealth of dynamic interaction of the solar plasma with magnetic fields. The literature on the solar chromosphere, and on stellar chromospheres, is numerous. We thus restrict here citations to the monographs by \citet{bray74} and \citet{athay76} and to the more recent proceedings from the conferences {\em Chromospheric and Coronal Magnetic Fields} \citep{2005ESASP.596.....I} and {\em The Physics of Chromospheric Plasmas} \citep{heinzel07}. With the latest technological advances we are able to scrutinize this atmospheric layer in great detail. The G-FPI in combination with post-processing techniques used in this work aims for the study of the temporal evolution  of the chromospheric dynamics with very high spatial, spectral and temoral resolution.

In this Chapter we present our investigations with the G-FPI inside the solar disc. The first Section focusses on data set ``mosaic'' and the presence of fast moving clouds. The subsequent Section presents the results of the investigation of fast events and waves from dataset ``sigmoid''. Finally we make a comparison between SI+AO and BD methods.

\section{Dark clouds\label{hr:darkclouds}}
As already noted in Sec. \ref{intro:chromo}, the chromosphere is highly dynamic. Within and in the vicinity of active regions the interaction of the plasma with the strong magnetic fields gives rise to specially complex phenomena with fast flows.  As an example we refer to a recent observation of fast downflows from the corona, observed in the XUV and in H$\alpha$ by \citet{2007A&A...472..633T}. Fast horizontal, apparent displacements of small bright blobs with velocities of up to 240~km\,s$^{-1}$ were observed in H$\alpha$ by \citet{2006ApJ...648L..67V}.

\subsubsection*{Observations and data processing}
\begin{figure}
\centering
\includegraphics[width=0.6\textwidth]{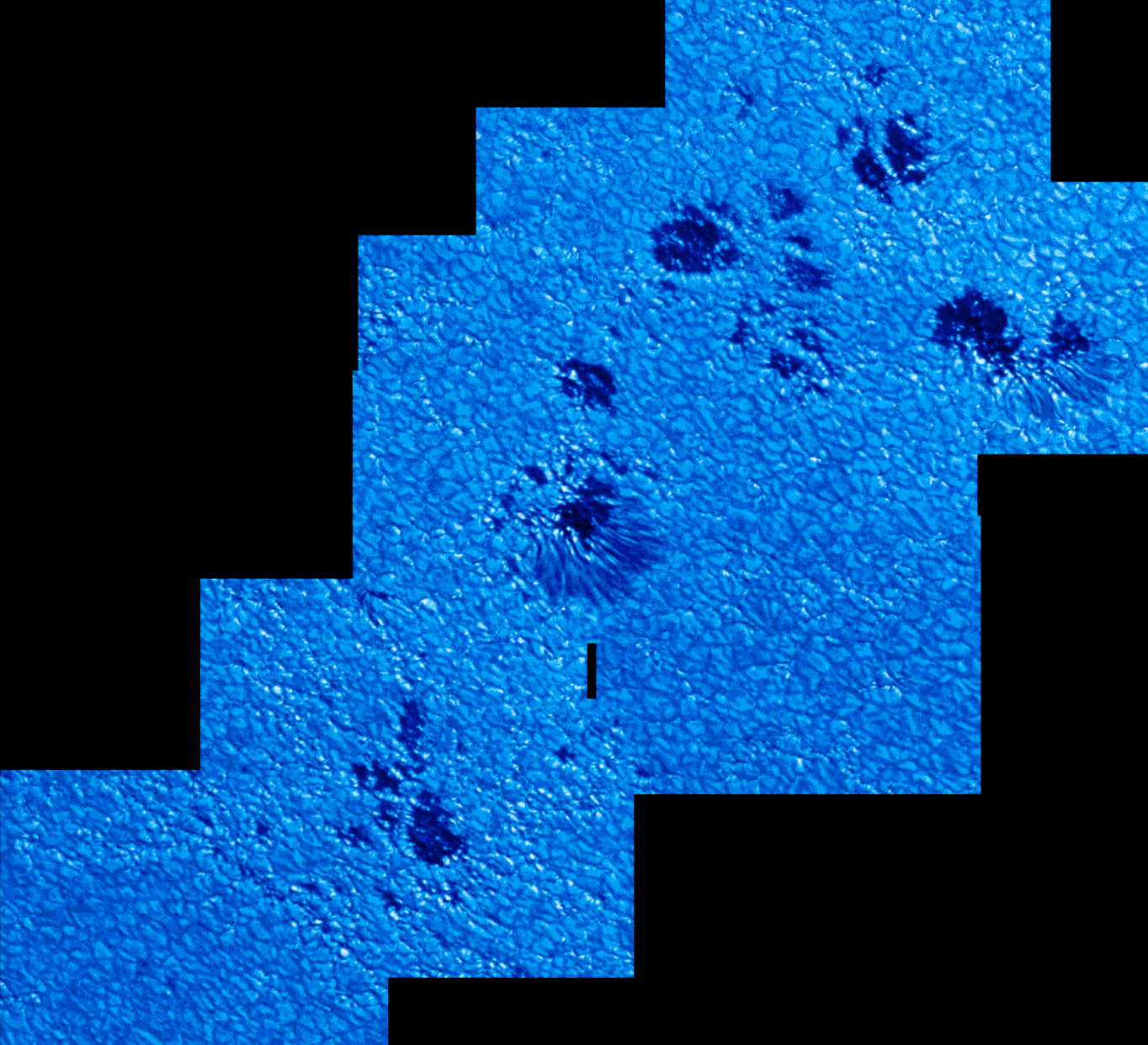}
\caption{Mosaic of speckle reconstructed broadband images of the active region NOAA AR0621, at $\mu$\,=\,0.68. The achieved high resolution by means of the adaptive optics and {\em post factum} reconstruction is $\sim0.2\arcsec$. The total area covered is $\sim$\,103\arcsec$\times$94\arcsec. Limb is located to the left lower corner.}
\label{fig:mos1}
\centering
\includegraphics[width=0.6\textwidth]{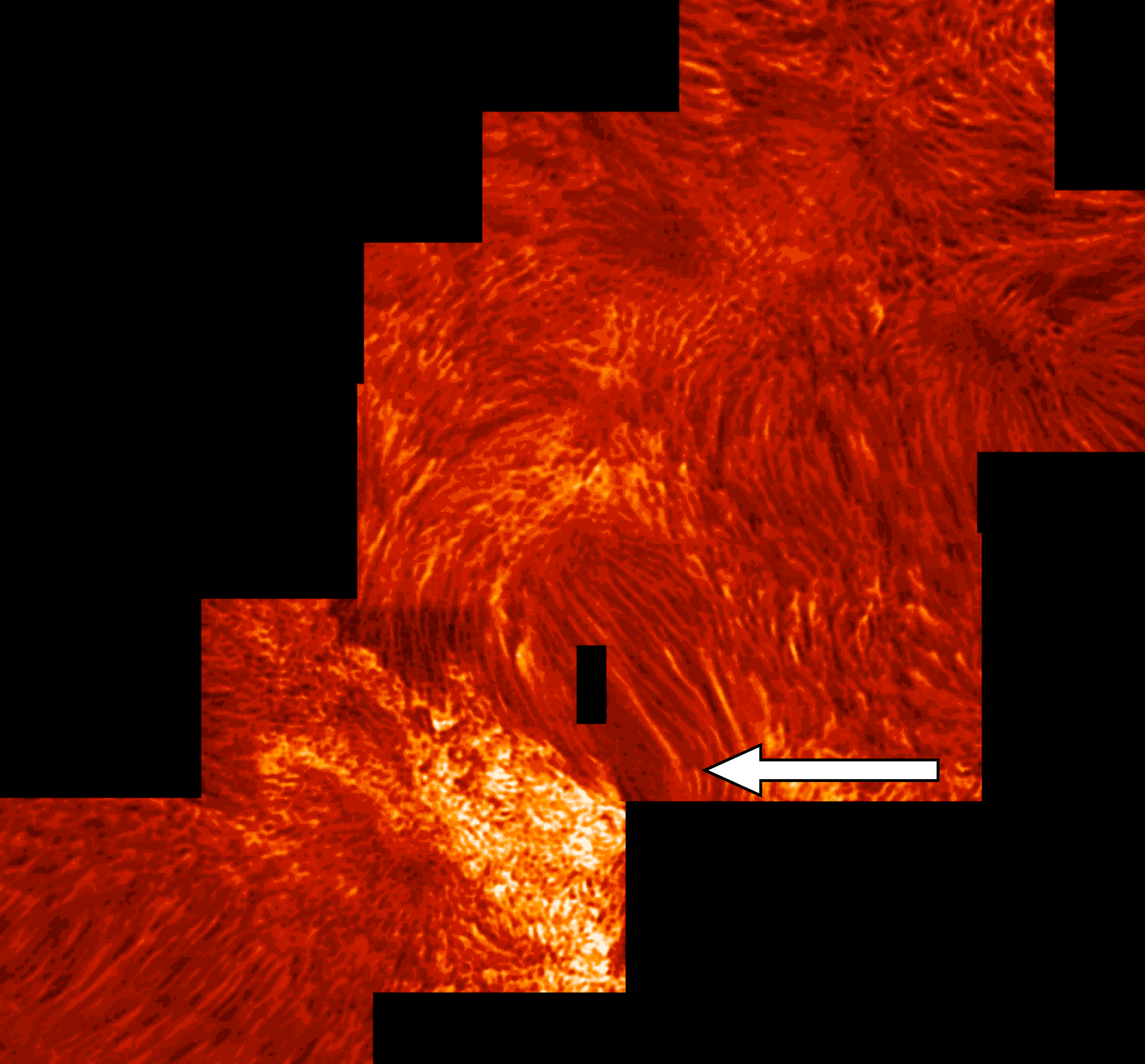}
\caption{H$\alpha$ line center filtergram. It corresponds to one of the 18 reconstructed images along the spectral line. The resolution in these narrow-band images is $<$\,0.5\arcsec. One notes the various chromospheric features: ubiquitous short fibrils with different orientation, a wide bright plage region full of facular grains on the lower central part, and dark fibrils packed together outlining the magnetic field lines between sunspots around the central data gap. White arrow indicates position and direction of the dark cloud in Fig. \ref{fig:cloud}}
\label{fig:mos2}
\end{figure}

In this Section we use the data set ``mosaic'' (See Table \ref{table:obs:HRb}) recorded on May, 31, 2004 by K. G. Puschmann, M. S\'anchez Cuberes and F. Kneer. It consists of a wide mosaiqued FoV around the active region AR0621. For each single FoV a series of five consecutive scans was performed, spanning a total of 4 min to study the temporal evolution. The FoV of a single exposure was $\sim$\,33$^{\prime \prime}\times$23\arcsec. To study a wide area the telescope was pointed consecutively to 13 overlapping contiguous areas. The resulting mosaic covers a wide region with a total FoV of $\sim$\,103\arcsec$\times$94\arcsec. In Figs. \ref{fig:mos1} and \ref{fig:mos2} we present the broadband image and narrow-band line core filtergram, respectively. In all mosaics, both in broadband and in all the narrow-band images there is a blank central area, that just corresponds to a small non-covered area. After dark subtraction and flat fielding, the data were processed using the SI approach (see Sec. \ref{datared}).

After the SI reconstruction, we have applied a destreching algorithm between the consecutive broadband images to remove residual \emph{seeing} effects. The deformation matrix for the destreching was calculated for the broadband channel using a mean image as reference. The same deformation matrix was then applied to the narrow-band spectrograms. To constrain the different frames of the mosaic of the broadband data, i.e. for connecting the individual subfields, a cross-correlation algorithm has been developed. The frames were smoothed by a boxcar of 5$\times$5 pixels to take into account only large structures for the destreching and to reduce noise. The overlapping regions between the individual subfields have been used to scale the intensities and the several areas have been connected after proper apodisation. The arrangement of the individual subfields inside the broadband mosaic have been directly applied to the narrow-band data. 

\subsubsection*{Data analysis and interpretation}
\begin{sidewaysfigure}
\centering
\includegraphics[width=0.47\textwidth]{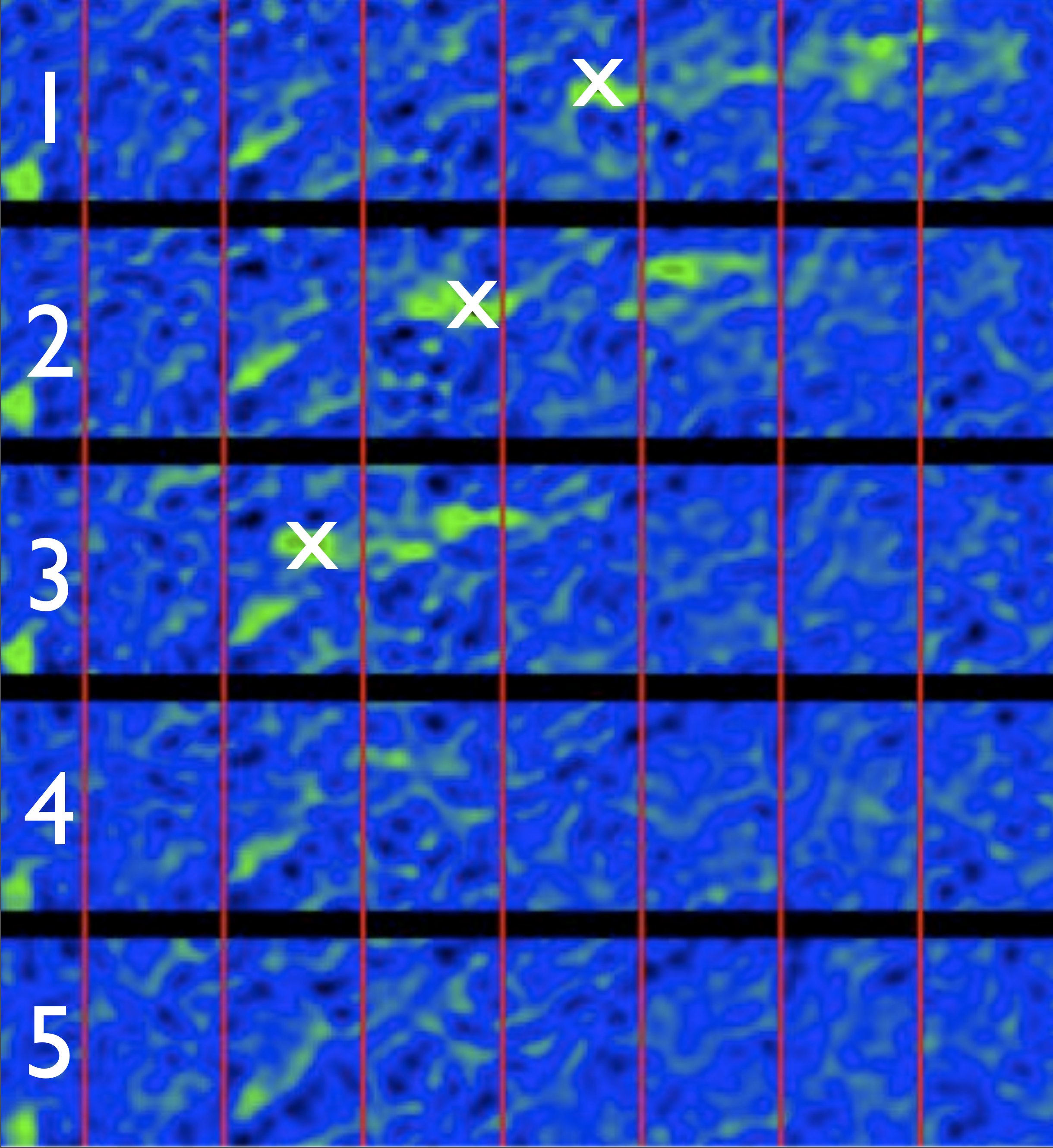}
\includegraphics[width=0.5\textwidth]{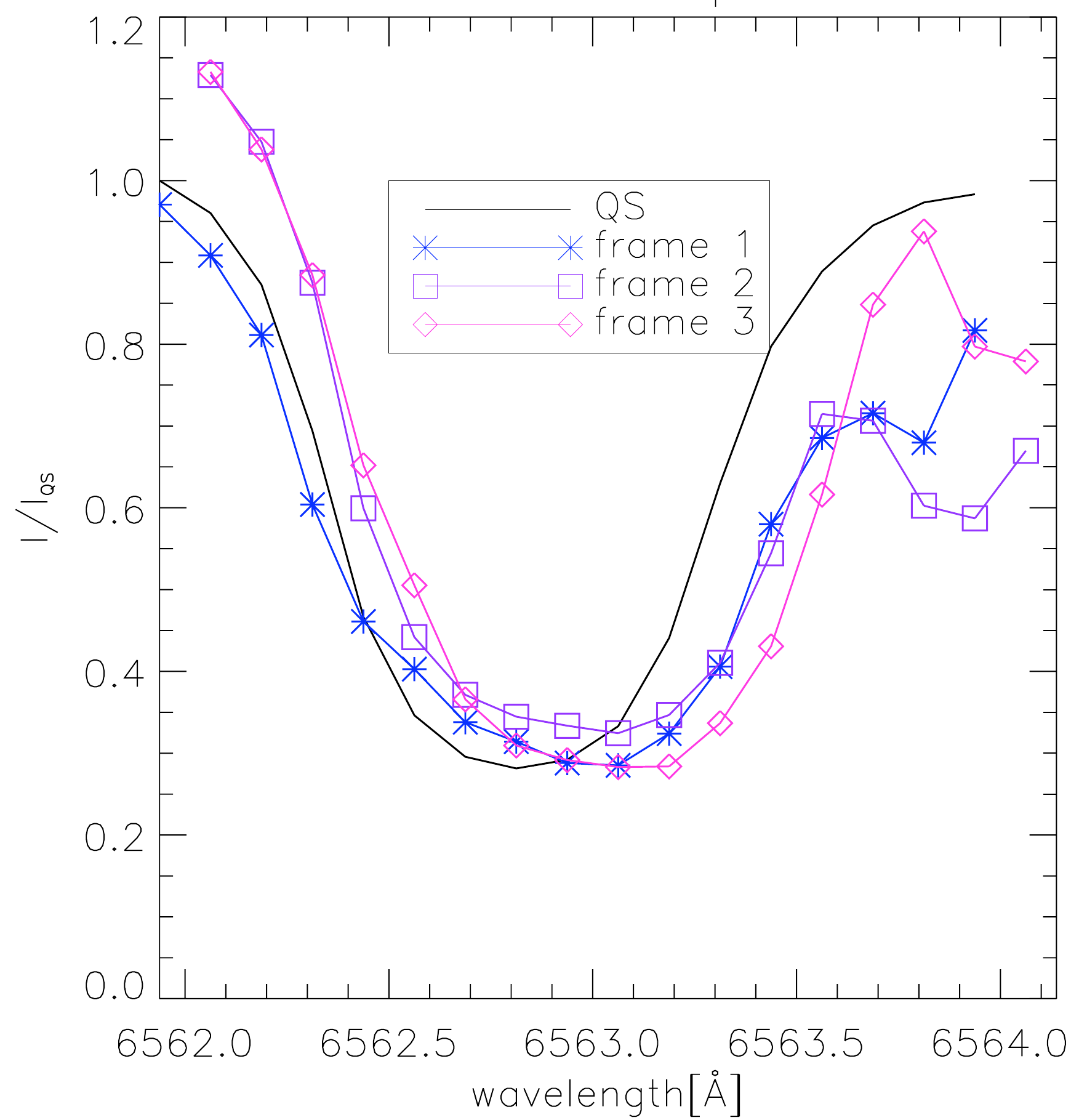}
\caption{\emph{Left}: Motion of dark feature seen in H$\alpha$ at +1 \AA\,off line center, presented in false color to increase contrast. Vertical red lines are separated by 3.15\arcsec ($\sim$\,2280 km). Time step between consecutive images $\sim$\,45 s. Horizontal tiles represent consecutive frames from the time sequence (from top to bottom). \emph{Right}: Spectral profiles, normalized to the quiet Sun spectrum at 6562 \AA, of the central part of one of the cloud members, marked by white crosses on the left image.  Black solid line is the mean profile of the surrounding quiet Sun.}
\label{fig:cloud}
\label{fig:sp}
\end{sidewaysfigure}

We report the observations of numerous fast moving dark clouds in the FoV. Dopplergrams reveal that these clouds correspond to downward motion. Here we show  a particular fast dark cloud. Neither the continuum image nor the line center exhibit strong activity. However, if we study the filtergram taken in the red wing of the H$\alpha$ line, a group of dark features becomes apparent (see panel 1 of Fig. \ref{fig:cloud}).

Successive spectrograms every 45 s of the same region (panels 2 to 5 of Fig. \ref{fig:cloud} ) reveal a fast differential motion of this dark cloud. The position and direction is marked by the white arrow in Fig. \ref{fig:mos2}. A horizontal surface velocity of $\sim90$ km/s is measured. Interestingly, the cloud has suddenly disappeared and was not longer seen in the last two observed frames.

In Fig. \ref{fig:sp} we display the corresponding spectral profiles for the central part of one of the cloud members (marked by white crosses in Fig. \ref{fig:sp}) at different times. 

We interpret the observed dark cloud, seen as a line depression in the red wing of the H$\alpha$ line, as a signature of the Doppler shifts related to the fast movement of the dark cloud. From the spectral distance between the line core of H$\alpha$ and the minimum position of the line depression we estimate a LOS downflow speed of $\sim$\,51 km/s. This, in combination with the observed horizontal velocity leads to an approximate total speed of $\sim103$ km/s directed downwards.  Further, the sudden disappearance of the cloud from the last 2 frames could be explained with a very strong related Doppler shift, thus the position of the line depression is displaced outside the scanned wavelength range.

\clearpage

\section{Fast events and waves}

We continue investigating the active chromosphere on the disc of the Sun. We report on fast phenomena and waves observed in the H$\alpha$ line with high spatial, temporal, and wavelength resolution.

\begin{figure}
\centering
\includegraphics[width=\textwidth]{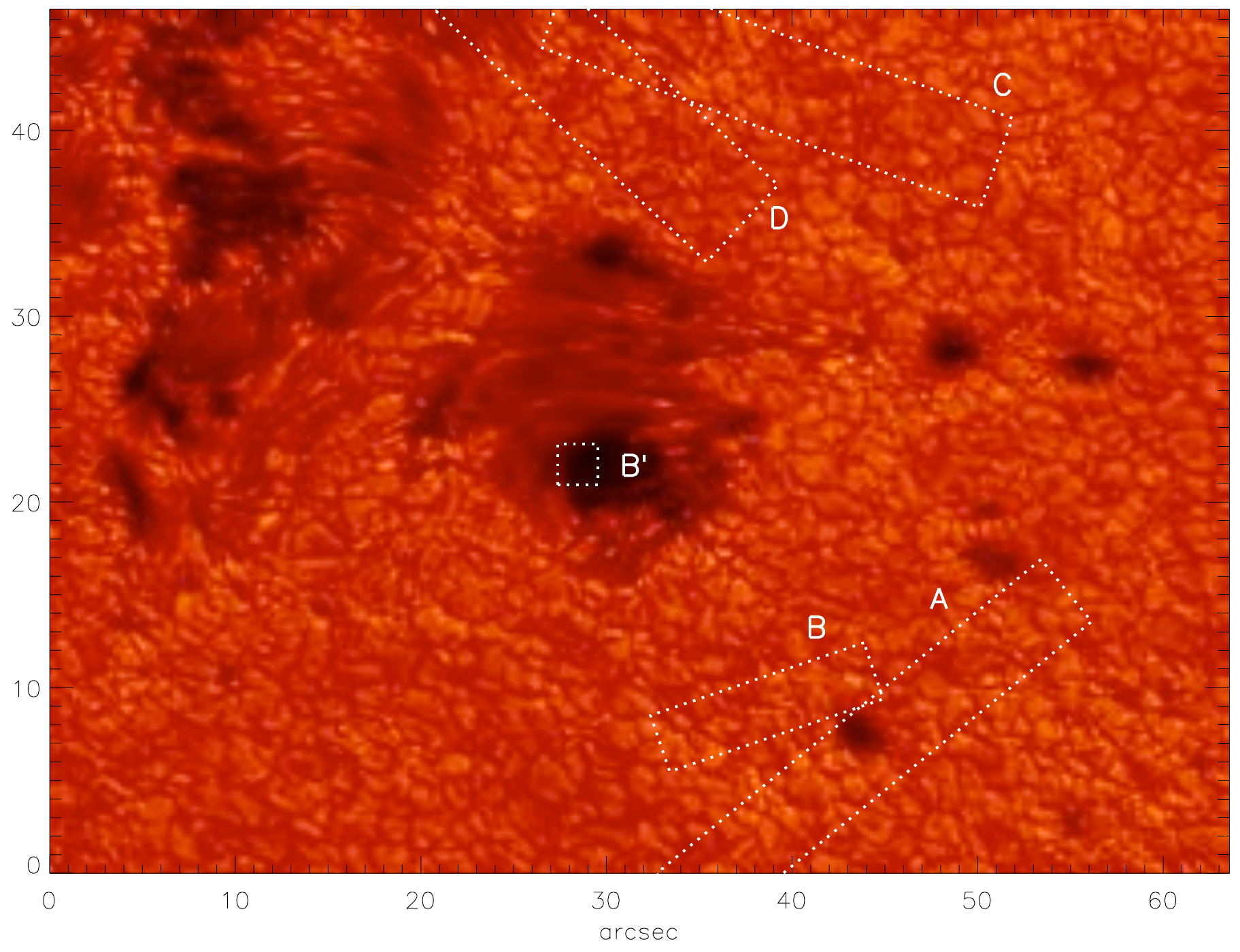} 
\caption{Broadband image of part of the active region AR10875 on  April 26,
  2006 at heliocentric angle $\vartheta=36\degr$. The rectangles, denoted by
  A, B, B\arcmin, C, and D, are the areas of interest (AOIs) to be analyzed and discussed below. }
\label{fig1}
\end{figure}



\subsection{Observations and data reduction\label{observations}\label{obser}\label{analysis}}

\begin{figure}
\centering
\includegraphics[width=\textwidth]{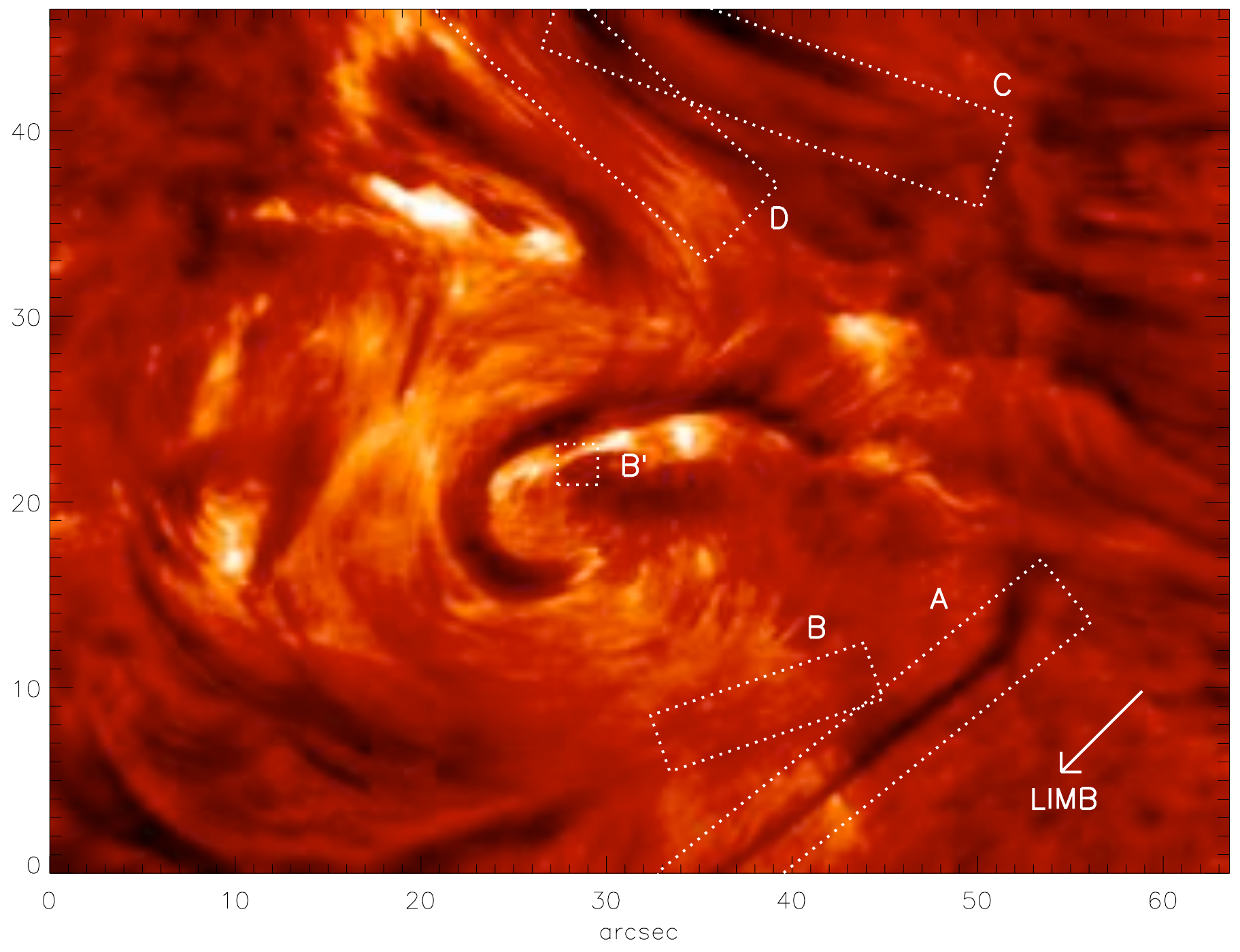} 
\caption{Narrow-band image corresponding to Fig.~\ref{fig1} in H$\alpha$ at +0.5\AA\ off line center. The same areas of interest are indicated as in Fig.~\ref{fig1} by the rectangles.}  
\label{fig2}
\end{figure}



The observations correspond to dataset ``sigmoid'' in Table \ref{table:obs:HRb}. They consist of a time sequence of 55\,min duration of H$\alpha$ scans with a mean cadence of 22\,sec from the  active region AR\,10875. The observations were supported by the Kiepenheuer Adaptive Optics system (KAOS, \citealt{2003SPIE.4853..187V}) under extremely good seeing conditions.

Due to a technical problem, an increasing delay between successive scans was noticed during the observations. When the accumulative delay reached around seven  seconds  a new scanning procedure was restarted to avoid higher gaps between frames. This operation needs around one minute. During the 55 minutes of this series, such an interrupt occurred  twice, at  08:10:19~UT and 08:29:46~UT. This programming bug was corrected afterwards for future observations.


The reduction process with SI+AO  is explained in Sec. \ref{datared}. We achieve a spatial resolution of $\sim$0\farcs25 for the broadband images at 630~nm and better than 0\farcs5 for each of the 21 narrow-band filtergrams.  Further, to follow the temporal evolution in time, both broadband and narrow-band images where cropped to the same common FoV, removing overall image shifts due to residual seeing effects. Afterwards, the speckle reconstructed broadband images were co-aligned to spatially and temporally smoothed images via a destretching code provided by \citet{1992lest.rept....1Y}. The destretching matrix from the broadband image was also applied to the simultaneous narrow-band scan.  To minimize the effects of the irregular sampling rate, the time sequences were interpolated to equidistant times with the cadence that leads to a minimum shift in time for each frame. This corresponds to a regular time step of 22~s. The data gaps at the times when observation was interrupted were filled by linear interpolation between closest observed images.

\vspace{2cm}
\pagebreak
Figures \ref{fig1} and \ref{fig2} give the broadband scenery at $t=40.9$ minutes during the series and the associated H$\alpha$ image at $+$0.5~\AA\ off line center, respectively. The whole region was very active with a flare during the observation of the time sequence \citep{2007msfa.conf..273S}. The data set is certainly rich of information on the dynamics of the active chromosphere, especially since the spatial resolution is high throughout the sequence. For the present study, we restrict further analyses and discussions to few regions. The areas of interest (AOIs) are indicated by rectangles and denoted by A, B, B\arcmin, C, and D. In the presentations below the images from the AOIs were rotated to have their long sides parallel to the spatial co-ordinate in space-time images. AOI A contains a region where a long fibril developed twice during our observations. It has the appearance of a small surge \citep{1977ASSL...69...97T}. AOIs B and B\arcmin\ show a simultaneous fast event, possibly `sympathetic' mini-flares with strong, small-scale brightenings in the H$\alpha$ line core which last only few tens of seconds. AOIs C and D, with their long fibrils, are suitable for the study of magnetoacoustic waves along magnetic field lines. Area C contains in its right part a region from which H$\alpha$ fibrils stretch out to both sides and which, at the beginning of the time sequence, contained a small pore that disappeared in the course of the observations. Note also from Fig.~\ref{fig1} that the fibrils on the upper left side of area D originate in the penumbra of a small sunspot. 


\subsection{Physical parameters\label{physpar}}
The possibility to extract information from a good part of the H$\alpha$ line profile in two dimensions and along the time series is highly valuable.
We are {interested} in the physical {parameters} of the H$\alpha$ structures. The line-of-sight velocities $v_\mathrm{LOS}$ can be retrieved using the lambdameter method, while also many other parameters like the temperature and the mass density can be inferred by means of the cloud model.

The lambdameter method \citep{1993A&A...271..574T} is a common procedure to measure line of sight (LOS) velocities. It compares the Doppler shift of a spectral line with the position of the quiet Sun profile. We measure the profile bisector at several line widths. The method consists in measuring the displacement between the bisectors of the spectral profile and the reference quiet Sun profile. As pointed out by \cite{1990A&A...230..200A} the resulting velocities give systematically lower LOS velocities than the cloud model (see below) by a factor of approximately $3$. However, the qualitative behavior of both methods are the same. The lambdameter method is therefore a fast method for a qualitative description of the velocity pattern of a region.

The cloud model yields a non-LTE inversion technique. The formation of line profiles is the result of a complex interaction between the plasma and the radiation. Among others, the formation of a spectral line depends on the local  temperature, velocity, chemical composition, magnetic fields, radiation fields, \dots Inversion techniques aim at retrieving this set of parameters from a given spectral profile.  These techniques modify, in an iteration scheme, the starting guesses of parameters based on certain assumptions until a converged solution, with modeled radiation and observations in close agreement, is obtained. We assume then that the calculated parameters producing the synthetic profile are the same as in the observed structure, as long as the assumptions are considered valid.

\pagebreak
\begin{wrapfigure}[19]{r}{0.4\textwidth}
\begin{center}
\includegraphics[width=0.5\textwidth]{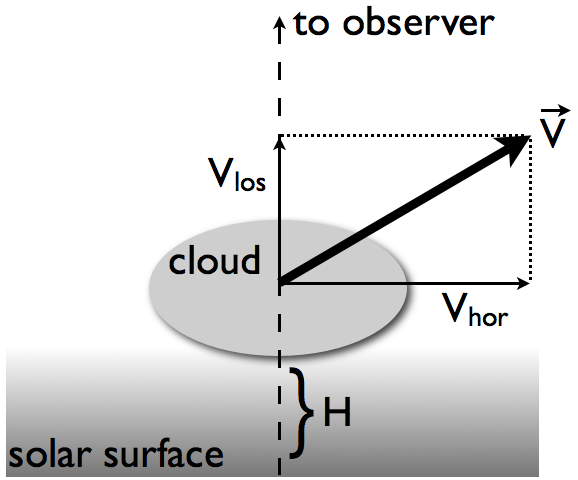}
\caption{Geometry of the cloud model.}
\label{cloud:geo}
\end{center}
\end{wrapfigure}

The cloud model allows the application of an inversion technique in cases when one can describe  the radiation transfer through structures located high above the unperturbed solar photosphere. This method was first described by \citet{1964PhDT........83B} and has been extensively used afterwards, e.g. by \citet{tsiropoula97,2004A&A...424..279T,2004A&A...423.1133T,2004A&A...418.1131A}. See also the recent review by \citet{2007ASPC..368..217T}.

Figure \ref{cloud:geo} depicts the geometry of the cloud model. The considered ``cloud'' is located above the underlying photosphere at a height $H$ and moving  at a speed $\vec V$. From the observer's position we can measure the projected proper motion relative to the background and the LOS velocity ($V_{los}$ in Fig. \ref{cloud:geo}) as Doppler shifts. The observed intensity $I(\Delta \lambda)$ is the combination of the absorption of the background intensity $I_{0}(\Delta \lambda)$ with the emission from the cloud, dependent on the optical thickness of the cloud $\tau(\Delta \lambda)$:
\begin{equation}
I(\Delta \lambda)= I_{0}(\Delta \lambda)\cdot e^{-\tau(\Delta \lambda)}+ \int_{0}^{\tau(\Delta \lambda)}  S_{t}e^{-t(\Delta \lambda)}dt \,,
\label{eq:cloud1}
\end{equation}
where $S$ is the source function, which depends on the optical thickness along the cloud. In the model we make the following assumptions:
\begin{enumerate}
\item The structure is well above the underlying unperturbed chromosphere.
\item Within the cloud, the source function $S$, velocity and  Doppler width are constant along the LOS.
\item The background intensity profiles entering the cloud from below and in the surroundings are the same.  
\end{enumerate}

These assumptions simplify Eq. \ref{eq:cloud1} to
\begin{equation}
I(\Delta \lambda)= I_{0}(\Delta \lambda)\cdot e^{-\tau(\Delta \lambda)}+  S \, (1-e^{-\tau(\Delta \lambda)}) \,.
\label{eq:cloud2}
\end{equation}

This, in terms of the \emph{contrast profile},
\begin{equation}
C(\Delta \lambda):= \frac{I(\Delta \lambda)-I_{0}(\Delta \lambda)}{I_{0}(\Delta \lambda)}\,,
\label{eq:contrast}
\end{equation}
can be rewritten as
\begin{equation}
C(\Delta \lambda)=\Bigg (\frac{S}{I_{0}(\Delta \lambda)}-1\Bigg ) \, \Big(1-e^{-\tau(\Delta \lambda)}\Big)\,.
\label{eq:contrast}
\end{equation}
Further, neglecting collisional and radiative damping of the H$\alpha$ absorption profile within the cloud, the optical depth can be given by a Gaussian profile, i.e.
\begin{equation}
\tau(\Delta \lambda)=\tau_{0}\,e^{-\Big(\frac{\Delta \lambda-\Delta \lambda_{I}}{\Delta \lambda_{D}}\Big)^2}\,,
\label{eq:cloud:gausian}
\end{equation} 
where $\tau_{0}$ is the line center optical thickness. Also, the  central wavelength of the profile can be displaced due to a LOS velocity $v$ of the cloud with a Doppler shift, $\Delta \lambda_{I}=\lambda_{0}v/c$, where $\lambda_{0}$ is the rest central wavelength and $c$ is the speed of light. The width of the profile $\Delta \lambda_{D}$ depends on the temperature $T$ and the microturbulent velocity $\xi_{t}$ trough the relation 
\begin{equation}
\Delta \lambda_{D}=\frac{\lambda_{0}}{c}\sqrt{\frac{2kT}{m}+\xi_{t}^{2}}\, ,
\label{eq:cloud:width}
\end{equation} 
where $m$ is the atom rest mass.

With these assumptions we end up with an inversion problem with 4 parameters: $S$, $\Delta \lambda_{D}$, $\tau_{0}$ and $v_{LOS}$. $\Delta \lambda_{D}$ is, in turn, the combination of 2 physical parameters (temperature and microturbulent velocity).  

More complex cloud models have recently been developed. These mainly focus on the nature of the source function $S$, allowing the parameters to vary along the LOS or multi-cloud models. However, as pointed out by \cite{1990A&A...230..200A}, a simple Beckers cloud model like the one described above and used here provides useful, reasonable estimates for a large number of optically not too thick structures, $\tau_{0} \lesssim 1$, for which the assumptions are adequate.

In this work we also used the inversion in H$\alpha$ structures where possible. The undisturbed reference profile $I_0(\lambda)$ is taken from a nearby area with low activity outside the FoV shown in Fig.~\ref{fig1}. As the region under study was `clouded out' in H$\alpha$, i.e. covered with {structures} to a large extent, the cloud model inversion failed often. In these latter cases, instead, the LOS velocity maps were determined with the lambdameter method or from difference images at H$\alpha\pm$0.5\,\AA\ off line center with appropriate scaling. Calibration curves to estimate from such Doppler-grams the true velocities were calculated by \citet{1990SoPh..129..277G}. From these we obtained that the velocities from the difference images were lower by a factor 2--4 than the true velocities, in agreement with those parts in the FoV where the cloud model inversion was successfully applied and with the results by \citet{2004A&A...423.1133T}.

With the application of the cloud model and the inferred values of $S$, $\Delta \lambda_{D}$, $\tau_{0}$ and $v_{LOS}$ we can derive other physical parameters. Following the approach by e.g. \cite{tsiropoula97} we can calculate the population densities of the hydrogen levels 1, 2, 3 ($N_{1}$,$N_{2}$,$N_{3}$), the total hydrogen density ($N_{H}$, including protons), the electron density ($N_{e}$), the total particle density ($N_{t}$) the electron temperature ($T_{e}$), gas pressure ($p_{g}$), total column mass ($M$), mass density ($\rho$) and degree of ionization of hydrogen ($x_{H}$):

\begin{eqnarray}
N_{1}=& \frac{N_{t}-(2+\alpha)N_{e}}{1+\alpha} & \\
N_{2}= & 7.26~10^{7}\frac{\tau_{0}\Delta\lambda_{D}}{d} & \mbox{ cm}^{-3} \\
N_{3}=& \frac{g_{3}}{g_{2}}N_{2}\Big ({\frac{2h\nu^{3}}{Sc^{2}}+1}\Big )^{-1}&  \\
N_{e}=& 3.2~10^{8}\sqrt{N_{2}} & \mbox{ cm}^{-3} \\ 
N_{H}=& 5~10^{8}\sqrt{N_{2}}&  \\
N_{t}=& N_{e}+(1+\alpha)N_{H}  & \\
p_{g}=& kN_{t}T_{e}\\
M=& (N_{H}m_{H}+0.0851N_{H}\cdot3.97m_{H})d& \\
\rho=& M/d& \\
x_{H}=& N_{e}/N_{H}
\end{eqnarray}
where $d$ is the path length along the LOS through the structure, $\alpha$ is the abundance ratio of helium to hydrogen ($\approx 0.0851$), $g_{2},g_{3}$ are the statistical weights of the hydrogen levels 2 and 3 respectively, $h$ is the Planck constant, $\nu$ is the frequency of H$\alpha$, $c$ the speed of light  and $k$ the Boltzmann constant.

\begin{figure}[]
\center 
\includegraphics[width=\textwidth]{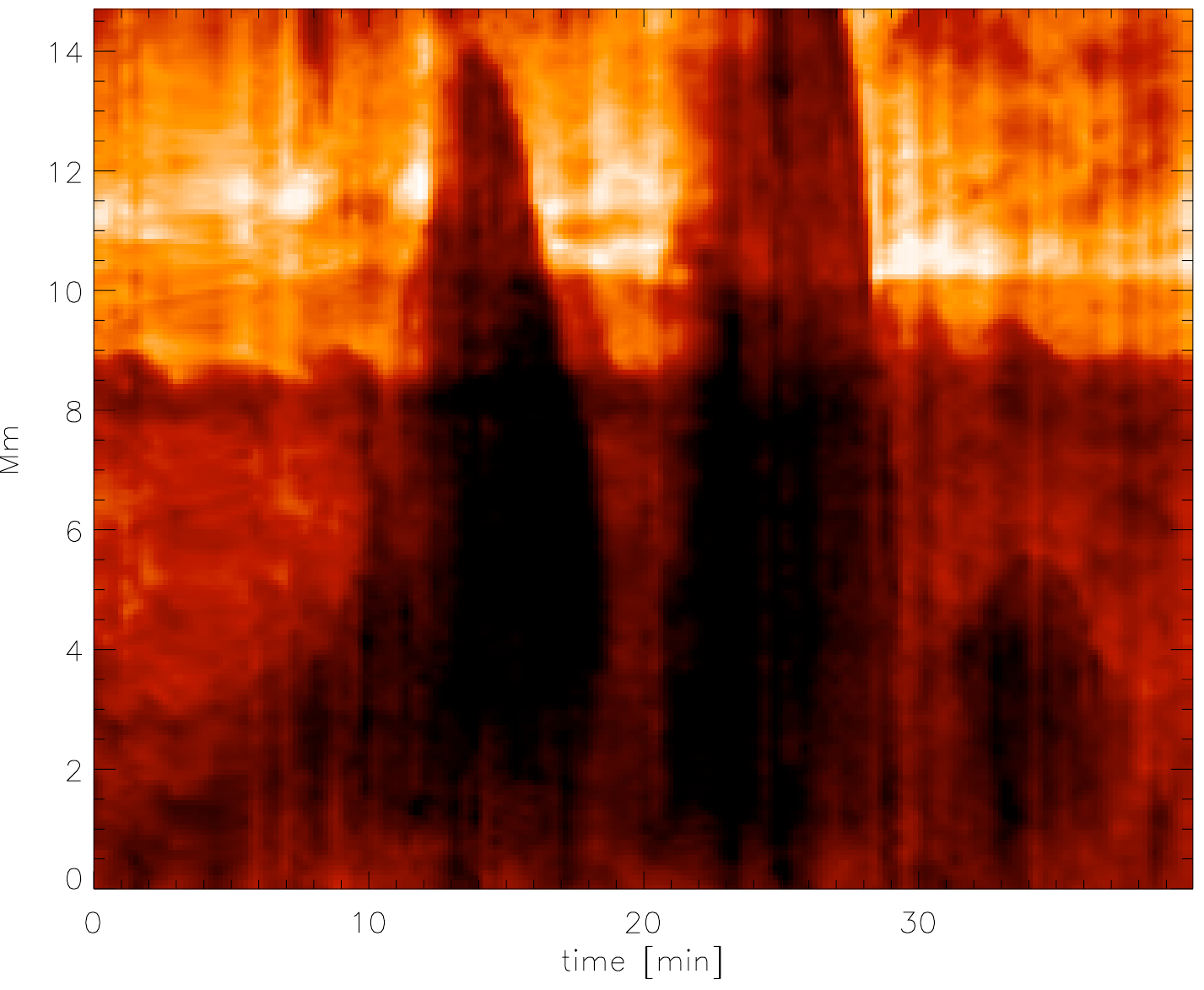}
\caption{Space-time image of surge in AOI A at H$\alpha$ + 0.5~\AA\ off line
  center, starting at 14.7~min after the beginning of the sequence. The spatial axis runs along the minima of the surge intensities at this wavelength.}  
\label{fig3}
\end{figure}

 Table \ref{tabla} summarizes average results from the cloud model and derived quantities for the long fibril in Fig.~\ref{fig3}:
\begin{table}[h]
 \begin{center}
  \normalsize
 \begin{tabular}{|c|c||c|c|}
 \hline
 \hline
Parameter & Av. value & Parameter & Av. value \cr
 \hline
v [km/s] &     11.7 & $\Delta\lambda_{D}$ [\AA]  &   0.34\cr
$S/I_{c}$ &     0.154 & $\tau$    &  1.05 \cr
\hline
N$_{2}$ [cm$^{-3}$]  &    $4.5\cdot10^{4}$ & N$_{e}$ [cm$^{-3}$]  & $6.8\cdot10^{10}$  \cr
N$_{H}$ [cm$^{-3}$]  & $1.1\cdot10^{11}$ & N$_{1}$ [cm$^{-3}$] & $3.8\cdot10^{10}$ \cr
N$_{3}$ [cm$^{-3}$]  &    $4.2\cdot10^{2}$ & $T_{e}$  [K]&    $1.51\cdot10^{4}$ \cr
$p$ [dyn cm$^{-2}$] & $0.38$ & M [g cm$^{-2}$] & $1.39\cdot10^{-4}$ \cr
$\rho$ [g cm$^{-3}$] & $2.3\cdot10^{-13}$ & $x_{H}$ &    0.64 \cr
$c_{s} [km/s]$  &  14.4 & & \cr
 \hline
 \hline
 \end{tabular}
\vspace{0.2cm}
\caption{Several derived parameters from the cloud model for the lower half section of the long fibril in Fig.~\ref{fig3} at $t=25$ min. We assume a LOS thickness equal to the width of the fibril ({cylindrical} shape) of $590$ km and a micro-turbulent velocity of  $10$ km/s. First two rows result from the inversion technique while the others are parameters derived from them.}
\label{tabla}
 \end{center}
 \end{table}



\newpage
\subsection{Fast events in H$\alpha$\label{fast}}

\subsubsection*{Small recurrent surge\label{surge}}
Ejecta from low layers of active regions, called {surges}, have been observed in time sequences of H$\alpha$ filtergrams since many decades \citep[e.g.,][]{1977ASSL...69...97T}.

In AOI A, a small surge occurred during the observed time series. It started near the pore at the upper right end of region A (cf. Figs.~\ref{fig1} and \ref{fig2}). It was straight and thin, with a projected length at its maximum extension of at least 15~Mm and with widths of approximately 2\arcsec\ at its mouth and 1\arcsec\ at its end. Figure~\ref{fig3} shows the temporal evolution of the surge in H$\alpha$ +0.5\AA\ off line center. The space-time image starts 14.7~min after the beginning of the series and goes to the end of it. Along the spatial axis in Fig.~\ref{fig3}, the minimum intensities along the surge are represented.

The surge consisted of very thin fibrils, at the resolution limit $<0$\farcs5, being ejected in parallel. It started with several small elongated clouds lasting for 1--2~min. Afterwards, it rose, reaching a projected length of around 14~Mm, and fell back after $\sim7$min. Then it suddenly rose again after two min reaching lengths out of the FoV (more than 15\,400 km) and lasted another five min before retreating again. And finally, the process recurred a third time, yet with lower amplitude in extension and velocity than for the first two times. The (projected) proper motion of the tip of the surge reaches a maximum velocity of approximately 100~km\,s$^{-1}$, for both the ascent and the descent phases. Especially the second rise and fall showed large velocities. It is unlikely that the rapid rise and appearance of the surge in H$\alpha$ are caused by cooling of coronal gas to chromospheric temperatures. The cooling times are much too long, of the order of hours \citep{hildner74}. Thus, the proper motions represent gas motions. The LOS velocities measured from Doppler-grams and corrected with the calibrations described above in Sect.~\ref{physpar}, amounted to +15~km\,s$^{-1}$ during the ascent of the surge and reached $-$45~km\,s$^{-1}$ at the mouth during retreat. These latter velocities are lower than the proper motions. It thus appears that the chromospheric gas is ejected obliquely into the direction towards the limb. 
Average physical parameters in the surge obtained with the cloud model inversion are listed in Table~\ref{tabla}. They are very similar to those of other chromospheric structures \mbox{\citep[see e.g.][]{tsiropoula97}.}

Surges are known to show a strong tendency for recurrence, but on time scales of $\sim$1~h. \citet{1989ApJ...343..985S} have treated numerically rebound shocks in chromospheric fibrils and presented results in which a single impulse at the base of the involved magnetic flux tube drives a series of shocks on time scales of approximately 5~min. This appears to be a viable mechanism for the small surge observed here, apart from the initial conditions. The small `firings' at the beginning of this surge suggest magnetic field dynamics that ultimately do cause a strong impulsive force, after some minor events.

\begin{sidewaysfigure}[t]
\center 
\includegraphics[width=\textwidth]{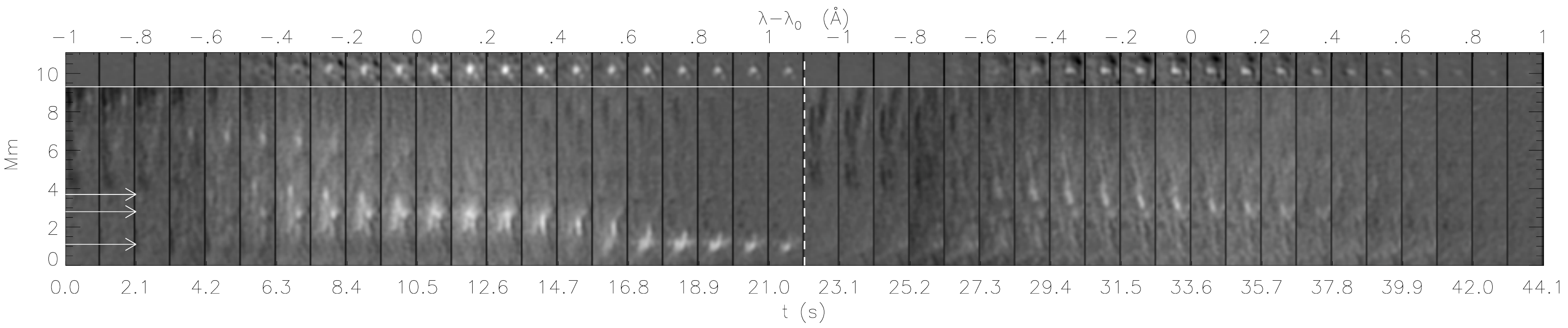} 
\caption{Simultaneous flash event on AOIs B and B\arcmin\ with projected distance $\approx$13.7 Mm. A pair of simultaneous, short, brightening was recorded at $t=52.2$ min. Top row from B\arcmin, bottom row from B. The tiles from left to right correspond to two successive H$\alpha$ scans. Upper x-axis is scaled to the wavelength of each 2-D filtergram tiles. Scanning time is numbered on the lower x-axis. $t=0$ corresponds to the beginning of the scan at 08:44 UT.  The integration time for each spectral position is $\approx 1$s, while the delay between two scans is $\approx 3$s (vertical dashed line). Each spectrogram on B is normalized with the background profile  (see Fig. \ref{fig5}) to emphasize the flash event. Neither the previous nor the following scan to the two presented exhibited any emission. The second scan (right half size of the figure) still shows some emission on the same positions. White arrows correspond to the position of the three different profiles in Fig. \ref{fig5}.}  
\label{fig4}
\end{sidewaysfigure}

\subsubsection*{Synchronous flashes\label{mini-flare}}
In the AOI pair (B, B\arcmin) with a projected distance of $\sim$14~Mm,
brightenings occurred 52.2~min after the start of the series in both sites at
least as simultaneously as we can detect with the observational mode of
scanning the H$\alpha$ line. AOI B\arcmin\ is located in the umbra of a small spot with a complex penumbra and AOI B next to a pore. In between the two AOIs the sigmoidal filament ended while more structures of the extended and active filament system crossed the region between the two AOIs. Figure~\ref{fig4} shows the temporal evolution of the brightenings. 

The upper row of this figure is from AOI  B\arcmin, the lower from B. Two scans through the H$\alpha$ profile are presented, of course without interpolation of the images to an identical time. The horizontal axes contain the run in both time and wavelength. 

The flash-like brightenings lasted only for less than 45~s, they were present neither in the scans before nor after the two scans shown in Fig.~\ref{fig4}. The simultaneity of the two flashes, or mini-flares, suggests a relation between them. Possibly, one sees here a kind of sympathetic flares. These were discussed earlier in the context of synchronous flares excited by activated filaments \citep{1977ASSL...69...97T}. Another interpretation is that one sees a mini-version of two-ribbon flares with a common excitation in the corona above them and simultaneous injection of electrons into the chromosphere.

In AOI B, the flash exhibited sub-structure and apparently moved during the first presented scan with speeds up to 200~km\,s$^{-1}$. This strong brightening between 15 and 22~s has disappeared in the following scan. 

\begin{figure}[t!]
\center 
\includegraphics[width=0.7\textwidth,angle=0]{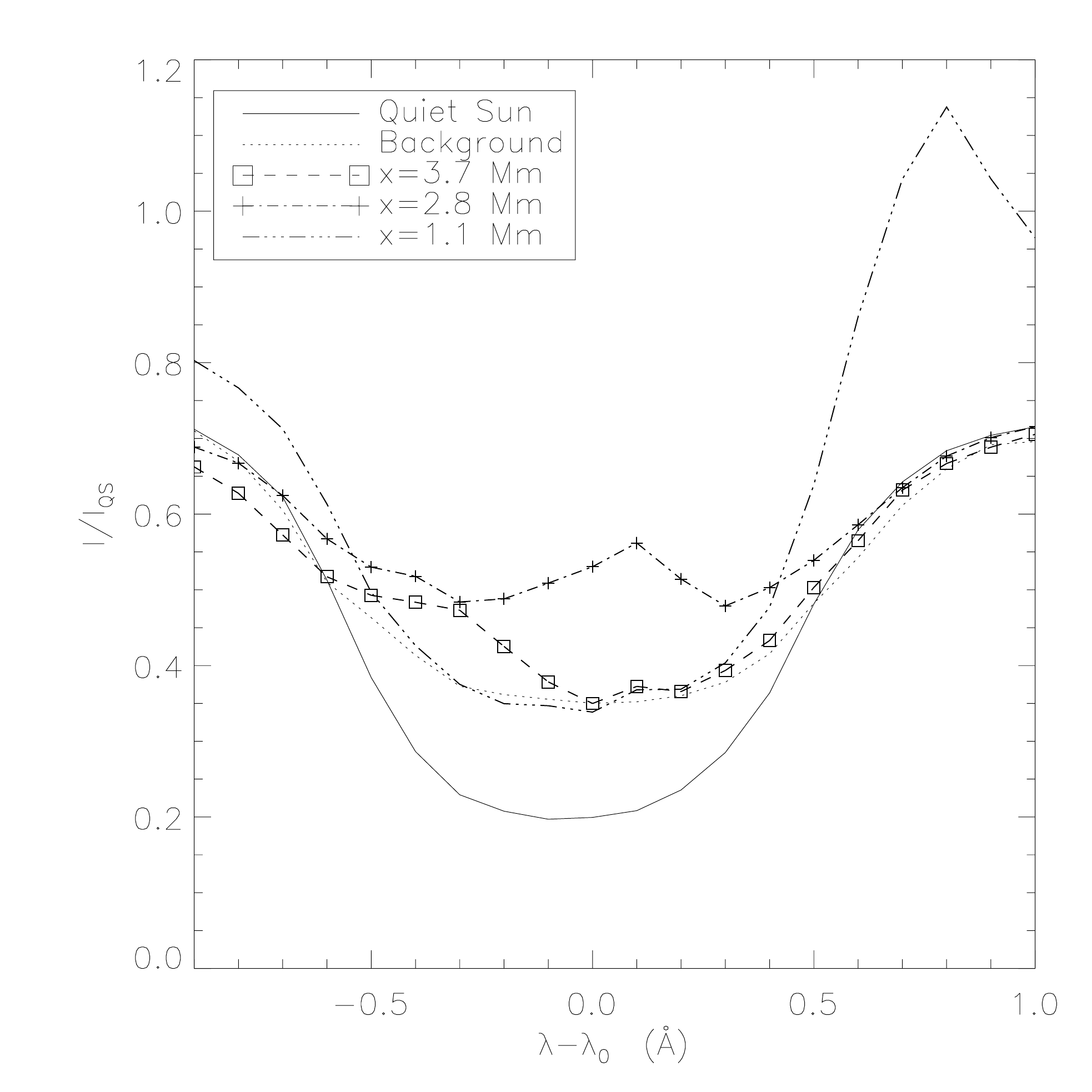} 
\caption{H$\alpha$ profiles from the flash event. Each profile corresponds to an average over three pixels around the three selected points where the emission is highest in the blue wing, at central wavelength, and in the red wing respectively, corresponding to the white arrows in Fig. \ref{fig4} (left half) at x=[3.7, 2.8, 1.1]~Mm, respectively. For comparison the quiet Sun profile is also shown. Background profile corresponds to the mean from previous and following scans, where no brightening  was found. The emission profile at 1.1~Mm reaches an intensity of 1.1 of the quiet Sun continuum intensity.}  
\label{fig5}
\end{figure}

Figure \ref{fig5} depicts the recorded H$\alpha$ profiles at the positions of the flash in AOI B, as indicated by the arrows on the left side of Fig.~\ref{fig4}. The profiles are compared with those from the quiet Sun and from the average background. The profile from the isolated bright blob at 3.7~Mm (see inset in Fig.~\ref{fig5}) shows a blue shifted emission above the background profile. This emission is still present in the following scan. At 2.8~Mm the line core is filled resulting in a contrast profile with strong emission (cf. Eq.~\ref{eq:contrast}). The profile at 1.1~Mm exhibits a strong emission beyond the continuum intensity in the red wing while the whole profile is enhanced above the background profile. the position of the emission peak would indicate a down flow with LOS velocity of 35~km\,s$^{-1}$. It was shown by \citet{2004A&A...418.1131A} that such emission (contrast) profiles can be understood if one assumes an injection, likely from the corona, of much energy and electrons to obtain a response of the H$\alpha$ line to temperature. These last two emissions at 2.8~Mm and 1.1~Mm have disappeared at the time of the following scan.

Obviously, such fast events as in AOIs B and B\arcmin\ lie beyond the observing capabilities of our consecutive scanning method. We could however retrieve high spatial resolution filtergrams at several wavelengths to follow the temporal evolution at time scales of few seconds. With the present data set at our hand, we cannot decide whether the apparent proper motion of the flashing structure in AOI B is indeed as high as 200~km\,s$^{-1}$ or whether the temporal resolution is too fast for the consecutive scanning. For example, the H$\alpha$ profile from 1.1~Mm could have been in emission over the whole profile, but only for few seconds. 
 It is, however, possible to design adequate observing sequences with duration of few seconds per scan, on the expense of taking filtergrams at fewer wavelength positions.
\clearpage


\subsection[Magnetoacoustic waves]{Magnetoacoustic waves\label{waves1}\label{waves2} \label{mhdapprox}}
In this Section, for the investigation waves in the chromosphere, we first refer to previous observations of magnetoacoustic waves and then outline the magnetohydrodynamic (MHD) approximation \citep{ferraro66,kippenhahn75,priest84}. From this, dispersion relations for sound waves, atmospheric waves, Alfv\'en waves, and magnetosonic, or magnetoacustic, waves are derived by linearization. We continue discussing the observations of waves in long H$\alpha$ fibrils. Finally we try an interpretation in the framework of waves in thin magnetic flux tubes.

\subsubsection*{Previous observations of magnetoacoustic waves}
Apart from oscillations in sunspot umbrae \citep{beckers69,wittmann69} and
running penumbral waves \citep[e.g.,][and references therein]{uexkuell83},
waves in the chromosphere were observed by many authors. E.g.,
\citet{1975SoPh...44..299G} describes waves along H$\alpha$ mottles and fibrils with
speeds of 70~km\,s$^{-1}$ and interprets them as Alfv\'en waves in magnetic
flux tubes with approximately 10~Gauss field strength. \citet{2006A&A...449L..35K},
from time sequences in H$\alpha$ off the limb, found kink waves in spicules
with periods of 35--70~s. \citet{hansteen06}, and \citet{rouppe07} observed
spicules and fibrils in the quiet Sun and in active regions with high spatial
and temporal resolution observations. They succeeded via numerical simulations
in explaining the dynamics of these chromospheric small-scale structures by
magnetoacoustic shocks, excited mainly by the solar 5-min {oscillations}
(see also the simulations by \citealt{1989ApJ...343..985S} and the review by \citealt{carlsson05}). 

Waves in the corona have as well been observed: E.g., \citet{robbrecht01} report on slow magnetoacoustic waves in coronal loops observed in high-cadence images from SoHO/EIT and TRACE. The speeds amount to 100~km\,s$^{-1}$. \citet{2002A&A...387L..13D}, also from high-cadence 171~\AA\ TRACE images, find that 3- and 5-min oscillations are common in coronal loops. They are also interpreted as magnetoacoustic waves. \citet{2007msfa.conf..265T}, from SoHO/SUMER data, study Doppler shift oscillations identified as slow mode standing waves in hot coronal loops. Fast-mode, transverse, incompressible Alfv\'en waves, with speeds of 2~Mm\,s$^{-1}$, in the solar corona were reported by \citet{2007Sci...317.1192T}.

\subsubsection*{Magnetohydrodynamic (MHD) approximation}
We use here the Gauss system of units. The MHD approximation is obtained from Maxwell's equations and the equation of mass conservation, the equation of motion, and an equation of state, under the following conditions:
\begin{enumerate}
\item
The gas velocities $v$ are small compared to the speed of light $c$, $v\ll c$.
\item
Any changes are slow, such that phase velocities $v_{ph}=L/t\ll c$, where $L$ is a typical length scale and $t$ a typical time scale.
\item
The electrical conductivity $\sigma$ is always very high such that the electric field is very small compared to the magnetic flux density, $\vert\vec E\vert\ll\vert\vec B\vert$. 
\item
One usually adopts in addition, to a good approximation, $\vec D=\vec E$ and $\vec B=\vec H$, i.e. magnetic flux density and magnetic field have the same strength, in Gauss units.
\end{enumerate}

With $\vec j$\, electrical current density, Maxwell's equations are then reduced to
\begin{equation}
\nabla\times\vec B = {4\pi\over c}\vec j\,,
\label{mhd1}
\end{equation}
\begin{equation}
\nabla\times\vec E = -{1\over c}{\partial\vec B\over\partial t}\,,
\label{mhd2}
\end{equation}
\begin{equation}
\nabla\cdot\vec B=0\,.
\label{mhd3}
\end{equation}

Ohm's law, conservation of mass, and the equation of motion read as
\begin{equation}
\vec j = \sigma\left(\vec E + {1\over c}\vec v\times\vec B\right)\,,
\label{mhd4}
\end{equation}
\begin{equation}
{\partial\rho\over\partial t} = \nabla\cdot\left(\rho\vec v\right)\,,
\label{mhd5}
\end{equation}
\begin{equation}
\rho{\partial\vec v\over\partial t} + \rho\left(\vec v\cdot\nabla\right)\vec v = \rho\vec g - \nabla p +{1\over c}\vec j\times\vec B\,.
\label{mhd6}
\end{equation}
Here, $\rho$ is the mass density, $p$ the gas pressure, and $\vec g$ the gravitational acceleration vector. The viscous, centrifugal, and Coriolis forces were omitted in the equation of motion, Eq.~\ref{mhd6}. The equation of state relating the gas pressure $p$ with mass density $\rho$ and temperature $T$, is
\begin{equation}
p = p\left(\rho,T\right)\,.
\label{mhd7}
\end{equation}
Furthermore, we assume for simplicity adiabatic motion
\begin{equation}
{\mathrm d\over\mathrm dt}\left({p\over\rho^\gamma}\right) = 0\,,
\label{mhd8}
\end{equation}
with the ratio of specific heats $\gamma=5/3$ for monoatomic gases.

The current density $\vec j$ can be eliminated by means of Ohm's law, Eq.~\ref{mhd4} which yields the equation of motion
\begin{equation}
\rho{\partial\vec v\over\partial t} + \rho\left(\vec v\cdot\nabla\right)\vec v = \rho\vec g - \nabla p +{1\over4\pi}\left(\nabla\times\vec B\right)\times\vec B\,,
\label{mhd9}
\end{equation}
and the induction equation
\begin{equation}
{\partial\vec B\over\partial t} = - \nabla\times\left({1\over\sigma}\nabla\times\vec B\right) + \nabla\times\left(\vec v\times\vec B\right)\,,
\label{mhd10}
\end{equation}
where the last term on the {\em rhs} of Eq.~\ref{mhd9} contains Maxwell's stress tensor.

In an atmosphere with constant temperature and with the gravitational acceleration opposite to the vertical direction ($\vec g=-g\vec e_z$, $\vec e_z$ unity vector into $z$ direction, $\vert\vec e_z\vert=1$) the hydrostatic equilibrium is 
\begin{equation}
{\mathrm{d}p_0\over\mathrm{d}z} = -\rho_0 g\,,
\label{mhd11}
\end{equation}
with the solution
\begin{equation}
p_0 = const1\cdot\mathrm{e}^{-z/\Lambda}\,;~~~\rho_0 = const2\cdot\mathrm{e}^{-z/\Lambda}\,,
\label{mhd12}
\end{equation}
where the scale height is $\Lambda=p_0/(\rho_0\cdot g)$.


\subsubsection*{Magnetoacoustic gravity waves\label{mhdwaves}}

We consider now, to arrive at a dispersion relation for magnetoacoustic gravity waves, small perturbations from the equilibrium
\begin{equation}
\vec B = \vec B_0 + \vec B_1\,;~~~\vec E = \vec E_0 + \vec E_1\,;~~~\vec j = \vec j_0 + \vec j_1\,;
\label{mhd13}
\end{equation}
\begin{equation}
p = p_0 + p_1\,;~~~\rho = \rho_0 + \rho_1\,;~~~\vec v = \vec v_0 + \vec v_1\,.
\nonumber
\label{mhd14}
\end{equation}

Assuming further that $\vec B_0$ is homogeneous, $\vec v_0=0$, and the conductivity is infinite, $\sigma\rightarrow\infty$, one obtains 
\begin{equation}
\nabla\times\vec B_0 = {4\pi\over c}\vec j_0=0\,~~~\mathrm{and}~~~\sigma\vec E_0=0\,,~~~\mathrm{i.e.}~~~\vec E_0 = 0\,.
\label{mhd16}
\end{equation}

Inserting then the perturbed quantities from Eqs.~\ref{mhd13} and \ref{mhd14} 
into the MHD equations and neglecting quadratic terms and terms of higher order we obtain the linearised MHD equations
\begin{equation}
{\partial\rho_1\over\partial t}+\left(\vec v_1\cdot\nabla\right)\rho_0+\rho_0\left(\nabla\cdot\vec v_1\right) = 0\,,
\label{mhd17}
\end{equation}
\begin{equation}
\rho_0{\partial\vec v_1\over\partial t} = -\nabla p_1 + {1\over4\pi}\left(\nabla\times\vec B_1\right)\times\vec B_0 - \rho_1g\vec e_z\,,
\label{mhd18}
\end{equation}
\begin{equation}
{\partial p_1\over\partial t}+\left(\vec v_1\cdot\nabla\right)p_0 -c_s^2\left[{\partial\rho_1\over\partial t}+\left(\vec v_1\cdot\nabla\right)\rho_0\right]=0\,,
\label{mhd19}
\end{equation}
\begin{equation}
\rho_0{\partial\vec B_1\over\partial t} = \nabla\times\left(\vec v_1\times\vec B_0\right)\,,
\label{mhd20}
\end{equation}
\begin{equation}
\nabla\cdot\vec B_1 = 0\,.
\label{mhd21}
\end{equation}
with the sound speed $c_s=[(\gamma p_0)/\rho_0]^{1/2}$. From these one arrives, after some algebra, at the wave equation for the velocity
\begin{eqnarray}
 \frac{\partial^2\vec v_1}{\partial t^2} = &c_s^2\nabla\left(\nabla\cdot\vec v_1\right)-\left(\gamma-1\right)g\, \vec e_z\left(\nabla\cdot\vec v_1\right) -g\nabla v_{1,z}\cr
&+{1\over\rho_0}\left[\nabla\times\{\nabla\times\left(\vec v_1\times\vec B_0\right)\}\right]\vec B_1/\left(4\pi\right)\,.
\label{mhd22}
\end{eqnarray}


\subsubsection*{Wave modes\label{wavemodes}}

With Eq. \ref{mhd22} we make the {\em ansatz} 
\begin{equation}
\vec v_1(\vec r,t)=\vec v_1\exp\left[i\left(\,\vec k\cdot\vec r-\omega t\right)\right]\,,
\label{mhd23}
\end{equation}
with wavevector $\vec k$.

For $\vec B_0=0$ and $\vec g=0$ one gets pure sound waves with phase velocity $v_{ph}=\omega/k=c_s$. With $\vec B_0=0$ and $g>0$ one obtains atmospheric waves \citep{bray74}.

When the gas pressure is negligible, $p=0$, and with $g=0$, but $\vert\vec B_0\vert>0$, the dispersion relation results
\begin{equation}
\omega^2\vec v_1/v_A^2=k^2\cos^2\hspace{-0.5mm}\alpha\,\vec v_1
-(\vec k\cdot\vec v_1)\,k\cos\alpha\,\vec{\hat{B}}_0+\left[(\vec k\cdot\vec v_1)-k\cos\alpha\,(\vec{\hat{B}}_0\cdot\vec v_1)\right]\vec k\,.
\label{mhd24}
\end{equation}
Here, $\vec{\hat{B}}_0$ is a unity vector parallel to $\vec{{B}}_0$, $\alpha$ is the angle between the wavevector $\vec k$ and $\vec B_0$, and $v_A$ is the Alfv\'en velocity with 
\begin{equation}
v_A^2={B_0^2\over4\pi\rho_0}=2{P_{m,0}\over\rho_0}\,,
\label{mhd25}
\end{equation}
where the magnetic pressure is $P_{m}=B^2/(8\pi)$. Scalar multiplication of Eq.~\ref{mhd24} with $\vec{\hat{B}}_0$ shows that $\vec{\hat{B}}_0\cdot\vec v_1=0$. This means that the (perturbed) velocity is perpendicular to $\vec B_0$ (since the Lorentz force on the perturbed gas is perpendicular to $\vec B_0$). 

Scalar multiplication of Eq.~\ref{mhd24} with $\vec k$ yields
\begin{equation}
\left(\omega^2-k^2v_A^2\right)\left(\vec k\cdot\vec v_1\right) = 0\,.
\label{mhd26}
\end{equation}
From this equation one can derive two magnetic wave modes:
\begin{itemize}
\item[(1)] Assuming $\nabla\cdot\vec v_1=0$ gives the so-called incompressible mode and from  the {\em ansatz} Eq.~\ref{mhd23} one gets $\vec k\cdot\vec v_1=0$. Thus, this mode is a transversal mode with the velocity perpendicular to the direction of propagation. From Eq.~\ref{mhd24} we have $\omega=\pm\cos\alpha\,v_A$. The waves are also called shear Alfv\'en waves. For $\alpha=0\degr$ one derives that $\vec B_1$ and $\vec v_1$ are parallel and the propagation is along $\vec B_0$.\\
\item[(2)] Another solution of Eq.~\ref{mhd26} is $\omega=kv_A$, independent of $\alpha$. These waves are compressional Alfv\'en waves, and for $\alpha=90\degr$ the velocity $\vec v_1$ is parallel to $\vec k$, i.e. we have longitudinal waves.
 
\end{itemize}

Finally, admitting that the gas pressure is not negligible, $p>0$, the phase velocity comes out as 
\begin{equation}
v_{ph}={\omega\over k}=\left[{1\over2}\left(c-s^2+v_A^2\right)\pm\left(c_s^4+v_A^4-2c_sv_a^2\cos2\alpha\right)^{1/2}\right]^{1/2}\,.
\label{mhd27}
\end{equation}
The `+' sign above gives the so-called `fast magnetoacoustic waves' and the `$-$'~sign the `slow magnetoacoustic waves'. Their phase speeds depend on $\alpha$ \citep[see the hodographs in][]{ferraro66,kippenhahn75,priest84}.

\newpage
\subsubsection*{Observational results\label{obswaves}}
\begin{figure}[t]
\center 
\includegraphics[width=\textwidth]{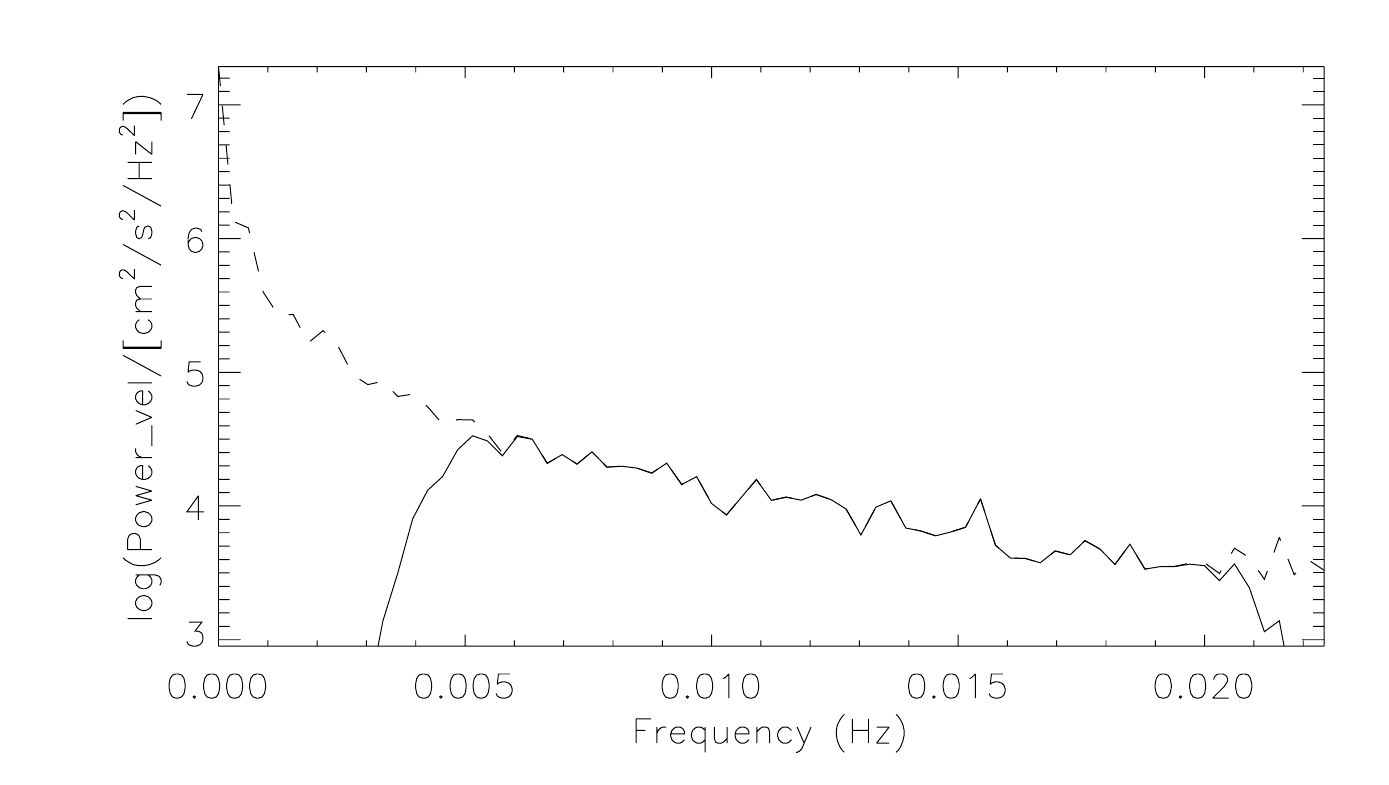} 
\caption{Average temporal power spectra of velocity from AOI C in Figs.~\ref{fig1} and \ref{fig2} before filtering (dashed) and after pass-band filtering (solid).}  
\label{fig6}
\end{figure}

The LOS velocities of the structures contain variations on long time scales of 10~min and longer as well as fluctuations with shorter time scales. To {distill} the latter, among them possibly magnetoacoustic waves, we applied a high-pass temporal filter and removed some high-frequency noise at the same time. The quantities then fluctuate about zero. Figure~\ref{fig6} depicts the average power spectra of the LOS velocities in AOI C in Fig. \ref{fig2} before and after filtering. We note that the 5-min oscillations are filtered out, while some oscillations at the acoustic cutoff (corresponding to periods of approximately 200~s) are partially retained. Yet the unfiltered and filtered power spectra in Fig.~\ref{fig6} do not show any predominant period.

\begin{figure}[]
\center 
\includegraphics[height=0.9\textheight,angle=0]{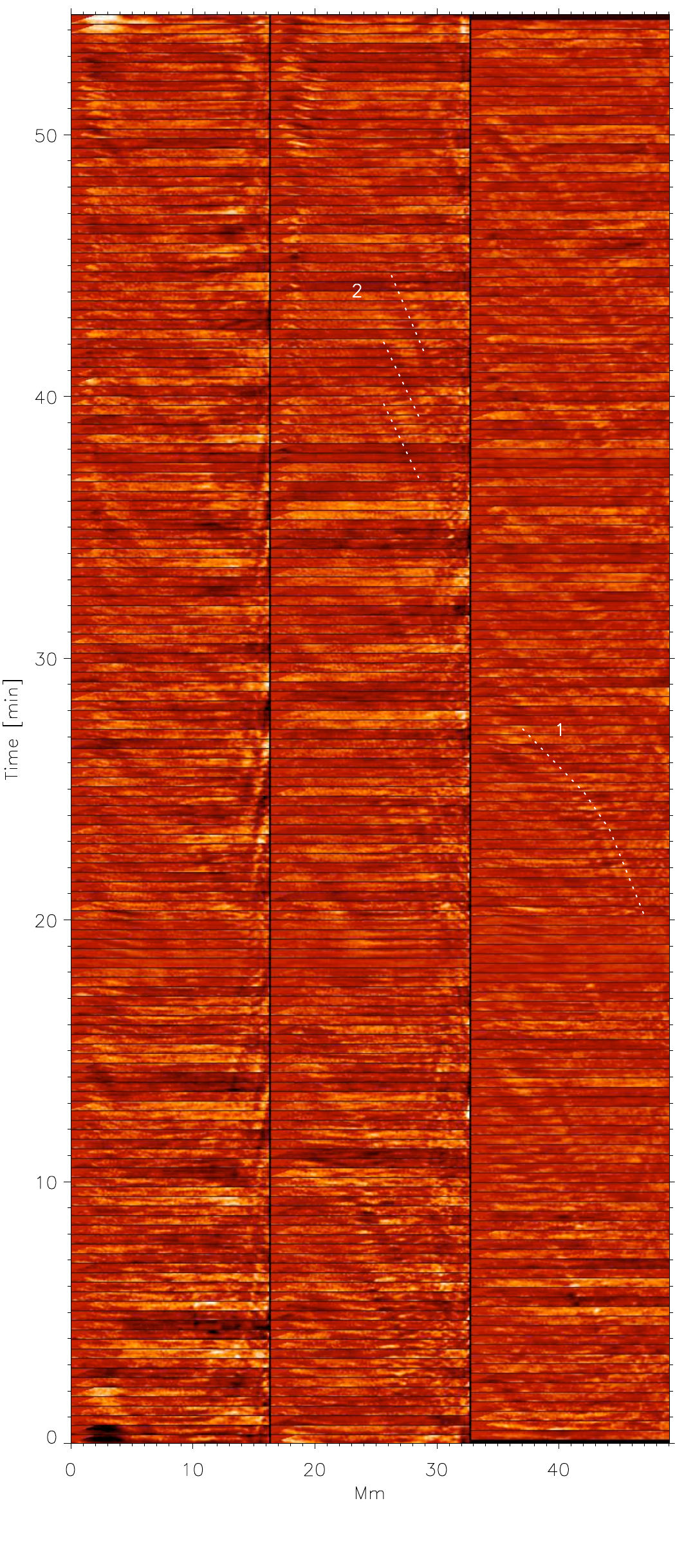} 
\caption{Example of space-time slices, of 1\farcs1 width, from AOI C in Figs.~\ref{fig1} and \ref{fig2}. From left to right: LOS velocity, H$\alpha$ line center intensity, and intensity differences at H$\alpha$~+0.5\AA\ off line center: $I_{0.5}(t_{i+1})-I_{0.5}(t_{i})$ with cadence of $\Delta t=t_{i+1}-t_{i}$\,= 22~s. The intensity differences in the right column are shifted up by 11~s. They are referred to as $\Delta I_{0.5}$ slices in the text.}  
\label{fig7}
\end{figure}

\begin{figure}[]
\center 
\includegraphics[width=0.67\textwidth,angle=0]{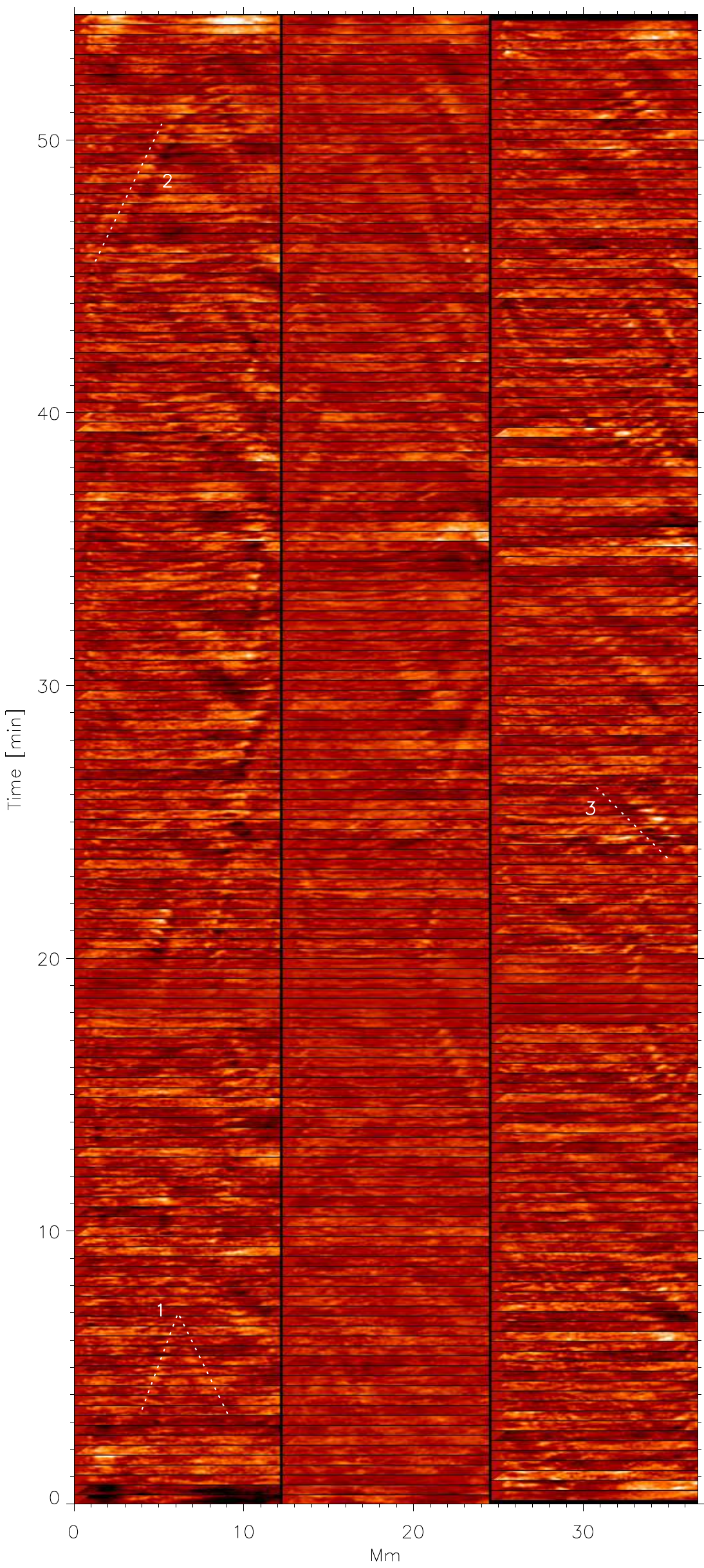} 
\caption{Example of space-time slices, of 1\farcs1 width, from AOI D. Same ordering as in Fig.~\ref{fig7}.}  
\label{fig8}
\end{figure}

\begin{figure}
\center 
\includegraphics[width=0.85\textwidth,angle=0]{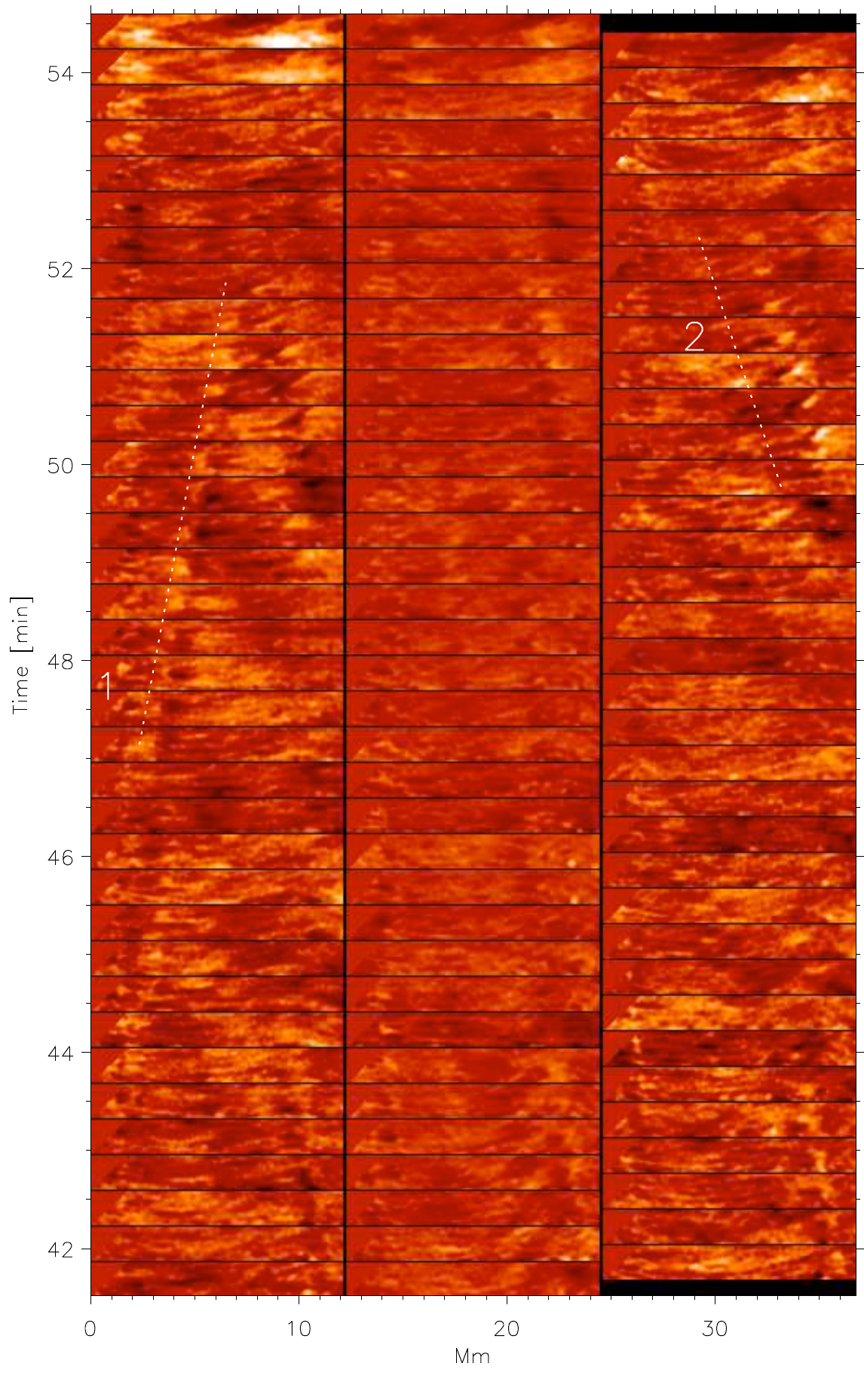} 
\caption{Selected part of space-time slices from AOI D with slice width of 2\farcs2. Same order as in Fig.~\ref{fig7}.}  
\label{fig9}
\end{figure}

\begin{figure}
\center 
\includegraphics[width=0.85\textwidth,angle=-0]{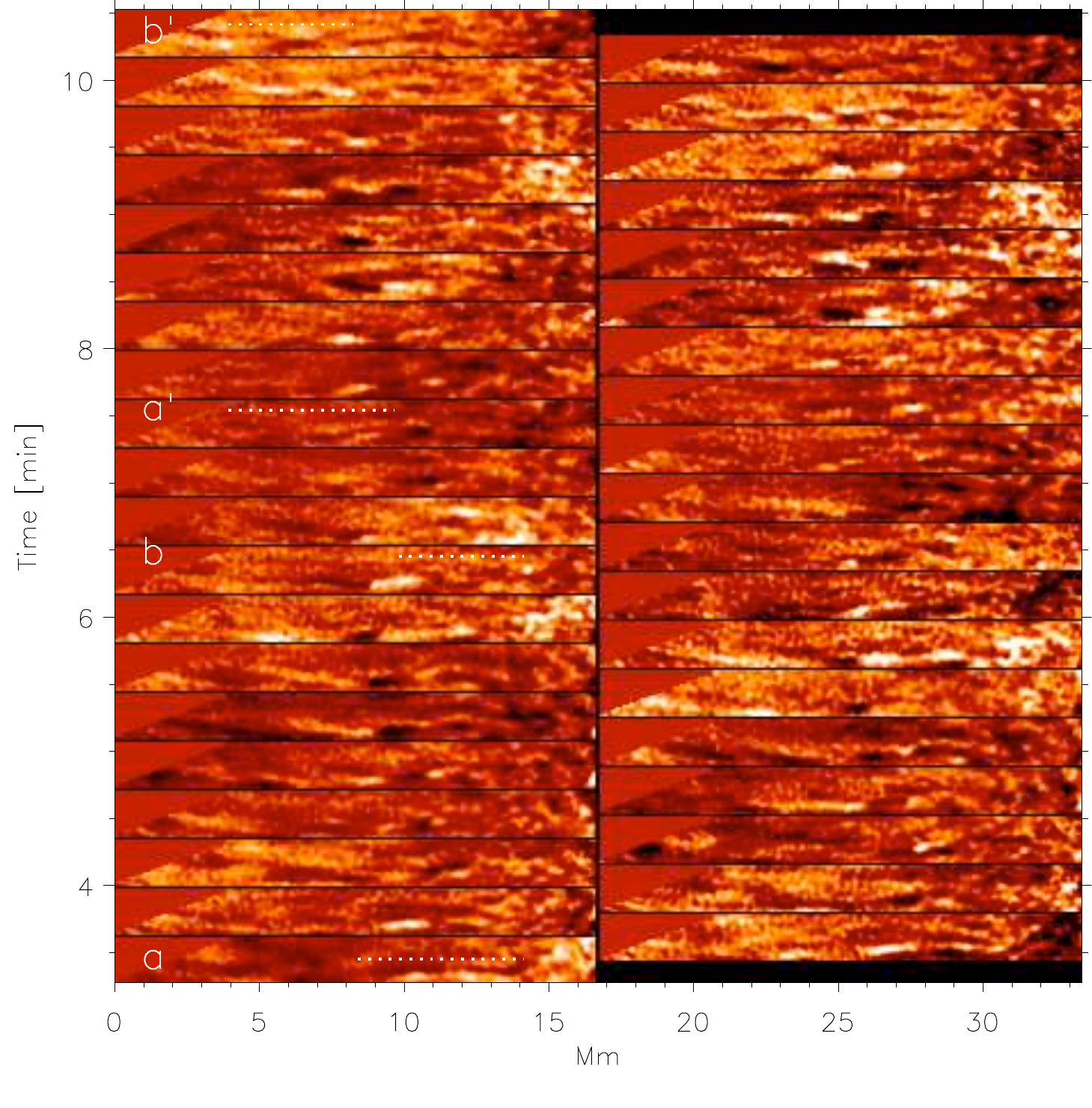} 
\caption{Selected part of space-time slices from AOI D with slice width of 2\farcs2. Left column: H$\alpha$ line center intensity fluctuations; right column: intensity differences at H$\alpha$ line center: $I_{LC}(t_{i+1})-I_{LC}(t_{i})$ with cadence of $\Delta t=t_{i+1}-t_{i}$\,= 22~s, referred to as $\Delta I_{LC}$ slices in the text.}  
\label{fig10}
\end{figure}

Figures \ref{fig7}--\ref{fig10} show examples of space-time slices from AOIs C and D. The {\em fluctuations} of several quantities are shown: 
\begin{enumerate}
\item
LOS velocities determined from differences of H$\alpha$ intensities at $\pm$0.5~\AA\ off line center, henceforth referred to as Doppler-gram slices (bright indicates velocity towards observer);
\item
H$\alpha$ line center intensities, henceforth LC slices;
\item
in Figs.~\ref{fig7}--\ref{fig9} differences of intensities at +0.5~\AA\ off line center $I_{0.5}(t_{i+1})-I_{0.5}(t_{i})$ with cadence $\Delta t=t_{i+1}-t_{i}$\,= 22~s, henceforth referred to as $\Delta I_{0.5}$ slices;
\item 
in Fig.~\ref{fig10} differences of intensities at line center $I_{LC}(t_{i+1})-I_{LC}(t_{i})$, henceforth referred to as $\Delta I_{LC}$ slices.
\end{enumerate}
Time runs from bottom to top with $t=0$ at the start of the series. The interruptions/interpolations at $t\approx$ 18.0--19.6~min and 37.5--38.5~min are obvious.


We focus attention to the oblique stripes in Figs.~\ref{fig7}--\ref{fig10}. These are the signatures of magnetoacoustic waves. From their slopes we can measure phase velocities projected on the plane {perpendicular} to the LOS. In Fig.~\ref{fig7} from AOI C, the waves appear to originate near the right edge of the AOI. This is one side at which the fibrils are rooted. Presumably, the waves are excited by the {buffeting} of motions at the photospheric foot points of the magnetic fields. As seen especially well in the Doppler-gram slices of Fig.~\ref{fig7}, but also in the LC slices, steep stripes originate from both sides of this region. The projected phase speeds are of the order of 8~km\,s$^{-1}$.

The stripes are often bent in the course of the temporal evolution, e.g. the wave parallel to the dashed line `1' in the $\Delta I_{0.5}$ slices of Fig.~\ref{fig7}. This wave starts off with a phase velocity of 14~km\,s$^{-1}$ and speeds up to approximately 40~km\,s$^{-1}$, one of the highest velocities measured.

A prominent period is not detected. Sometimes, the waves appear to be {repetitive}, with two or three, at most, wave trains in sequence with periods between 90~s and 180~s. An example of consecutive wave trains is indicated by the three dashed lines `2' in the LC slices of Fig.~\ref{fig7}. Yet most time, the waves are solitary, with one single wave package traveling across the FoV. Many of the waves appear to spread out along the direction of propagation and to fade after having traveled a distance of 5--10~Mm.

The amplitudes of the LOS velocities in the Doppler-gram slices are measured to approximately 1~km\,s$^{-1}$, be it in the waves with low phase speeds or in those with high phase speeds. With the calibration discussed above in the context of the cloud model (see Sect.~\ref{physpar}) these amplitudes have to be multiplied with a factor of approximately 3. The resulting {amplitudes} are thus of the order of 3~km\,s$^{-1}$, which is not a small perturbation compared with the sound speed (c.f. below the discussion on the magnetoacoustic waves).

Figure~\ref{fig8} from AOI D shows similar space-time slices as those from
AOI~C in Fig.~\ref{fig7}. Yet here, the waves are excited at both sides and
travel into the AOI, {sometimes} crossing from left and right
and possibly colliding as in {the example} parallel to the dashed lines `1' in
the Doppler-gram. The long lasting (more than 7~min), solitary wave train
(parallel to dashed line `2' in Fig.~\ref{fig8}) has a phase velocity of
approximately 13~km\,s$^{-1}$, a typical speed of the `slow' waves in this
AOI. A correction for foreshortening, i.e. that we see only the projection of
the phase speed on the plane perpendicular to the LOS is to be excluded since
the fibrils in AOI~D, as well as those in AOI~C, are oriented almost perpendicularly to the direction to the limb (see Fig.~\ref{fig2}), thus perpendicularly to the LOS. The wave parallel to the dashed line `3' in the $\Delta I_{0.5}$ slices of Fig.~\ref{fig8} gives a phase speed of 30~km\,s$^{-1}$, again not to be corrected for foreshortening. This is a typical phase speed of the fast waves.

Figure~\ref{fig9} gives another example of space-time slices from AOI~D, with wider slices of 2\farcs2 width and shorter time span of 13 min duration than in Figs.~\ref{fig7} and \ref{fig8}. The long lasting wave train in the Doppler-gram slices (parallel to dashed line `1') gives again the typical phase speed of 13.3~km\,s$^{-1}$ with (calibrated) LOS velocity amplitudes of approximately 2~km\,s$^{-1}$. These LOS velocities are transversal, in the sense that they are perpendicular to the propagation and to the H$\alpha$ fibrils. The `fast' wave in the $\Delta I_{0.5}$ slices (parallel to dashed line `2') exhibits also the typical phase speed of 32~km\,s$^{-1}$ with calibrated LOS velocities of approximately 1.5~km\,s$^{-1}$.

Figure~\ref{fig10} gives a 7.25~min long section of the temporal development of fluctuations in AOI C with slice widths of 2\farcs2, but this time the LC slices and the $\Delta I_{LC}$ slices only. Note that dark and bright features in H$\alpha$ LC indicate increased and decreased absorption, {respectively}, {\em not} enhanced and reduced temperature \citep[see][]{2004A&A...418.1131A,2006ApJ...648L..67V}. The two solitary waves between the pairs of horizontal dashed lines (a, a\arcmin) and (b, b\arcmin) have phase speeds of approximately 25~km\,s$^{-1}$. Inspection of the LC slices shows that the waves consist of elongated, thin blobs with length of 1\arcsec--2\arcsec\ and width of approximately 0\farcs5. Apparently, the waves do not travel in the spatial direction along straight lines, but along sinuous lines with deviations  from straight lines of approximately 0\farcs5 in amplitude. This suggests that on these small scales the magnetic field is not straight and homogeneous but entangled. 

The presentation of the difference slices $\Delta I_{LC}$ in Fig.~\ref{fig10} is prepared to study temporal displacements of absorption features. These are only seen if the displacements have a strong component perpendicular to the LOS. The bright and dark small-scale features, lying parallel and next to each other, in the upper part of the $\Delta I_{LC}$ slices, between the dashed line pair (b, b\arcmin) are suggestive of such displacements perpendicular to the direction of propagation.

We summarize in short the observational findings on magnetoacoustic waves:
\begin{enumerate}
\item
Generally, we find two kinds of waves: slow waves with phase velocities of 12--14~km\,s$^{-1}$ and fast waves with phase velocities of 25--33~km\,s$^{-1}$ (maximum velocity found 42~km\,s$^{-1}$). The waves appear to develop from low phase speed to high phase speed waves and vanish after having traveled a distance of 5--10~Mm.
\item
Irrespectively of the wave mode, the LOS gas velocities are of the order of 2--4~km\,s$^{-1}$.
\item
The waves are mainly solitary. They consist of short (1\arcsec--2\arcsec) and thin ($\approx0\farcs5)$ blobs of compressed gas.
\item
The waves appear to follow wiggly, entangled magnetic field lines with possible lateral displacements. 
\end{enumerate}


\subsubsection*{Interpretation -- waves in thin magnetic flux tubes\label{tubewaves}}

For the interpretation of the observations from AOI C and D, we adopt the picture of waves in thin magnetic flux tubes, whose radius is small compared to the pressure scale height. The propagation of waves in magnetic flux tubes were treated by, among others, \citet{defouw76}, \citet{wentzel79}, \citet{1982SoPh...75....3S}, and recently by \citet{mousielak07}. Spruit assumes a thin, cylindrical magnetic flux tube parallel to the $z$ axis, with radius $R$, magnetic field along the tube of strength $B$, gas pressure $p$, mass density $\rho$, and temperature $T$. The gravity is neglected. The tube is embedded in an external medium with properties $B_e$, $p_e$, $\rho_e$, and $T_e$. Inside and outside the tube the magnetic and atmospheric parameters are constant. In Spruit's \citeyearpar{1982SoPh...75....3S} work, the MHD equations are linearized and a mode analysis is performed, with proper conditions at the {interface} between flux tube and surrounding medium.

Incompressible Alfv\'en waves ($\nabla\cdot\vec v_1=0$, with small velocity perturbation $v_1$) are also possible in flux tubes. They are torsional Alfv\'en waves. The compressive solutions lead to 
\begin{equation}
\nabla\cdot\vec v_1=A\,{\cal B}_m(nr)\exp[i\,(\omega t+m\phi+kz)]\,,
\label{mhd28}
\end{equation}
with amplitude $A$, ${\cal B}_m(nr)$ Bessel functions of order $m$, $r$ the distance from the axis of the tube, and $\phi$ the azimuthal angle. Inside the tube, the waves propagate along the $z$ direction. For $n$ the relation holds
\begin{equation}  
n^2 = (\omega^2-v_A^2k^2)\,(\omega^2-c_s^2k^2)/[(\omega^2-c_t^2k^2)\,(v_A^2+c_s^2)]\,.
\label{mhd29}
\end{equation}
Here, the tube speed $c_t$ is introduced with
\begin{equation}
c_t^2={v_A^2c_s^2\over v_A^2+c_s^2}\,,
\label{mhd31}
\end{equation}
 which shows that the tube speed is smaller than both the Alfv\'en and the sound velocity.

\citet{1982SoPh...75....3S} showed that in the limit $k\,R\rightarrow0$ the mode with $m=0$ is a longitudinal mode with $v_{ph}=c_t$, which is {approximately} the sound speed $c_s$ for $v_A\gg c_s$. This mode is often referred to as the `sausage mode', with velocity inside the tube parallel to the magnetic field.

In the same limit and for $m>0$ one obtains the so-called `kink waves', with phase speeds related to the magnetic fields and densities through
\begin{equation}
v_{ph}^2={\rho v_A^2+\rho_e v_{A,e}^2\over\rho+\rho_e} = {1\over4\pi}\cdot{B^2+B_e^2\over\rho+\rho_e}\,.
\label{mhd32}
\end{equation}
These waves are transversal waves, and Spruit's \citeyearpar{1982SoPh...75....3S} analysis takes into account the dragging by the ambient medium. The phase speeds are obviously $v_{ph}^2=v_A^2$ for $\rho_e=\rho,\,\,B_e=B$; $v_{ph}^2=v_A^2/2$ for $\rho_e=\rho,\,\,B_e=0$, and $v_{ph}^2=2\cdot v_A^2$ for $\rho_e=0,\,\,B_e=B$.


We now compare the observations of waves with the expectation from this linear wave theory. We adopt that the waves propagate along the magnetic field and that the influence of gravity on the wave properties is {negligible}. The period at the acoustic cutoff of 200~s is longer than the periods, actually seen only rarely, in our data. Likewise, the period for the cutoff of kink waves \citep{spruit81,choudhuri93} is approximately 400~s, for small plasma $\beta$, which is the ratio of gas pressure to magnetic pressure, $\beta=(8\pi p)/B^2$.

With the parameters in Table~\ref{tabla} for the surge discussed above in Sect.~\ref{surge}, i.e. with gas pressure $p=0.38$~dyn\,cm$^{-2}$ and mass density $\rho=2.3\times10^{-13}$~g\,cm$^{-3}$, the sound velocity is $c_s=16.6$~km\,s$^{-1}$. From the determination of parameters in a wide range of chromospheric H$\alpha$ structures by \citet{tsiropoula97}, \citet{tsiropoula00}, and  \citet{2004A&A...424..279T} we obtain values of the sound speed in the range of 13.5--16.7~km\,s$^{-1}$. The widely found temperatures of $T=10^4$~K and the mean molar mass of 0.8 from the ionization equilibrium of hydrogen found from Table~\ref{tabla} and from the above works give a sound velocity of 14.4~km\,s$^{-1}$. The phase velocities of the slow waves observed here are compatible with these values, if one accounts for possible small projection effects and for a small reduction for the velocity of tube waves (cf. Eq.~\ref{mhd31}).

\citet{2004Natur.430..536D} adopted magnetic field strengths of the order of 100~Gauss in the chromosphere of active regions. With this value and the commonly found mass densities of 0.8--2.3$\times10^{-13}$~g\,cm$^{-3}$, the Alfv\'en velocity is $v_A=1\,000\dots600$~km\,s$^{-1}$, much higher than the velocities of the fast waves in the present observations. We believe, that 100~Gauss is an upper limit of the field strengths in the chromosphere of AOIs C and D. From high spatial resolution (approximately 0\farcs35) data from a plage region by \citet{Gonzalez:2007fk} we find an average field strength in the photosphere of 60--90~Gauss. This may possibly be reduced by a factor of 2 in chromospheric fibrils as in AOIs C and D by spreading out of the field lines over areas which possess little field in the photosphere. Otherwise the fibrils would not be so elongated. Yet still this yields to Alfv\'en velocities of $v_A\approx200$~km\,s$^{-1}$, as a minimum value.

\citet{1975SoPh...44..299G} has measured velocities of 70~km\,s$^{-1}$ in chromospheric H$\alpha$ structures. With a magnetic field strength of 10~Gauss and with reasonable particle densities he arrived at the Alfv\'en velocity in agreement with these measured phase velocities. In the present work, one would need field strengths as low as 5~Gauss for an Alfv\'en velocity of 32~km\,s$^{-1}$ as observed. We note that even with 5~Gauss the motions are still dominated by the magnetic field, i.e. $\beta\ll1$ holds.

We estimate the maximum phase speed measurable from our data to 250--300~km\,s$^{-1}$. Such velocities would still be detectable. The phase speeds found here are in the range 25--35~km\,s$^{-1}$. (The highest measured speed amounts to 42~km\,s$^{-1}$). These are obviously incompatible with Alfv\'en waves in a homogenous magnetic field with 30--100~Gauss. We mention several {possibilities} to reconcile our measurements with the picture of fast mode magnetoacoustic waves along the magnetic field, i.e. of Alfv\'en waves.
\begin{enumerate}
\item
The magnetic field strength in the fibrils of AOIs C and D is indeed as low as 5~Gauss, which is not very probable considering the very high activity in the whole area observed. AOIs C and D are not especially located at the outskirts of this activity.
\item
Propagation of a fast mode wave in a flux tube surrounded by a medium with low or zero field strength but with high gas density would reduce the phase speed (cf. Eq.~\ref{mhd32}).
\item
Apparently, the waves start as slow mode waves with phase velocities of the order of 10--14~km\,s$^{-1}$ and then are transformed into fast mode waves propagating with Alfv\'en velocity. Yet the transformation does not occur immediately. Examples are seen in Fig.~\ref{fig7}. While the solitary waves evolve into fast mode waves their wave packages get dispersed and they decay by spreading out along the direction of propagation.
\item
We do not measure phase velocities but group velocities of solitary wave packages. We have calculated for the slow mode the cusped surface of the wave front according to \citet[][cf. their Fig.~13]{ferraro66}, which is rotationally symmetric about the direction of the magnetic field. The adopted Alfv\'en and sound velocities were 200~km\,s$^{-1}$ and 16~km\,s$^{-1}$, respectively. The maximum velocity of this surface is only marginally larger than the sound speed by 3.2\%, and the maximum deviation from the direction of the magnetic field is 0\fdg01. Thus, the propagation of such slow mode pulses is practically along the magnetic field with the sound velocity. 

\item
The picture is actually more complicated: The waves with low phase speed seen here are not pure longitudinal waves. The gas velocities of the waves have a strong transversal component of the order of 3~km\,s$^{-1}$. Furthermore, the propagation of the fast waves deviates from straight lines, their motion appears more wiggly, possibly because the magnetic fields are entangled. Under the aspect of these observations the linear theory of small perturbations of straight flux tubes appears to be not sufficient.
\end{enumerate}

\newpage
\subsection{Summary on some properties of the active chromosphere\label{conclusions}}

Thanks to the good resolution, we could follow the evolution of small-scale chromospheric structures of an active region. From the rich dynamical processes in the observed, very active, flaring region some areas were selected for detailed investigation in the present work:
\begin{enumerate} 
\item
A small surge: It showed repetitive occurrence with a rate of some 10 minutes. The surge developed from initial small active fibrils to a straight, thin stucture of approximately 15~Mm length, then retreated back to its mouth to reappear again two times. The gas velocities reach approximately 100~km\,s$^{-1}$. The rebound shock model by \citet{1989ApJ...343..985S} seems to be a viable explanation.
\item
Two small-scale, synchronous, possibly sympathetic flashes, or mini-flares: In
a pair of small areas, two brightenings occurred simultaneously and
disappeared during two H$\alpha$ scans with total duration of
45~s. Presumably, the evolutionary time scale is much shorter, few to
10~s. Yet we could follow the evolution with a temporal resolution of 2~s by analysing H$\alpha$ filtergrams at different wavelengths. One of the two flashes showed an apparent proper motion with a speed up to 200~km\,s$^{-1}$, while it was developping a high emission, above the continuum intensity, in the red part of the H$\alpha$ profile. However, the cadence of the scanning was too slow to decide whether the temporal evolution consisted in a rapid horizontal proper motion with a final fast down flow or in a rapid change of emission at fixed local postitions.
\item
Magnetoacoustic waves in long fibrils: In two areas with long fibrils, the structures exhibited many magnetoacoustic waves running parallel to the fibrils, thus presumably also parallel to the magnetic field. The waves are mostly solitary. Few times, two or three repetitive wave trains could be seen with periods of 100--180~s. The waves start at the footings of the fibrils with a speed of 12--14~km\,s$^{-1}$, which is not much lower than the sound speed estimated for such structures and similar to the tube speed. Most of the waves get accelerated to reach phase speeds of approximately 30~km\,s$^{-1}$. Then they spread out along the fibrils and fade. The final phase speed is much lower than the Alfv\'en speed of $\ge200$~km\,s$^{-1}$, estimated from reasonable magnetic field strengths in the active region chromosphere of 30--100~Gauss and reasonable mass densities in the fibrils of 2$\times10^{-13}$~g\,cm$^{-3}$. Furthermore, we observe that the slow waves have strong transversal (LOS) velocity components with $\sim$3~km\,s$^{-1}$, i.e. are not purely longitudinal, and that the fast waves consist of short (1\arcsec--2\arcsec), thin ($\sim$0\farcs5) blobs and apparently move along sinuous lines. We conlude from these findings that a linear theory of wave propagation in straight magnetic flux tubes is not sufficient. 
\end{enumerate}

\clearpage

\section{Comparison between speckle interferometry and blind deconvolution \label{sec:comp}}

In Sec. \ref{seeing} we introduced the image degradation problem due to atmospheric distortions for all ground based solar observations and astronomical observations on general. In Sec. \ref{obs:kaos} we explained the adaptive optics approach used at the VTT to reduce the image distortions in real time. Finally, in Sec.  \ref{datared} we described the basis of two different \emph{post factum} image reconstruction techniques,  the Speckle Imaging (SI) and one type of Blind Deconvolution (BD) with  simultaneous Multiple Objects and Multiple Frames (MOMFBD).
 
In the case of data from the G-FPI, the SI approach reconstructs separately the broadband images and uses both the original recorded frames (after dark subtraction and flat fielding) and the reconstructed image from this channel to obtain the reconstructed narrow-band images at the various positions along the spectral line.

The MOMFBD code applied to our G-FPI data uses at the same time, for each spectral position, the various (15) pairs of simultaneously recorded broadband and narrow-band frames. Thus, at 21 wavelength positions along the spectral line, two different objects were observed and their images were reconstructed with 15 frames per object.

In this thesis work we used the SI method for fields of view on the solar disc. The last version of the code takes into account the field dependence of the PSF around the AO lockpoint, so we will refer to this version as SI+AO. However, data frames on and off the limb cannot be reconstructed with the current code. Near the limb the contrast is lower than near disc center, which makes any reconstruction more difficult. Moreover, KAOS can lock on the low-contrast structures near the limb only under very good seeing conditions.  Also, the limb darkening at large heliocentric angles makes it difficult to determine the STF on the rings around the lockpoint for the SI+AO. Beyond the problems inside the limb, off-the-limb emission features seen in the narrow-band images lack broadband counterparts, and therefore there exist no simultaneous data from which we could apply the second part of the SI to reconstruct the off-limb parts of the images.

Dataset \emph{limb}  (results in Sec. \ref{sec:limb:ha}) was recorded under very good seeing conditions, with KAOS locked on a nearby facula correcting 27 (Zernike polynomial) modes most of the time. For the {\em post factum} image reconstruction we used the MOMFDB, for which the limb darkening presents no problems. Spicules above the limb do not posses a simultaneously observed  broadband object, so it is expected that their spatial resolution is lower, since there are no multiple objects, just the multiple frames for each spectral position in the narrow-band channel.

In order to compare results from both approaches we have reconstructed with both methods the same field of view on the disc, a frame of the \emph{sigmoid} data set. In this Section we present the comparison. As we show, for the BD case we made two different reconstructions with different limits of the expansion of the aberration in Karhunen-Loeve (K-L) modes. In the first case the expansion of the wavefront aberration is done until de $17^{th}$ K-L polynomial, while on the second run we expanded to the first 100 modes. The running time of the code is highly sensitive to the number of modes, being slower with more modes.

\begin{figure}[p]
\begin{center}
\includegraphics[width=0.85\textwidth]{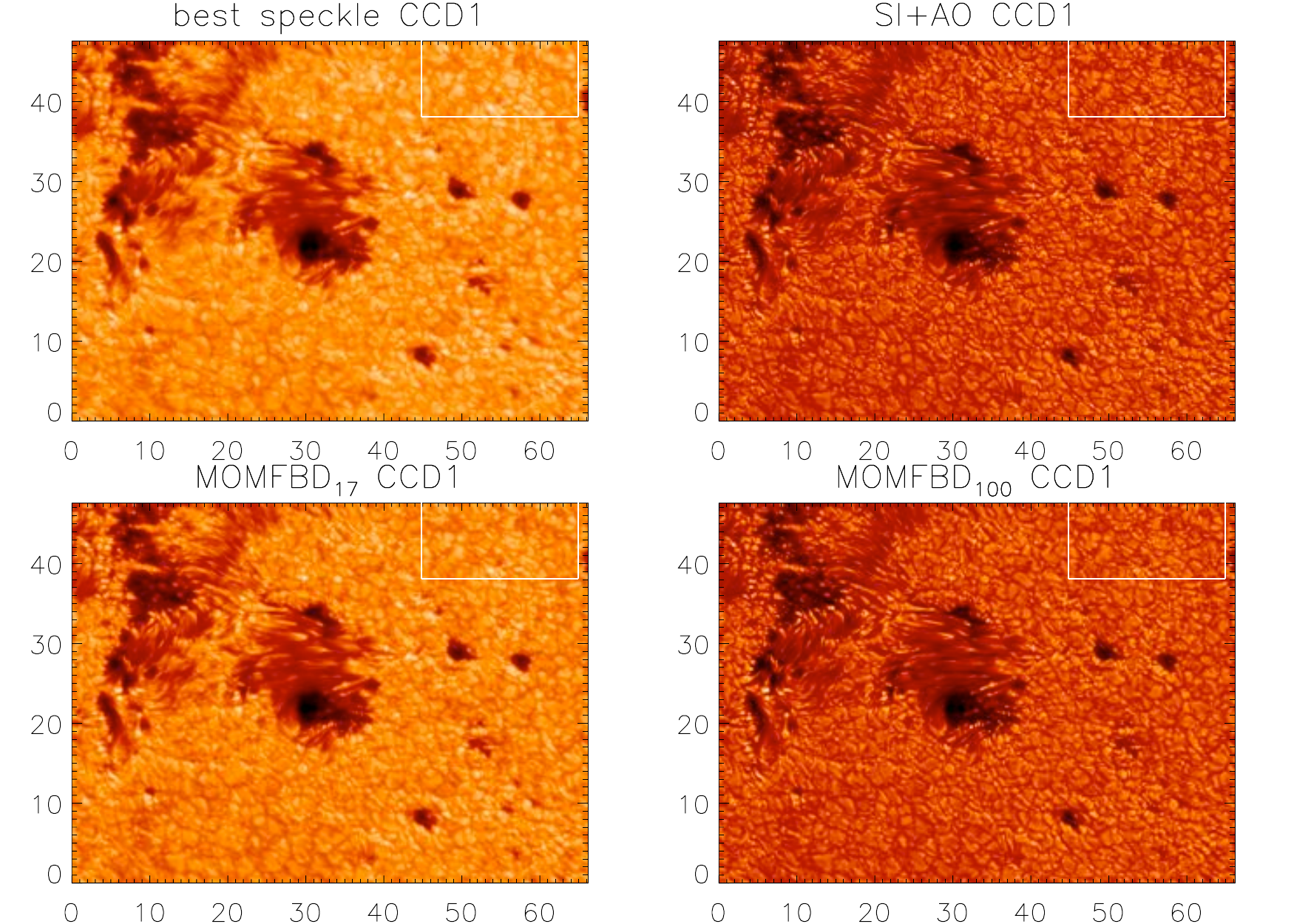}
\caption{Results for different image reconstruction techniques. Axes are in arcseconds. White rectangles enclose the region where the RMS contrast is calculated. \emph{Upper left}: Best speckle raw image, contrast is 5.9\%. \emph{Upper right}: SI+AO result, contrast is 11.1\%. \emph{Lower left}: MOMFBD running 17 modes, contrast is 6.7\%. \emph{Lower right}: MOMFBD running 100 modes, contrast is 10.0\%.}
\label{com:ima1}
\end{center}
\begin{center}
\includegraphics[width=0.85\textwidth]{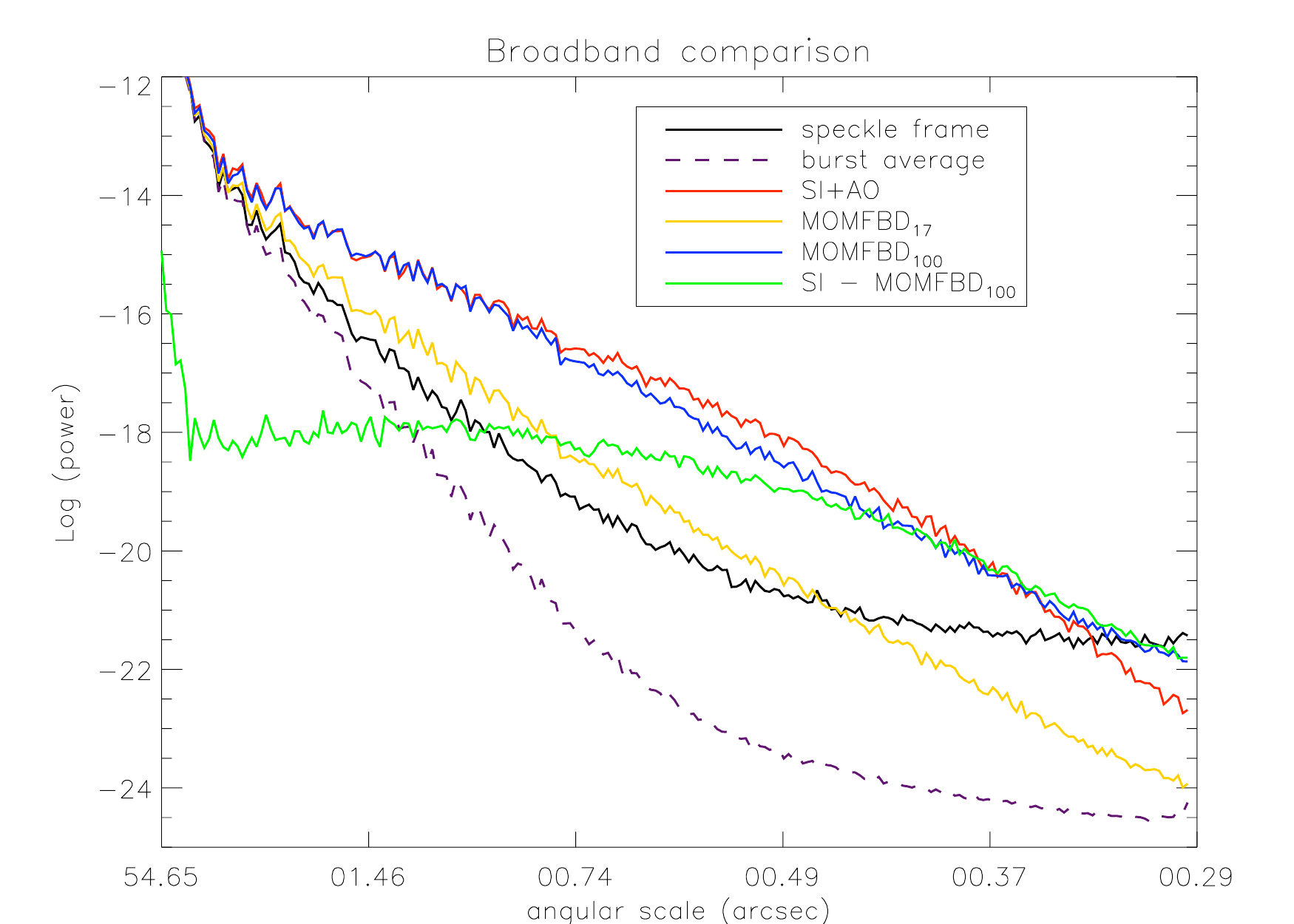}
\caption{Comparison of the power spectra (in arbitrary units) of the same image for different  image reconstruction techniques.} 
\label{com:pow1}
\end{center}
\end{figure}  

\begin{figure}[p]
\begin{center}
\includegraphics[width=0.9\textwidth]{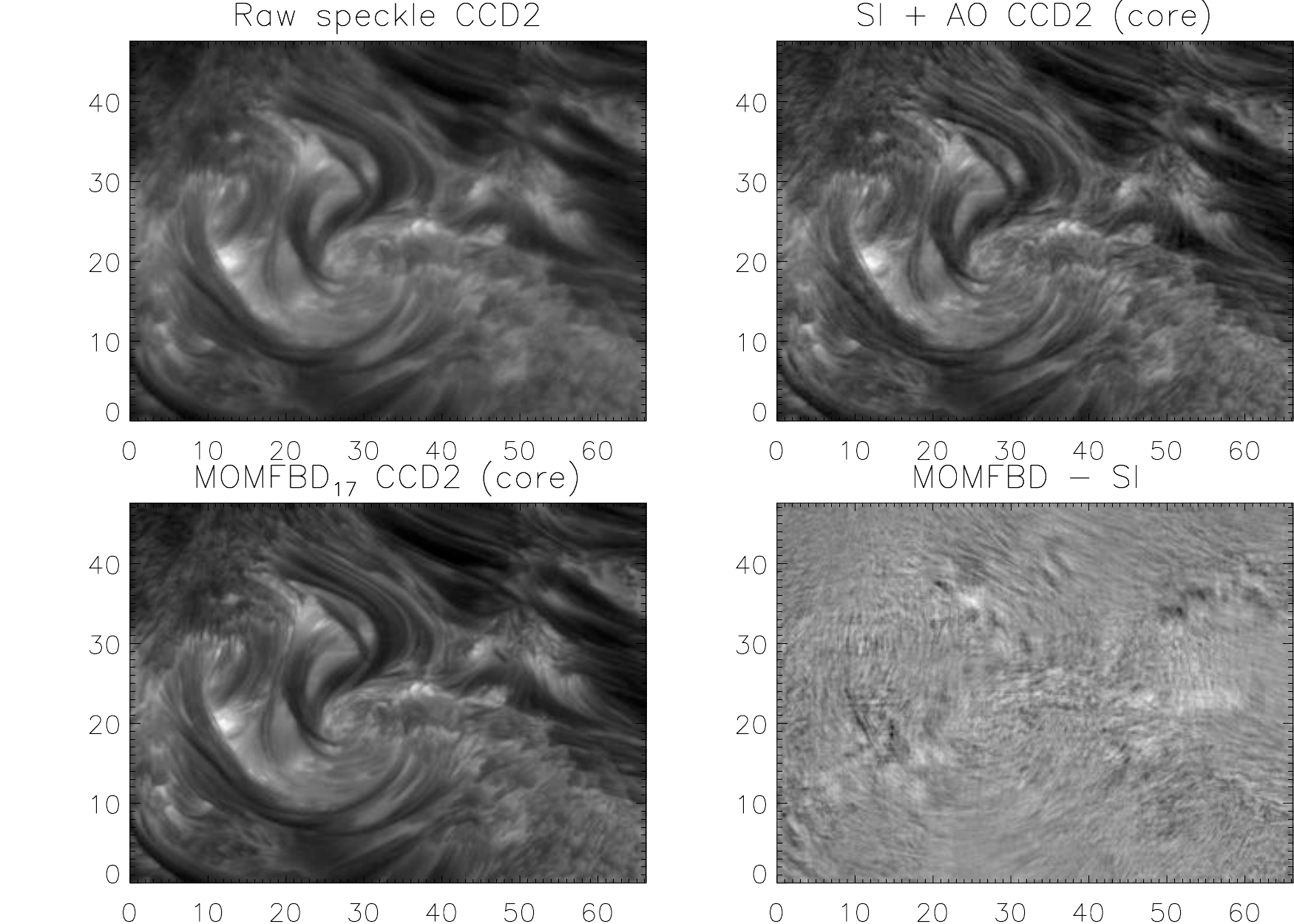}
\caption{Results for different image reconstruction techniques for the line center narrow-band filtergram. Scales on the axes are in arcseconds. \emph{Upper left}: Best speckle raw image.\,\emph{Upper right}: SI+AO result. \emph{Lower left}: MOMFBD running 17 modes. \,\emph{Lower right}: Image difference. In this case the differences reach 45\% of the fluctuations in the reconstructed frames.}
\label{com:ima2}
\end{center}
\vspace{-1cm}
\begin{center}
\includegraphics[width=0.9\textwidth]{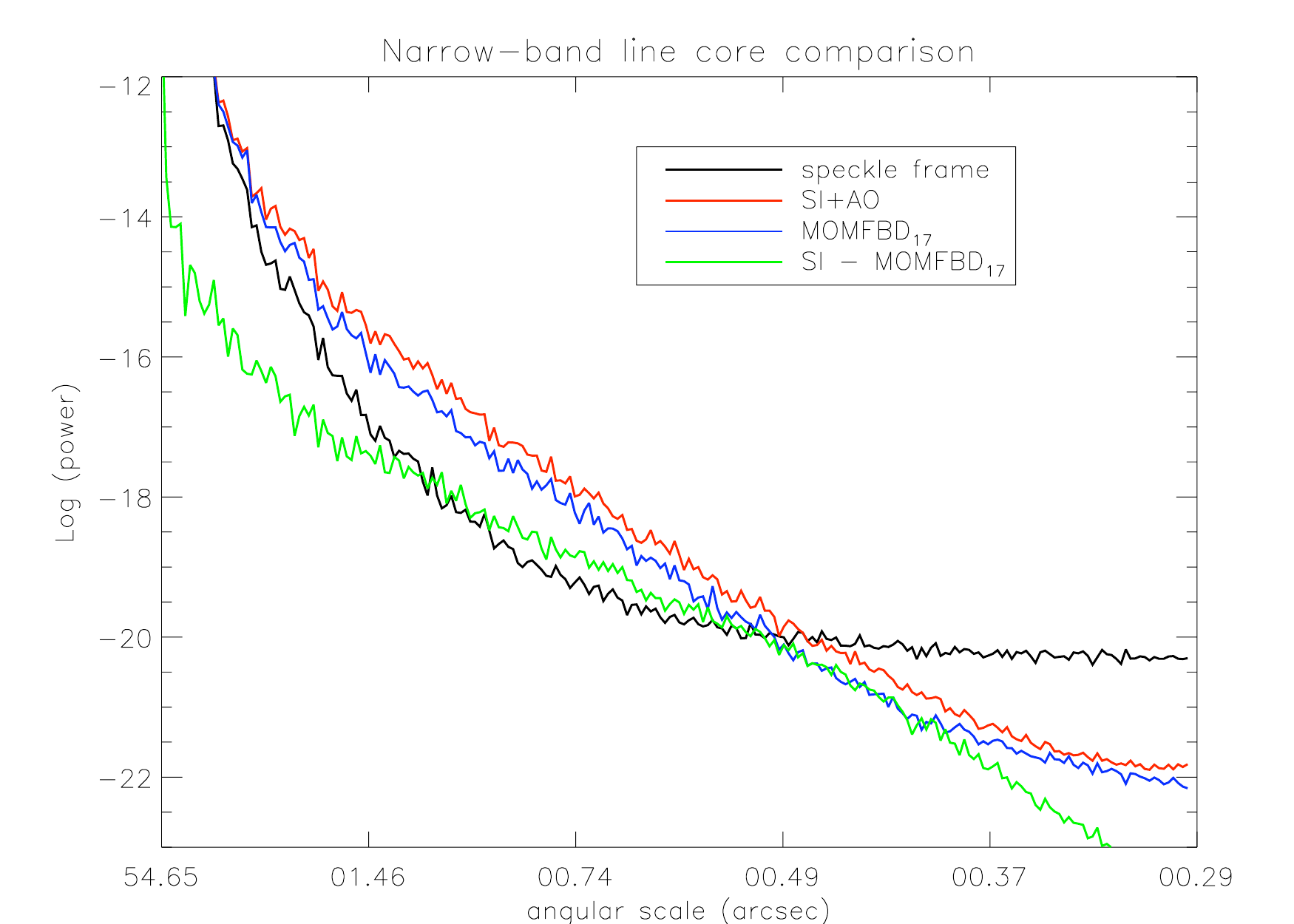}
\caption{Comparison of the power spectra of the same image for different image reconstruction techniques.}
\label{com:pow2}
\end{center}
\end{figure}  

\subsubsection*{Broadband}

Figure \ref{com:ima1} compares the full FoV image, while Fig.  \ref{com:pow1} shows the corresponding power spectra. The speckle frame corresponds to the one with highest rms contrast (5.9\% inside the white rectangle). The SI+AO reconstruction shows a much higher resolution, with more power at all frequencies than the speckle frame for angular scales larger than  $\sim 0\farcs32$. The reconstructed image has also less noise than the speckle frame (small-scale end of the power spectra). The rms contrast of granulation for this image is 11.1\%. In the case of the MOMFBD with 17 modes the power of the reconstructed image is significantly lower than for SI+AO, albeit having a lower noise level (comparable even with the burst average). We have to run up to 100 modes to arrive at a similar contrast as for SI+AO. The noise threshold for the last run, coincides with that of the SI+AO approach. The rms contrast of granulation for these images are 6.7\% with 17 modes and 10.0 \% with 100 modes.

Figure \ref{com:pow1} shows also the power spectrum of the difference between SI+AO and MOMFBD$_{100}$ (green line), which is many orders magnitude lower than one of the power spectra themselves. Only at scales smaller than $\sim0\farcs4$, the difference becomes of the same order as the power spectra. Taking the differences of the reconstructed images shows that $99.8\%$ of the pixels in the FoV have intensity fluctuations lower than $15\%$ of the intensities in the images themselves. 
%

\subsubsection*{Narrow-band}
\begin{figure}[b]
\begin{center}
\includegraphics[width=\textwidth]{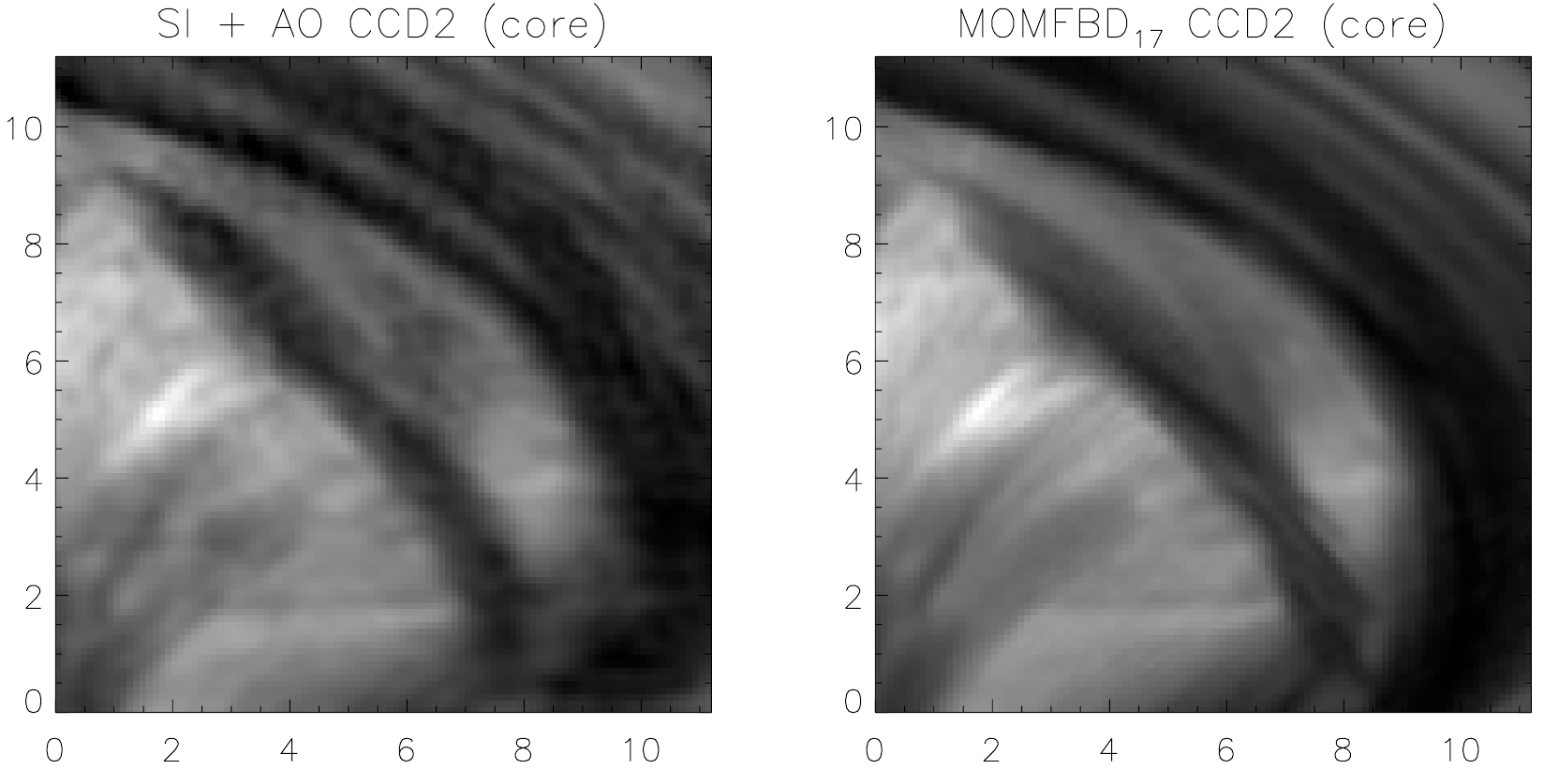}
\caption{Close-up subfield from the narrow-band spectrogram at the H$\alpha$ core. Axes are in arcseconds.}
\label{com:ima2b}
\end{center}
\end{figure}

The narrow-band images have lower intensities than the broadband images, especially at the core of the H$\alpha$ line. Much less images are used for reconstruction, so a lower resolution is expected. Figure \ref{com:ima2} compares the images at the H$\alpha$ line center, while Fig.  \ref{com:pow2} shows the corresponding power spectra. The SI + AO reconstruction shows a higher resolution than the speckle frame, with more power at scales  larger than  $\sim 0\farcs5$. At smaller scales, the speckle frame is dominated by noise. The MOMFBD with only 17 modes gives already a similar resolution than the SI+AO and better treatment of the noise. Fig \ref{com:ima2b} shows a close-up region where the better noise treatment of the MOMFBD is clearly visible.

The difference between the methods is bigger than in the broadband case, as expected since the intensity and resolution are lower. Nonetheless the agreement is very high, 92.8\% of the pixels in the difference image have amplitudes smaller than $0.15$ of the average intensity in the reconstructed images (similar results are found for other spectral position, reaching 99\% in the wings, at wavelengths $\pm 1 $ \AA\, off the line center).

\subsubsection*{Conclusions}
In this Section we have shown the good convergence  of both post-processing approaches. Using different techniques we arrive at similar results and spectral profiles. The amplitudes of the difference images are lower than $0.15$ of the average image intensity in more than 99\% of the broadband and above $\sim90$\% for the narrow-band images. In the case of the broadband reconstruction it was necessary to use 100 modes for the MOMFBD method to reach similar results as for SI+AO, while, in the narrow-band case, already with 17 modes the MOMFBD gives better images than SI+AO.

The main disadvantage of BD methods is the computational load. The reconstruction of the single data set from broad and narrow-band and only 17 modes takes $\sim7$ hours to process with 20 CPU cores of $3.2$GHz. For the 100 modes run, given the limited resources, we only used the broadband frames (Multi Frame BD). If the data set ``sigmoid'' were reconstructed with BD methods, even with only 17 modes, it would have taken around 130 days on our computing resources.

The main advantage of the BD is its ability to reconstruct an image even with only few frames. This is of special importance when observing fast evolving targets. The SI needs much more frames. The property of reconstructing \emph{simultaneously} recorded images from different ``objects'' (e.g. broadband and the H$\alpha$ narrow-band) leads to a perfect sub-alignment of the results, which avoids spurious signals in derived quantities. Note however that, not simultaneously observed objects, like in the several consecutive scans with the G-FPI, are not aligned since they are not recorded under identical \emph{seeing} conditions.

The SI+AO method is considerably much faster, around 10 and 15 minutes for the broadband and narrow-band images, respectively, with the same computers used for the MOMFBD reconstruction and gives better results for the broadband reconstruction, even using 100 modes in the latter method. However, with MOMFBD, the resolution and treatment of the noise is better in the narrow-band case. The main current advantage of the BD methods for our work and data is the possibility of reducing narrow-band limb and off-the-limb data scans.

Anisoplanatism is an issue common to both approaches. In both cases the large FoV is divided into smaller subfields, where the assumption of isoplanatism is valid. It is therefore important to address this point for both cases. The image difference does not show any subfield pattern. However, there can still be some small effects. For this reason we have used the integrated contrast profile of the difference, defined as
\begin{equation}
\mathbb{CI}=\sum_{\lambda} \Big | \frac{I_{SI+AO}(\lambda)-I_{MOMFBD_{17}}(\lambda)}{I_{SI+AO}(\lambda)} \Big|\, ,
\end{equation}
where $I_{SI+AO}(\lambda)$ and  $I_{MOMFBD_{17}}(\lambda)$ correspond to the reconstructed images using the SI+AO method, and  to the images using MOMFBD with 17 modes, respectively.

$\mathbb{CI}$ qualitatively measures the total difference between the profiles. If they were equal, then $\mathbb{CI}$ would be 0, while an increasing difference in the profiles increases the value of $\mathbb{CI}$. Since the subfield locations are the same for all the spectral positions, this information is added along the scan, while the intensities of the structures at each position are essentially subtracted out. The subfield pattern does not disappear with the subtraction of images reconstructed with different methods since they do not necessarily coincide.

\begin{figure}[]
\begin{center}
\includegraphics[width=\textwidth]{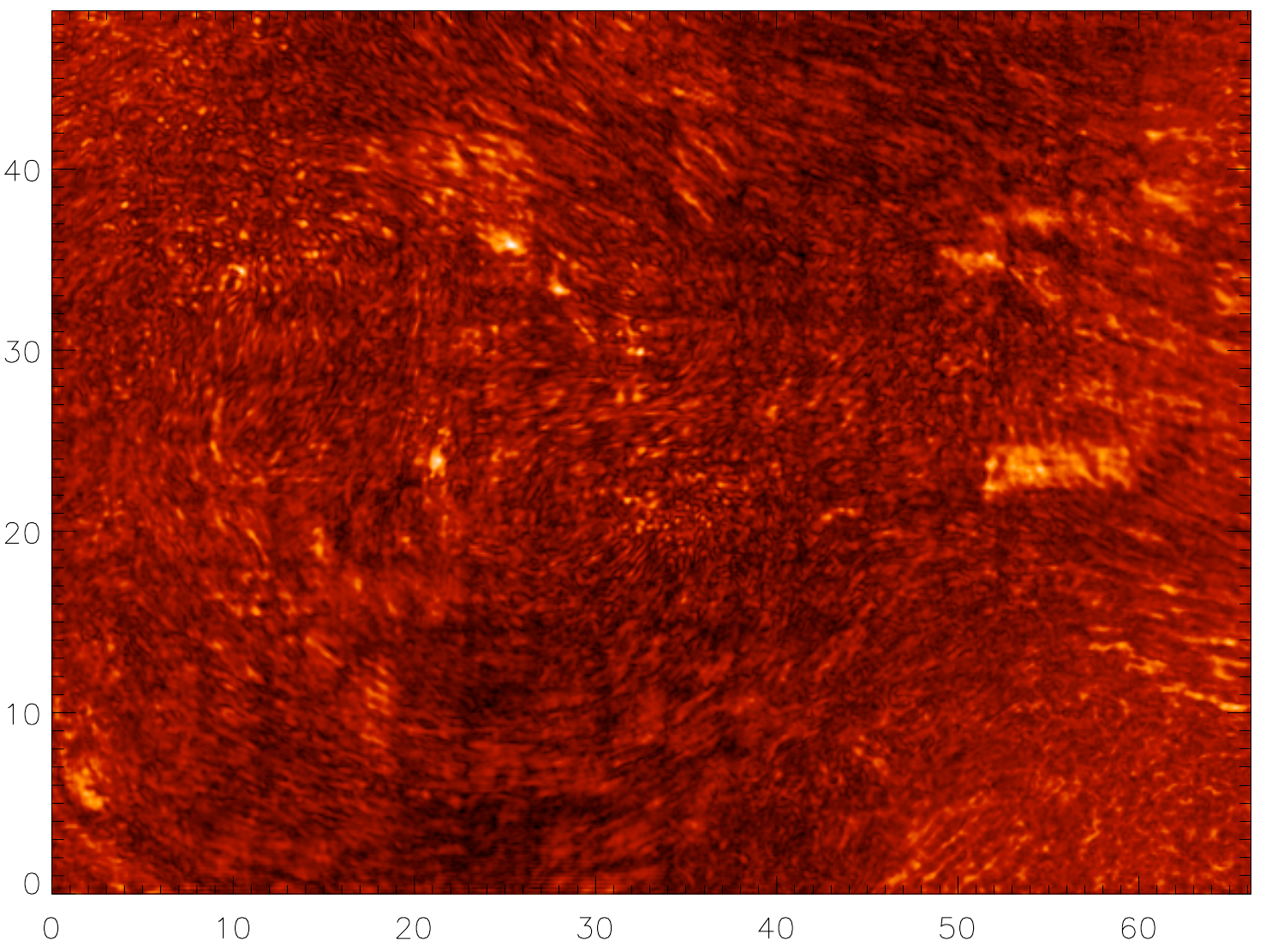}
\caption{Isoplanatic subfield array pattern when calculating $\mathbb{CI}$. The mean edge length of the squares is approximately 32 pixels. Axis scale is in arcseconds.}
\label{com:prof}
\end{center}
\end{figure}

Figure \ref{com:prof} shows the calculated $\mathbb{CI}$. The weak subfield pattern is revealed, especially in regions where the difference is low (dark background). The mean edge length of the squares is approximately around 32 pixels. 

The amplitude of the grid pattern is very low, only revealed after the calculation of $\mathbb{CI}$. Presumably this comes from the apodization. When joining common regions on overlapping subfields, the common parts are overlaid in the final image. This, while preserving the structures, reduces the noise, which leads to slightly smaller noise levels in these overlapping lanes. This grid is common for all wavelength positions. The difference between the methods is low enough to reveal this small decrease of the noise level (leading to darker areas in Fig. \ref{com:prof}) when the total effect is calculated, by using the $\mathbb{CI}$ parameter. Therefore, regions with more contrast, where the difference between SI and BD is bigger, the presence of this pattern is masked, as shown in the figure. 

\chapter[Spicules at the limb]{Spicules at the limb \label{ch:spicules}\Large{\protect\footnote{Contents from this Chapter have been partially published as \cite{2007A&A...472L..51S}}}}

Spicules, known for more than 130 years \citep[see the hand drawings by][]{secchi1877}, represent a prominent example of the dynamic
chromosphere. We refer the reader to reviews by \citet{beckers68,beckers72} and to the paper by \citet{wilhelm00} on UV properties. According to these works, spicules are seen at and outside the limb of the Sun as thin, elongated features. They develop speeds, measured from both proper motion and Doppler shifts, of 10--30 km\,s$^{-1}$ and reach
heights of 5--9 Mm on average, during their lifetimes of 3--15 minutes. Recent observations from HINODE \citep{2007SoPh..243....3K} and own results presented below in Sec. \ref{sec:limb:ha} have changed the traditional picture. Some spicules live for only few seconds, and spicules may be much more inclined with respect to the vertical than adopted hitherto.  

As pointed out by \citet{sterling00}, a key impediment to develop a satisfactory understanding has been the lack of reliable observational data. 
Many theoretical models have been proposed to understand the nature of spicules, using a wide variety of motion triggers and driving mechanisms. 
In this Chapter we focus on the \ion{He}{i} 10830\,\AA\ triplet emission line (see Sec. \ref{sec:limb:he}), using recent technical improvements in observational facilities, and on the results from the limb observations in H$\alpha$.

\section{Spicule emission profiles observed in \ion{He}{i} 10830 \AA\label{spicules}}

The energy levels that take part in the   \ion{He}{i} 10830 \AA\, triplet are basically populated via
an ionization-recombination process \citep{1994isp..book...35A}. The EUV coronal irradiation (CI) at 
wavelengths $\lambda<504$~\AA\ ionizes the neutral helium, and subsequent recombinations of singly ionized helium with free electrons lead to an overpopulation of all ortho-helium levels. Alternative theories suggest other mechanisms that might also contribute to the formation of the helium lines relying on the collisional excitation of the electrons in regions with higher temperature \citep[e.g.,][]{1997ApJ...489..375A}.  We are able to provide observational evidence of the link between the corona and the infrared emission of this line, in the frame of the current theoretical models of the solar atmosphere.
 
\citet{Centeno06} modelled the ionisation and recombination processes using various   
 amounts of CI, non-LTE radiative transfer, and different atmospheric
 models {  \citep[see also][]{cente07}}.
 They have simulated limb observations for different heights, obtaining
 synthetic emission profiles in spherically symmetric models of the solar atmosphere. One conclusion of their  study is that the ratio of intensities $({\cal R}=I_{\rm blue}/I_{\rm red})$ of the `blue' to the `red' components of the \ion{He}{i} 10830~\AA\ emission
 is a very good candidate for diagnosing the CI. The population of the metastable level depends on optical thickness, whose variation with height governs the change in the ratio $\cal R$ as a function of the distance to the limb.
 
\citet{truj05} measured the four Stokes parameters of quiet-Sun chromospheric spicules and could show evidence of the Hanle effect by the action of inclined magnetic fields with an average strength of the order of 10 G. They modelled the \ion{He}{i} 10830~\AA\
profiles assuming the medium along the integrated line of sight as a slab of constant properties and with its optical thickness as a free parameter. \citet{truj05} showed that the observed intensity profiles and their ensuing $\cal R$ values can be reproduced by choosing an optical thickness significantly larger than unity. \citet{Centeno06} demonstrated that this optical thickness is related to the coronal irradiance (through the ratio $\cal R$), thus providing a physical meaning to the free parameter in the slab model (see also Centeno et al. 2007).
\begin{figure}[t]
\center \includegraphics[width=\textwidth]{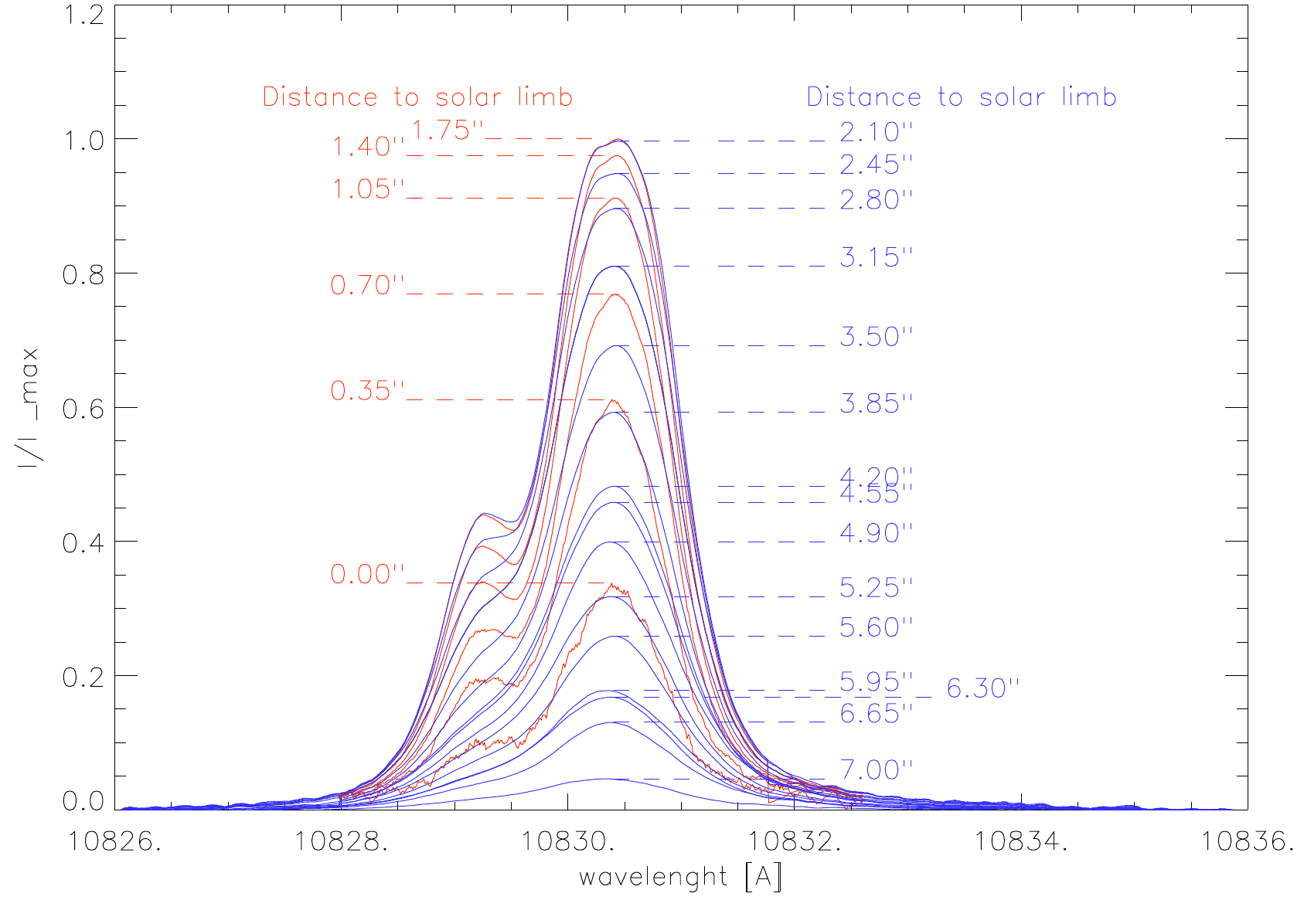} 
\caption{Measured \ion{He}{i} 10830~\AA\ emission profiles for increasing
  distances to the solar limb, scanning a broad range of the height extension of the 
  spicules. Each profile is the average of the 312 pixels along the slit (which was always kept parallel to the limb).}  
\label{fig:spe}
\end{figure}

\subsection{Observational  
 intensity profiles and intensity ratio}

We present novel observations showing the spectral emission of \ion{He}{i} 10830~\AA\ and its dependence on the height of the
spicules above a quiet region. We compare the deduced observational $\cal R$ with that obtained from detailed non-LTE numerical calculations using available atmospheric profiles. 

\begin{figure}[t]
\center \includegraphics[width=\textwidth]{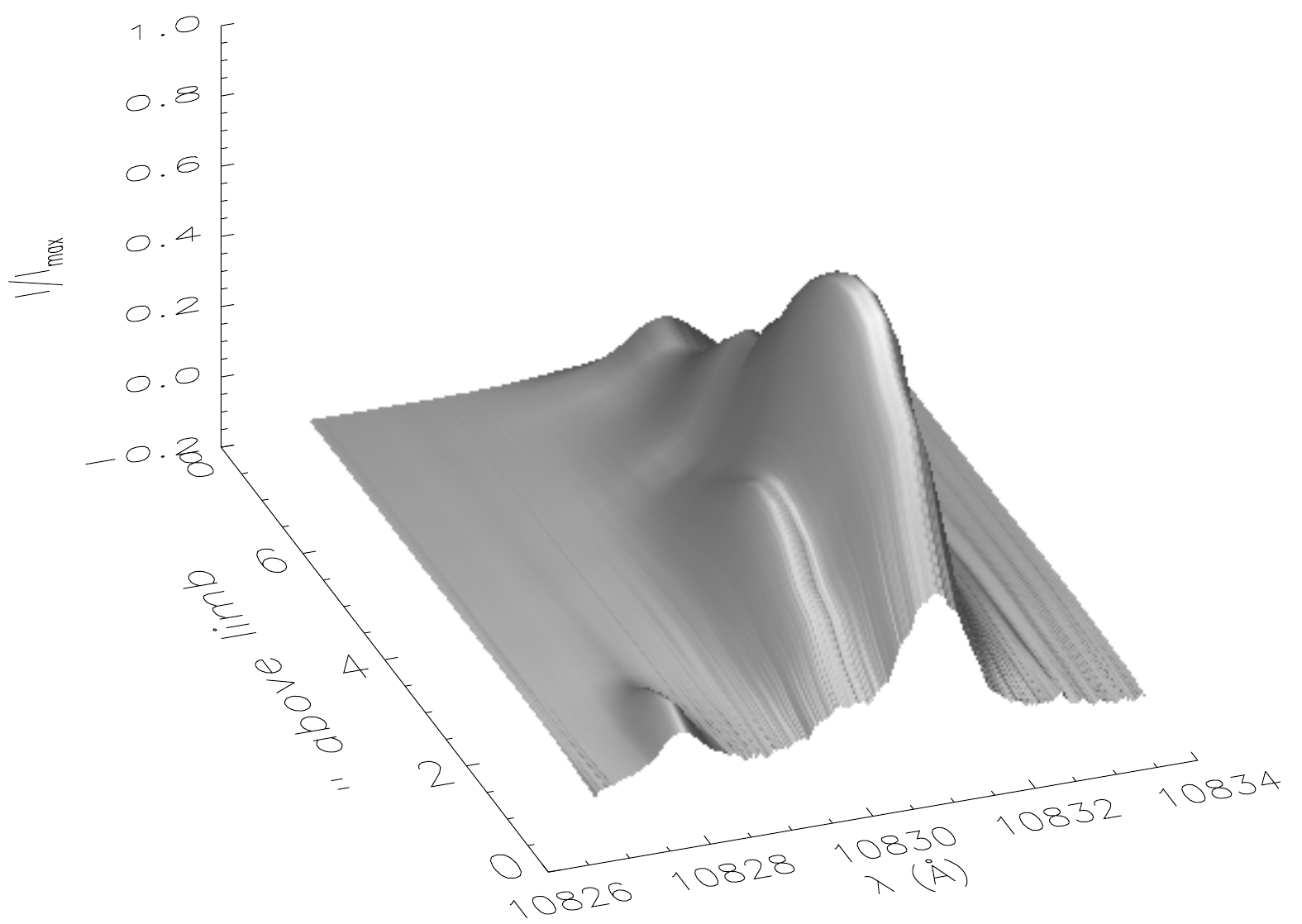} 
\caption{3D representation of the measured \ion{He}{i} 10830~\AA\ emission
  profiles for increasing distances to the solar visible limb. Note that the  x-axis is
  wavelength, the y-axis the height above solar limb and the z-axis the intensity
  normalised to the maximum emission in the line center of the red component.}
\label{fig:3d}
\end{figure}

These data correspond to the data set described in Table \ref{table:obs:tip}. After the standard reduction process (Sec. \ref{tip:reduc} ) we obtain 21 intensity profiles above the infrared limb, with a step size of $0\farcs35$. Figure~\ref{fig:spe} shows the emission profiles of the \ion{He}{i} 10830~\AA\, (after the reduction process) for different heights above the limb. 
Figure~\ref{fig:3d} illustrates this in three dimensions, as a function of
wavelength and the distance to the solar limb, clearly showing 
a change in the intensity ratio of the blue and red components of the 
multiplet $({\cal R}=I_{\rm blue}/I_{\rm red})$ with height. 

\begin{figure}[t]
\center \includegraphics[width=\textwidth]{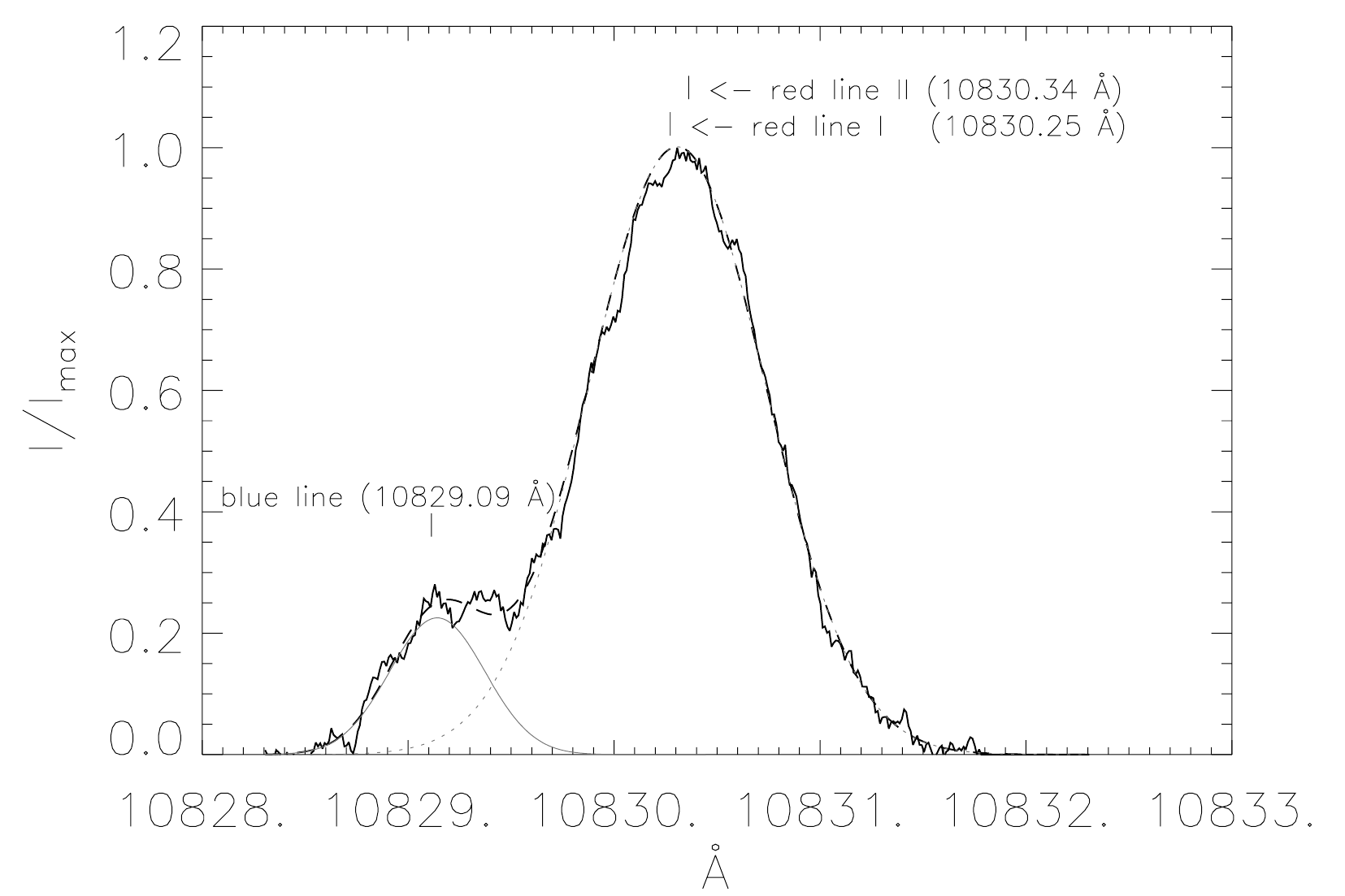} 
\caption{Determination of the blue and red components of the \ion{He}{i} 
10830~\AA\ triplet from the observed emission profiles. In this example the
 slit was placed at 1\farcs4 off the solar visible limb. See text for details. 
  The solid line {  represents} the average emission profile. The dotted line is the Gaussian fit to the symmetrised red
  component. Subtraction of this from the observed profile leaves the blue
  component, which is also fitted by a Gaussian profile (thin solid line). 
The sum of both Gaussians (dashed line)  gives the fit to the 
observed profile.}
\label{fig:ajuste}
\end{figure}

For the calculation of $\cal{R}$ we need to determine the amplitudes of the 
blue and red components of the emission profile (as shown in Fig.  
\ref{fig:ajuste}).

To determine the core wavelength of the red 
component of the triplet we fitted a Gaussian profile to its core, in a 1.3 \AA\ 
range around the maximum. After symmetrising the observed profile around this 
maximum, using the values on the red side of the red component, we fitted another Gaussian 
function to the resulting symmetric profile. Subtraction of the fitted 
symmetric profile from the data leaves the emission profile of the blue 
component, which was also approximated by a Gaussian to determine its central wavelength.  Our tests trying to fit directly both profiles using two Gaussians failed in a number of cases, probably due to the following reasons: (a) the red component is in fact the result of two blended lines, (b) the much weaker blue component was almost hidden in the broadened red component, and (c) the presence of noise. Our technique determines first the red component and then, after subtraction of the fitted profile, the blue one.

We have thus separated the helium emissions into their red and blue
components  assuming only that both are present and that they are both symmetric. We can now measure their widths and intensities and also check that the line core positions coincide with the theoretical ones. After the fitting process the residuals
between measured and observed profiles were small, the largest errors occurring
in the determination of the core intensities of the red line. This happens because the red component consists of two blended lines (with a separation of 0.09 \AA), a fact that flattens the emission profile near the core as opposed to a more peaked Gaussian function. Nevertheless, the differences between fitted profiles
and data are only significant in the red core and are always lower than
+0.08 of the maximum normalised intensity, with a mean difference of $\sim$0.02. To avoid systematic errors, we used the observational values for the center of
the red component when calculating $\cal R$.

\subsection{Results}
The chromospheric temperature and density are too low  to populate the
ortho-helium levels via collisions \citep{1994isp..book...35A}. The EUV irradiation from
the corona (CI) ionises the para-helium, and  
the subsequent recombinations lead to an overpopulation of all the 
ortho-helium levels, in particular those involved in the 10830~\AA\
transitions. \citet{Centeno06} and Centeno et al. (2007) have modelled the off-the-limb emission profiles and concluded that the ratio $\cal{R}$\,=\,$I_{blue}/I_{red}$ is a function of the height and a direct
tracer of the amount of CI. Here we compare the results from the theoretical modelling with observations.

\begin{figure}[t]
\center \includegraphics[width=\textwidth]{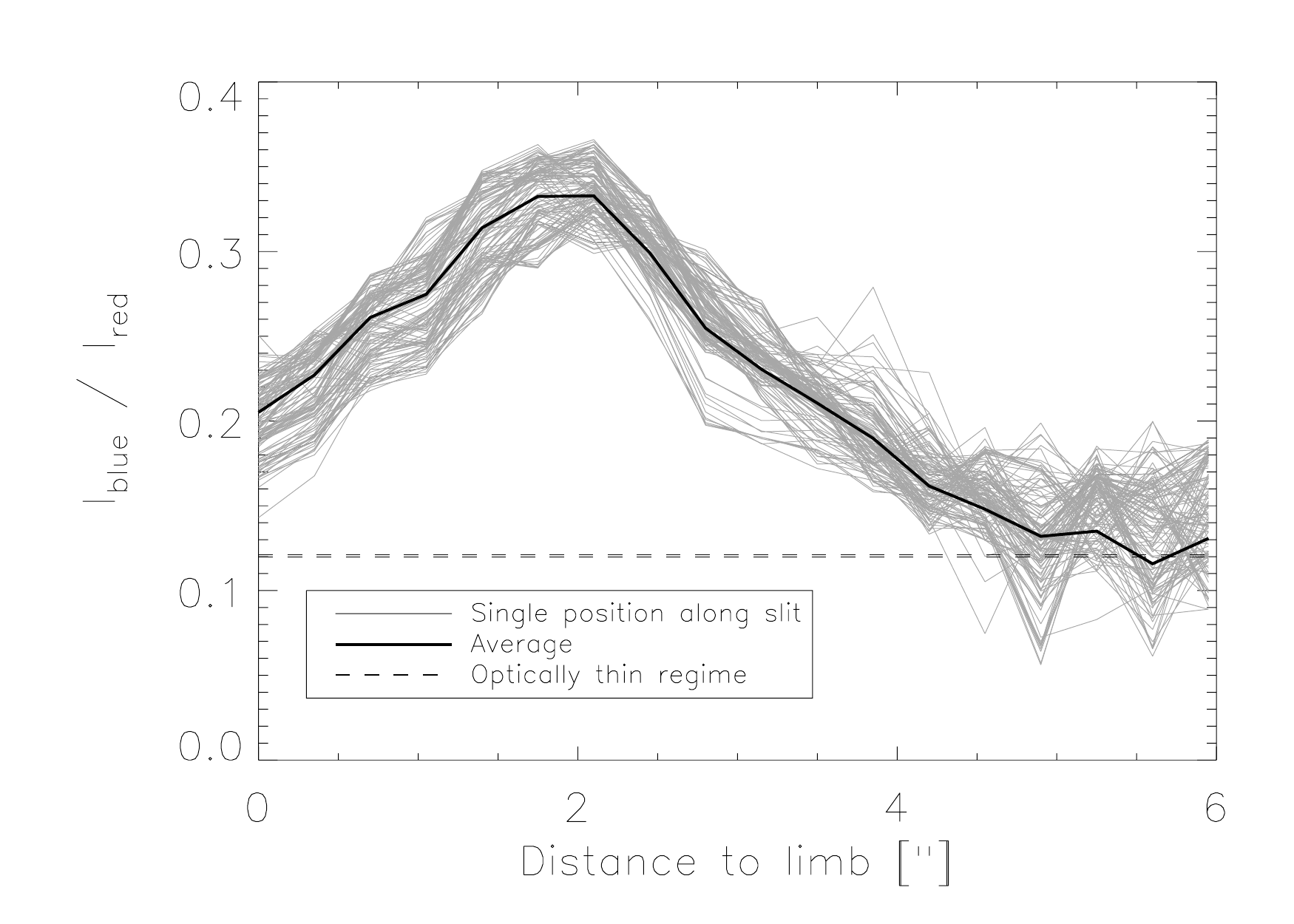}
\caption{Measured ratio $\cal R$\,=\,$I_{blue}/I_{red}$ as a function of
  distance to the solar limb. Thin lines come from
  each pixel along the slit. The thick solid line represents the average and the dashed line the value of the optically thin regime.} 
\label{fig:ratios}
\end{figure}

\citet{truj05} {modelled their spectropolarimetric observations assuming a slab with constant physical properties with a given optical thickness}. In the optically thin regime $\cal R$\,$\sim$\,0.12, which is the ratio of the relative oscillator strengths of the triplet. As the optical
thickness (at the line-center of the red blended component) grows, this ratio also increases 
until it reaches a saturation value slightly larger than 1 for $\tau\sim10$. (This type of calculation can be done and improved as explained in Trujillo Bueno \& Asensio Ramos 2007). To reproduce the observed emission profile \citet{truj05} had to choose $\tau \sim 3$. Interestingly, {  the values of $\tau$ yielded by this modelling strategy are consistent} with the more realistic 
approach of Centeno (2006), where non-LTE radiative transfer calculations in 
semi-empirical models of the solar atmosphere are presented, using spherical 
geometry and taking into account the ionising coronal irradiation.
With our data we are able to test such theoretical calculations by comparing 
the measured values of $\cal R$ with those resulting from various chromospheric
models. This way we may eventually trace the amount of CI inciding on the 
spicules. The analysis described above yielded the 
values of $\cal R$ for the observed profiles. The resulting dependence on 
the distance to the solar
limb, for each pixel along the slit and each position of the slit above the
limb, are presented in Fig.~\ref{fig:ratios}. The solid black line gives the average value of $\cal R$.

The dependence of $\cal R$ with height can be understood in a qualitative way as
follows: In the outer layers of the chromosphere the density is so low 
that the transitions occur in the optically thin regime. 
With decreasing altitude the ratio $\cal R$ increases (proportionally with 
density) until a maximum optical thickness is reached.  At even lower 
layers, although the density still continues to rise,  the extinction of the coronal
irradiance leads to a reduction in the number of ionizations, which results 
in a decrease of the optical thickness in the core wavelength of the red 
component, {  and thus in a decrease of $\cal R$.}


For a quantitative comparison with theoretical modelling we have 
used the results from \citet{Centeno06} and \cite{cente07} where they calculated the ratios $\cal R$ for different
standard model atmospheres: FAL-C and FAL-P \citep{fontenla91} and 
FAL-X \citep{avrett95}. The FAL-C and FAL-X models may be considered as illustrative of the thermal conditions in the quiet Sun, while the FAL-P model of a plage region. The FAL-X model has a relatively cool atmosphere in order to explain the molecular CO absorption at 4.6~$\mu$.

\begin{figure}[t]
\hspace{-0.5cm}\includegraphics[width=\textwidth]{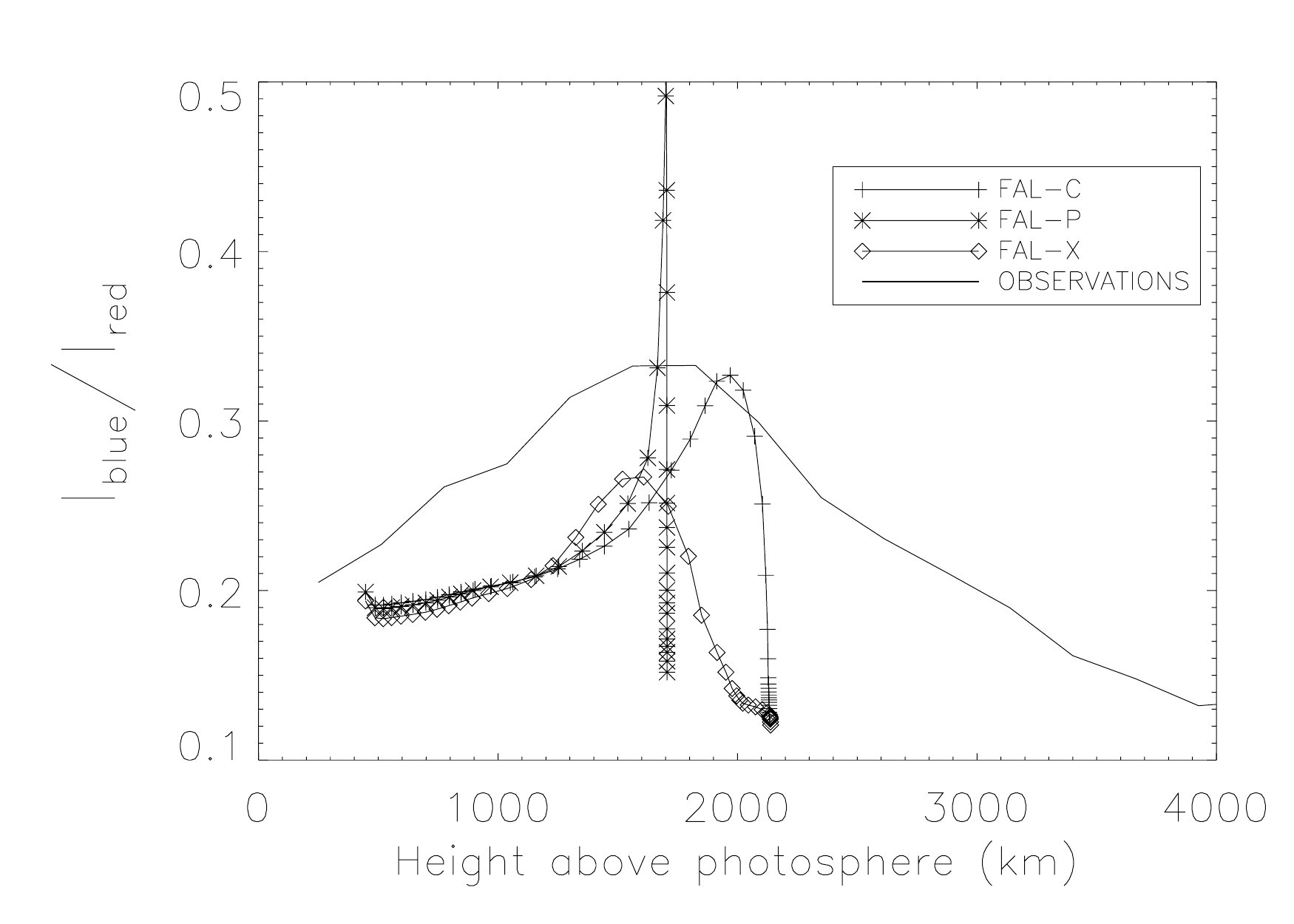} 
\caption{Observed (average) vs. theoretical variation of the ratio $\cal R$$ =
  I_{blue}/I_{red}$ 
  with height.}
\label{fig:comp}
\end{figure}

The comparison is shown in Fig.~\ref{fig:comp}. We notice that the modelled
height variations of $\cal{R}$ agree only in a qualitative manner with what is 
found in our observations. However, the calculations from different models of 
the solar atmosphere are unable to reproduce the measured ratio. Higher values of the coronal irradiance lead to an increase of the optical thickness (at the line centers of the \ion{He}{i} multiplet) and an upward shift in the run of $\cal{R}$ vs. height. Yet the shape of the height dependence is mainly given by the atmospheric density profile and the attenuation of the ionising radiation as it reaches the lower layers of the chromosphere.
It is also clear from Fig.~\ref{fig:comp} that the models do not extend high enough. 

\subsection{Conclusions\label{conclusion}}
The theoretical behaviour of the ratio $\cal R$ agrees
qualitatively with observations. Yet, a quantitative comparison shows poor
agreement. Also, the simulated ratios are highly model dependent. 
As already explained, {  the failure to reproduce the observed profiles is very likely due to the density stratification not being adequate for spicule modelling and to the limited vertical extension of the atmospheric models.}
Modelling of the intensity ratio $\cal R$ in the
\ion{He}{i} infrared triplet should account for the fact that the solar
chromosphere is inhomogeneous on small scales and that the spicules are
small-scale intrusions of chromospheric matter into the hot corona.

\section{High resolution imaging of spicules\label{sec:limb:ha}}

\cite{2007arXiv0710.2934D} recently published high resolution observations of spicules with the Solar Optical Telescope on board Hinode  \citep{2007SoPh..243....3K} in the \ion{Ca}{II} H line at 3968 \AA. They find at least two types of spicules that dominate the structure of the magnetic solar chromosphere: Type I with 3-7 minute timescales that correspond to the hitherto known spicules, and the new Type II spicules, developing in  $\sim10$ s, and lasting 10-150 s. These are also very thin, with widths down to the spatial resolution (120 km).  Also, \cite{truj05} used spectropolarimetric observations and a theoretical modeling accounting for radiative transfer effects. They find that the magnetic field in spicules is aligned with the visible structures and measure field strengths of up to 40 G with an inclination of 35$^{\circ}$ with respect to he local vertical.

\begin{figure}
\center \includegraphics[width=\textwidth]{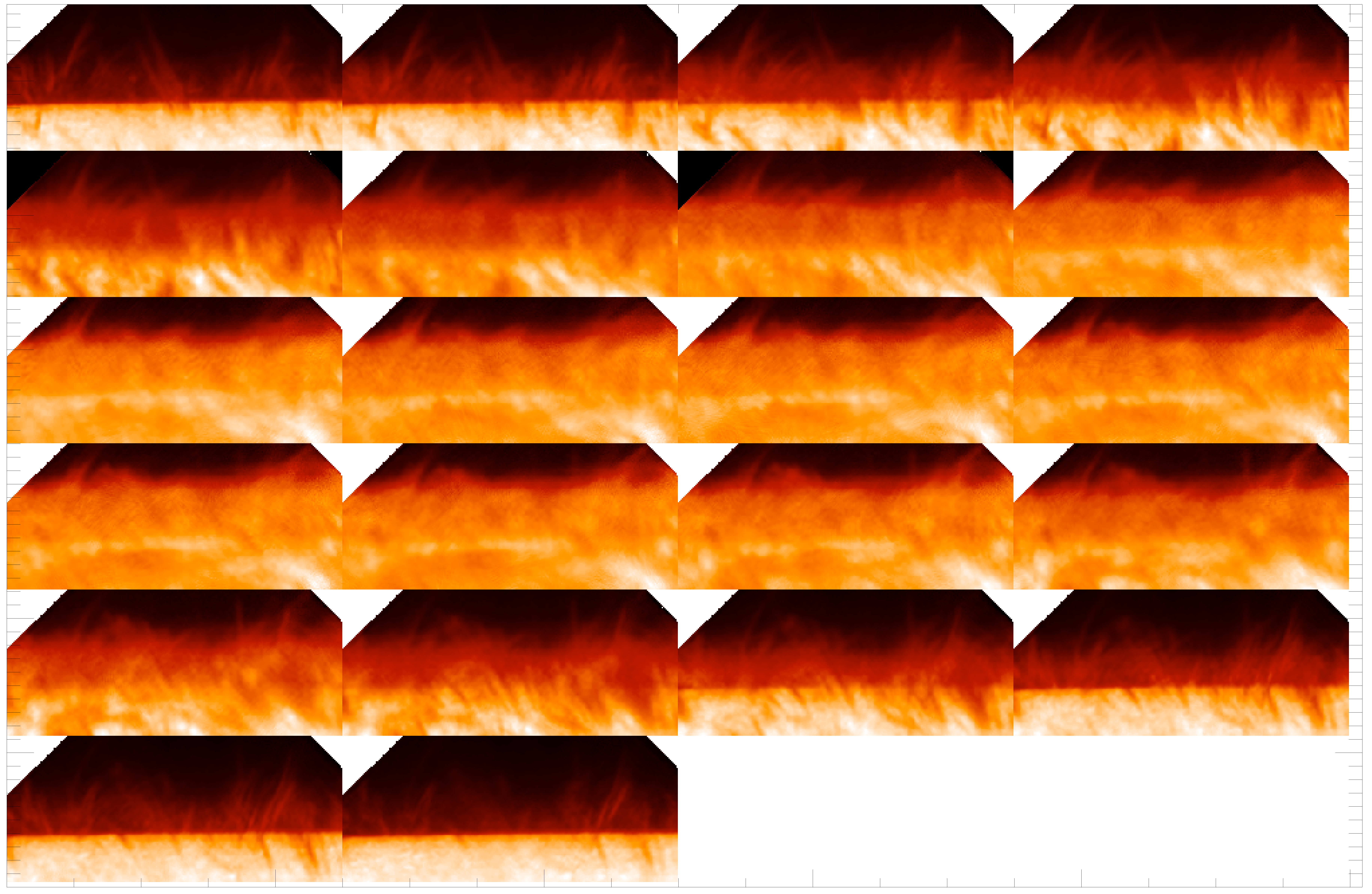} 
\caption{Reconstructed narrow-band scan observed near the solar north pole. The size of each image is $56\farcs1\times19\farcs1$. The wavelength of the filtergrams increases by $0.1$\AA\, from top left to bottom right row by row. The third image in the third row is closest to the center of the mean line profile. The images have been rotated to have the limb parallel to the horizontal axis.}  
\label{fig:spicules}
\end{figure}

\begin{sidewaysfigure}
\center \includegraphics[width=\textwidth]{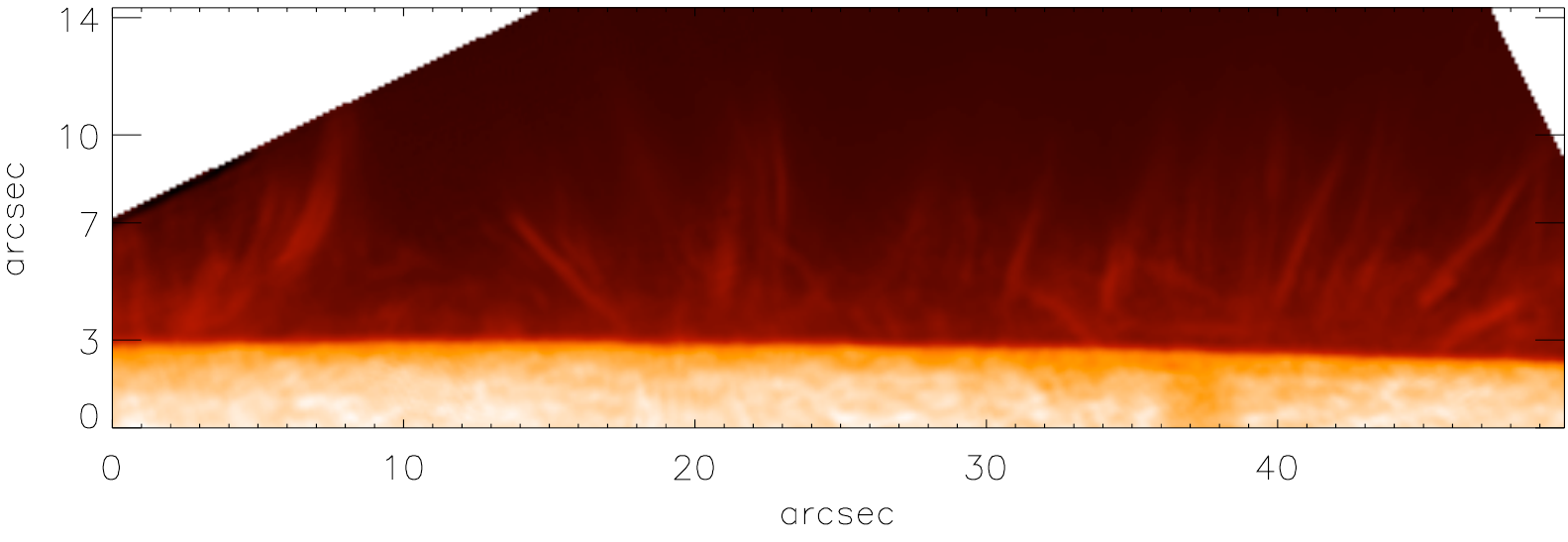} 
\caption{Limb H$\alpha$ 2D filtergram at $\lambda_{0} +1.1$ \AA\, near the south pole, where a coronal hole was present. This region shows much stronger emission and more variation of spicule width, height and inclination as Fig. \ref{fig:spicules}. A background of thin vertical spicules can be seen overlaid with wider and more inclined spicules, including nearly horizontal jets. Some of the spicules appear to be bent and show internal structure such as splitting into parallel jets. The maximum projected height above the limb is $\approx 8\,250$~km, while the mean height at this wavelength is $\approx$3700 km. The image has been rotated to show a horizontal limb in the presentation.}  
\label{fig:spicules2}
\end{sidewaysfigure}

This Section compares these observational and theoretical properties with our high spatial resolution observations with the G-FPI. We use the dataset ``limb''. It consists of several consecutive H$\alpha$ scans with a field of view that includes the limb and a region outside the disc up to a height where no emission from spicules in the H$\alpha$ core is observed. The \emph{seeing} conditions were extremely good during the observations and the AO system could lock on a nearby facula. After usual dark subtraction and flat fielding we have used the BD method (see Sec. \ref{momfbd}) to achieve highest spatial resolution. We were observing the limb near both poles. In Figures \ref{fig:spicules} and \ref{fig:spicules2} we present some examples of the reduced data. 

Our time series of four minutes duration already shows a wide range of dynamics. We observe both long lasting spicules and fast evolving phenomena. Measuring the inclination of the projected spicules to the local vertical we find angles up to 30\degr\ for the north pole, as it has been reported \citep{beckers68} . The projected  height above the limb varies from the wings to the core, from 2770 to 3750 km at $\pm 0.5$ and $\pm 1$\AA\, respectively. Near the south pole we find, however, much stronger emission and higher inclinations. The maximum angle is close to 70\degr\ from the local vertical, while the maximum height reaches up to 8250 km. We also find one horizontal fibril/spicule, as well as the presence of kinks or bends in some spicules. The width of single resolved spicules varies from a maximum width of 1\,000 km at the spicule footing to a minimum size of 250 km, almost down to the resolution limit of the images, both in faint spicules and in others with strong emission. We also can retrieve the spectral profile at each pixel. 

Figure \ref{fig:spicules} demonstrates an important contribution to the understanding of spicules. It solves the long-standing question about the counterparts of spicules on the disc \citep{1992A&A...264..236G}. The first and last four filtergrams of the scan across H$\alpha$ in this figure show that spicules outside the limb continue as dark fibrils inside the disc.

\begin{figure}
\center 
\includegraphics[width=\textwidth]{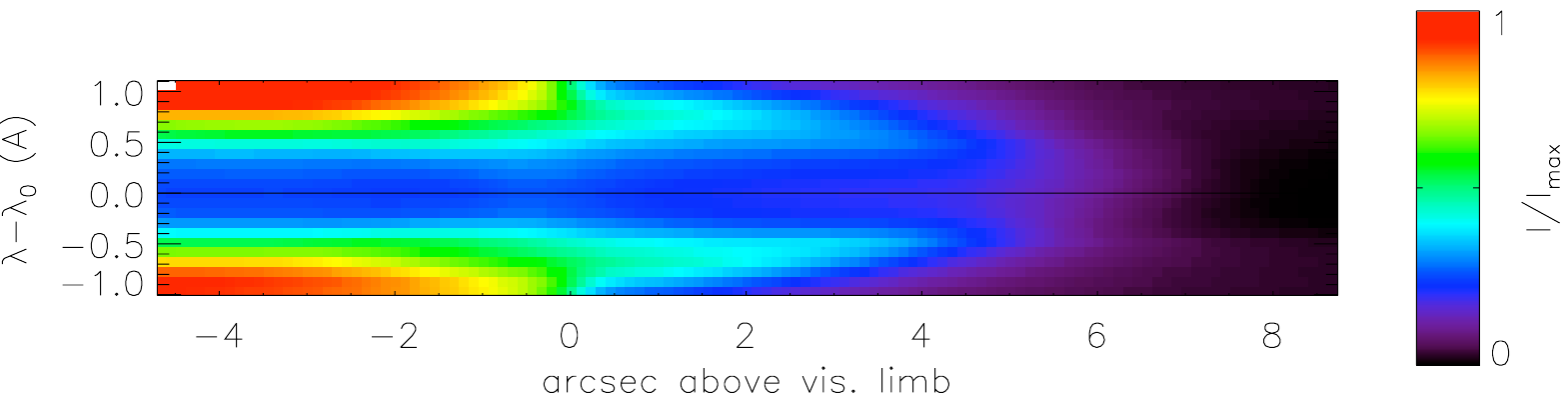} 
\caption{Image representation of the mean measured spicular profiles from image \ref{fig:spicules}. the x-axis is the height above limb, while the y-axis is the wavelength around H$\alpha$ line center (black horizontal line). Horizontal cuts at $\lambda-\lambda_{0}=[0,\pm 0.5, \pm 1]$\AA\, are shown in Fig. \ref{fig:spiprof}.} 
\label{fig:spiprofima}
\end{figure}

\begin{figure}
\center \includegraphics[width=\textwidth]{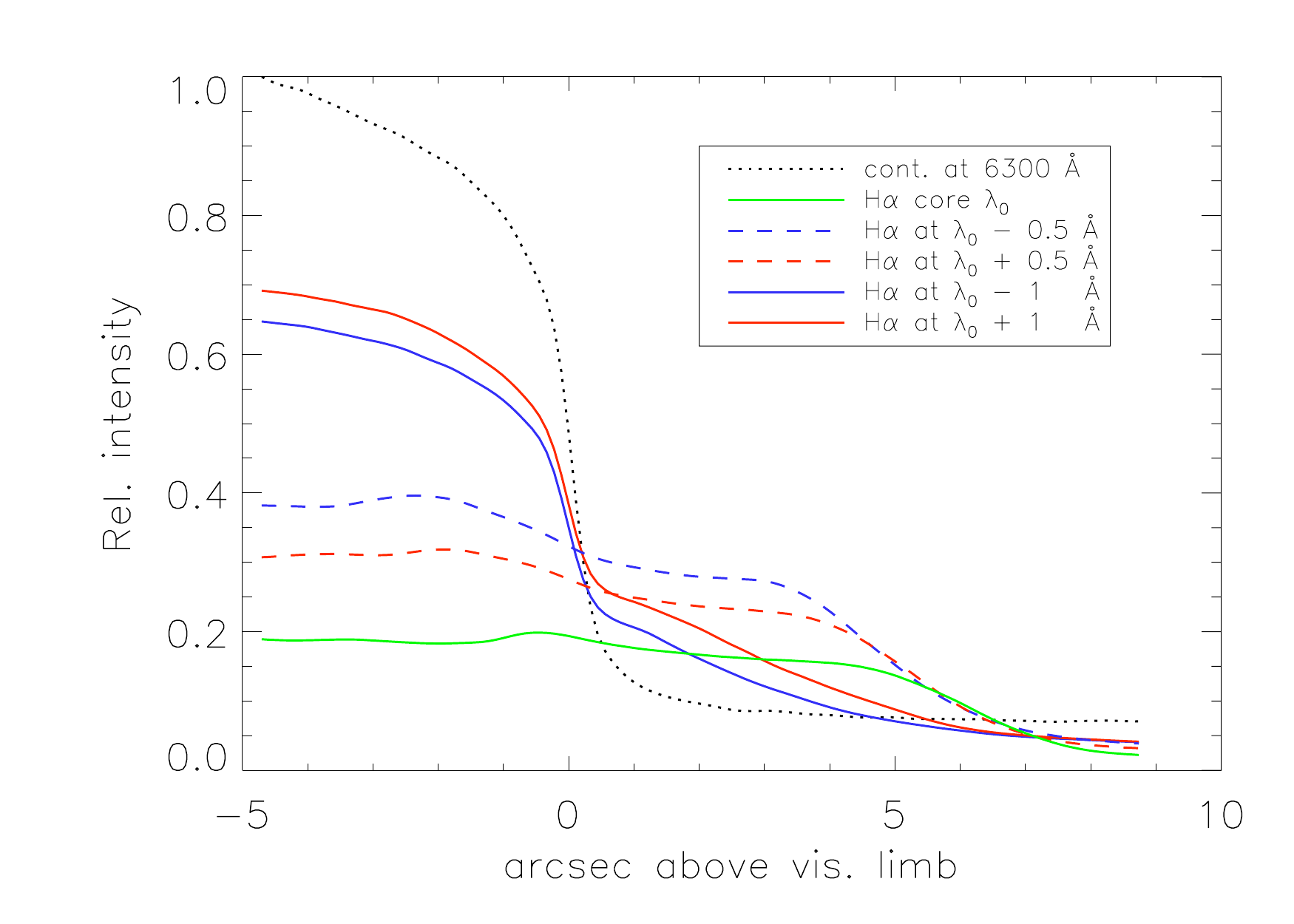} 
\caption{Average over $11\farcs2$ of H$\alpha$ intensity profiles inside and outside the limb, for several line positions, observed near the solar north pole. \emph{Dotted line}: broadband intensity at 6300 \AA, the inflection point defines the position of the solar limb; \emph{green thick line}: intensity at H$\alpha$ line center, which is nearly constant till $\approx$5\arcsec\ above the limb and then decreases outwards; \emph{dashed lines}: intensities at $-0.5$ \AA\,(blue) and $+0.5$ \AA\,(red) off line center, with height of spicular visibility decreasing at $\approx3\farcs7$; \emph{solid lines}: intensities at $\pm1$\AA\, off line center. It is seen that the H$\alpha$ line turns from absorption inside the limb into an emission line (line intensities higher than the continuum intensity) above the limb.}  
\label{fig:spiprof}
\end{figure}

In Fig. \ref{fig:spiprofima} we show the mean variation from the disc to the limb of the intensity in H$\alpha$  around the north pole. Further, Fig. \ref{fig:spiprof} presents mean intensity variations from the disc to the limb for several wavelengths around the H$\alpha$ line center. The emission at the line center is almost constant from the disc up to a height of around 5\arcsec\ above the limb, where the intensity starts to decrease.


\chapter{Conclusions and outlook \label{ch:conclusions}}
We have studied the dynamics of the solar chromosphere, both on the disc and above the limb, using two spectral regions (H$\alpha$ in visible light and the infrared \ion{He}{i} 10830 \AA\, multiplet). By means of real-time correction and different post-processing techniques we have reduced the image degradation induced by the Earth's atmosphere achieving resolutions in H$\alpha$ up to $0\farcs5$ and better. This Chapter summarize the main conclusions of this work.

\subsubsection*{Observations and analysis}
The basic results from the observations taken from the disc are:
\begin{itemize}
\item Data taken in the combination with the ``G\"ottingen''-Fabry Perot Interferometer (G-FPI), adaptive optics and speckle interferometry have high quality. We have obtained a time sequence in H$\alpha$ of 55 min from the active region AR 10875 at  heliocentric angle $\vartheta\approx36\degr$. The time cadence is 22 seconds, and its field of view  $77\arcsec\times94\arcsec$. For each interpolated time step we can retrieve 23 filtergrams along the H$\alpha$ spectral line with 45 m\AA\, FWHM and spatial resolution better than $0\farcs5$ . Simultaneous broadband images at 6300\,\AA\,were also obtained, with spatial resolution of $0\farcs25$, close to the telescopic diffraction limit.

\item We have observed the dynamics of a small surge in detail: It showed repetitive occurrence with a rate of some 10 min. The surge developed from initial small active fibrils to a straight, thin structure of approximately 15~Mm length, then retreated back to its mouth to reappear again two times. The gas velocities reach approximately 100~km\,s$^{-1}$. The rebound shock model by \citet{1989ApJ...343..985S} seems to be a viable explanation.

\item The region was very active during the observations. We studied two small-scale, synchronous, possibly related flashes, or mini-flares. The simultaneity is within seconds, while their total evolution time was $\sim45$s. The brightenings were separated by $\sim14$ Mm. The used scanning parameters of the G-FPI were slow for this fast evolution, yet we could follow it with a temporal resolution of 2~s by analysing filtergrams taken consecutively at different wavelengths across the H$\alpha$ line. One of the two flashes showed an apparent proper motion with a speed up to 200~km\,s$^{-1}$, while developing a high emission in H$\alpha$, above the continuum intensity.

\item For the observations of waves we restricted our study to two areas exhibiting long fibrils. Yet the results likely represent the typical behavior of chromospheric magnetoacoustic waves within this active region.  By means of high-pass frequency filtering, we observe waves running parallel to the fibrils, thus presumably also parallel to the magnetic field. They were mostly solitary waves, although sometimes repetitive wave trains could be seen with periods of 100--180~s. Most pulses start with velocities on the order of 12--14~km\,s$^{-1}$ and get accelerated to reach phase speeds of approximately 30~km\,s$^{-1}$.  Furthermore, we observe that the slow waves have strong transversal (LOS) velocity components with $\sim$3~km\,s$^{-1}$, i.e. are not purely longitudinal, and that the fast waves consist of short (1\arcsec--2\arcsec), thin ($\sim$0\farcs5) blobs and apparently move along sinuous lines. 
\end{itemize}

Further, we have analyzed observations of spicules inside the disc and above the limb with the G-FPI data. Given the properties for this kind of observations we could not use the speckle interferometry method to reduce the atmospheric distortions. Instead we have used the blind deconvolution approach, in particular the version developed at the Swedish Institute for Solar Physics for multiple simultaneous objects with multiple frames per object \citep{2005SoPh..228..191V}. The observations and analysis yielded the following main results:
\begin{itemize}
\item It is possible to successfully use multi-object multi-frame blind deconvolution methods with the G-FPI to reduce atmospheric distortions. This is specially important for on-limb observations, where the current speckle interferometry method is not applicable.
 
\item We have observed spicules in H$\alpha$ at both solar polar caps. Compared with the solar north pole, we find much stronger spicular emission at the south pole that could be related to the presence of a coronal hole. The maximum projected height reaches 8250 km, while we see inclinations of the spicules up to 70\degr\, form the local vertical. We can resolve the detailed structure of the spicules as well as the presence of kinks or bends in some cases. The width of a single spicule ranges from 1\,000 km down to the resolution limit of around 250~km.
\item Using the retrieved spectral profile we can observe that spicules outside the limb continue as dark fibrils inside the disc, as shown in Fig.  \ref{fig:spicules}. This answers a long standing question, e.g. cited by \citet{1992A&A...264..236G}.

\end{itemize}
Thus, we have used two different post-processing approaches to reduce the image degradation with the H$\alpha$ spectral line data. Since both methods are based on different approaches, we have reduced the same observational data with both techniques and compared the results. These are the main conclusions:
\begin{itemize}
\item The agreement of the images from both approaches is high. The achieved resolution comes close to the diffraction limit in both cases. Even though both methods split the image into isoplanatic subfields for individual reconstruction, there is no difference of subfield re-composition when comparing the results.
\item In general, the biggest advantage of speckle interferometry over blind deconvolution is the small computational time required. A complete restoration of one full H$\alpha$ scan like the ones used in this work is roughly $\sim400$ times faster with speckle interferometry than with blind deconvolution methods.
\item The main advantage of blind deconvolution methods is their versatility. It can be applied with only few frames, with one or few simultaneous objects, or both at the same time. This method is highly advisable when aiming for e.g., fast evolving targets, or limb observations. The perfect sub-alignment of simultaneous objects avoids spurious signals on deduced quantities, like magnetograms.
\item For the broadband channel we find that the speckle interferometry gives images with high contrast. Only when forcing the blind deconvolution method to reconstruct up to high  ($100$) Karhunen-Loeve modes, the results are similar. We note that the speed of the reconstruction is proportional to the number of modes.
\item The narrow-band images are clearly better reconstructed with the blind deconvolution, even with only 17  Karhunen-Loeve modes. The noise treatment gives smoother images with details at smaller scales (of the order of $\approx 0\farcs3$).
\end{itemize}
Further, we have obtained and analysed spectroscopic measurements in the infrared. We have centered our studies here on the intensity profiles of the \ion{He}{i} 10830 \AA\, multiplet above the limb.
\begin{itemize}
\item Recent work, like e.g. by \citet{truj05, Centeno06}, has demonstrated the importance of the intensity ratio between the blue and red component of this triplet as tracer of the coronal irradiance. In this work we present novel observations showing the variation of this parameter with distance to the limb with a resolution of $0\farcs35$ up to 7\arcsec\, above the solar visible limb (See Fig. \ref{fig:ratios}).
\end{itemize}

\subsubsection*{Interpretation of observations}

For the interpretation of the observed data we have used several models and previous theoretical results to compare with the presented data. The main results from this analysis are:
\begin{itemize}
\item From the intensity profiles of the H$\alpha$ spectral line inside the disc we can infer many physical parameters. We have applied the lambdameter method as a fast and easy way to retrieve qualitative velocity maps. Also we have used Beckers's \citeyear{1964PhDT........83B} cloud model. Our simple non-LTE inversion code provides the possibility to infer many physical parameters, e.g. hydrogen and electron density, mass density, temperature,~\dots The results are in agreement with the data given in the current literature.

\item From the linearization of the MHD problem, we discuss the interpretation of the observed waves as magnetoacoustic waves.  We assume estimates with reasonable magnetic field strengths  in the chromosphere of the active region of 30--100~Gauss and reasonable mass densities in the fibrils of 2$\times10^{-13}$~g\,cm$^{-3}$. The observed wave speeds are much lower than the expected Alfv\'en velocities. We conclude from these findings that a linear theory of wave propagation in straight magnetic flux tubes is not sufficient.

\item From the infrared observations we have calculated the ratio of amplitudes in the two main components of the \ion{He}{i} 10830 \AA\, multiplet. \cite{Centeno06} has modelled synthetic limb observations according to the current theories of formation of this triplet and chromospheric models. The agreement is only qualitative. The failure to reproduce the observed profiles is very likely due to the density stratification not being adequate for spicule modelling used 
and to the limited vertical extension of the atmospheric models. Modelling of the intensity ratio should account for the fact that the solar chromosphere is inhomogeneous on small scales and that the spicules are small-scale intrusions of chromospheric matter into the hot corona. Future models of the solar chromosphere should be constrained by the observational evidences presented within this work.

\end{itemize}

\subsubsection*{Outlook}

The solar chromosphere represents a lively and exciting field of research. The wealth of structures, its dynamics and the wide range of evolution timescales are the consequence of the peculiar properties of this atmospheric layer. Current instruments like the ones used here, are able to observe and study in great detail new phenomena, that test current models and, as a consequence, helps our understanding of the solar atmosphere. This thesis aimed at contributing to the understanding. Yet, work to extend this research is already in progress. Here we shortly describe some of this work and give an outlook to further steps to be undertaken next.
\begin{itemize}
\item
The blind convolution method provides a practical way to study the spicules in H$\alpha$ near and above the limb. Data from a short time sequence, taken under very good seeing conditions, are currently under reconstruction with phase diversity methods. The analysis will shed light onto the dynamic phenomena in spicules.
\item
We have learned that the sequential scanning, with the G-FPI, with cadence of 22~s is not fast enough in some cases. For future
observations, we can design scanning modes of 2--3 second resolution taking images at fewer wavelength positions in a spectral line, like H$\alpha$.

\item New infrared data of spicules near the solar poles and the equator, below coronal 
holes or in coronal active regions will help us to understand the detailed 
formation process of the \ion{He}{i} 10830~\AA\ lines.

\item Full Stokes spectropolarimetric data of the \ion{He}{i} 10830 \AA\, multiplet are available from an earlier  observing campaign. Scans above the limb were performed under very good seeing conditions. We can therefore extend our present analysis and study the polarization. We aim to investigate the Hanle effect as suggested by \cite{truj05} and make use of available inversion techniques like e.g. from \cite{2004A&A...414.1109L}.

\item The new Gregor telescope  \citep{2007ASPC..368..605B} will host the G-FPI from the coming year on. The combination of this new facility with other instruments like Hinode  \citep{2007SoPh..243....3K}, will provide new exciting resources for further research.
\end{itemize}


\bibliographystyle{thesismp}
\bibliography{thesismp}

\chapter*{Publications\markboth{Publications}{Publications}}
\addcontentsline{toc}{chapter}{Publications}

\begin{enumerate}
\item[] \emph{Refereed papers}
\item
\textbf{B.~{S{\'a}nchez-Andrade Nu{\~n}o}}, R.~{Centeno}, K.~G. {Puschmann},
  J.~{Trujillo Bueno}, J.~{Blanco Rodr{\'{\i}}guez}, and F.~{Kneer}.
\newblock {Spicule emission profiles observed in \ion{He}{i}~10830~\AA}.
\newblock {\em \aap}, 472:L51--L54, Sept. 2007.
\item
\textbf{B.~{S{\'a}nchez-Andrade Nu{\~n}o}},  N.~Bello Gonz\'alez, J.~{Blanco Rodr{\'{\i}}guez}, F.~{Kneer},   and K.~G. {Puschmann}.
\newblock {Fast events and waves in an active region of the Sun
       observed in H$\alpha$ with high spatial resolution}.
\newblock {\em \aap}, submitted, Dec. 2007.

\item
J.~{Blanco Rodr{\'{\i}}guez}, O.~V. {Okunev}, K.~G. {Puschmann}, F.~{Kneer},
  and \textbf{B.~{S{\'a}nchez-Andrade Nu{\~n}o}}.
\newblock {On the properties of faculae at the poles of the Sun}.
\newblock {\em \aap}, 474:251--259, Oct. 2007.

\item[]  \emph{Conference contributions}
\item
\textbf{B.~{S{\'a}nchez-Andrade Nu{\~n}o}}, R.~{Centeno}, K.~G. {Puschmann},
  J.~{Trujillo Bueno}, and F.~{Kneer}.
\newblock {Off-limb spectroscopy of the He I 10830 {\AA} multiplet:
  observations vs. modelling}.
\newblock In F.~{Kneer}, K.~G. {Puschmann}, and A.~D. {Wittmann}, editors, {\em
  Modern solar facilities - advanced solar science},
  pages 177--180, 2007.

\item
\textbf{B.~{S{\'a}nchez-Andrade Nu{\~n}o}}, K.~G. {Puschmann}, and F.~{Kneer}.
\newblock {Observations of a flaring active region in H$\alpha$}.
\newblock In F.~{Kneer}, K.~G. {Puschmann}, and A.~D. {Wittmann}, editors, {\em
  Modern solar facilities - advanced solar science},
  pages 273--276, 2007.

\item
\textbf{B.~{S{\'a}nchez-Andrade Nu{\~n}o}}, K.~G. {Puschmann}, M.~{S{\'a}nchez Cuberes},
  J.~{Blanco Rodr{\'{\i}}guez}, and F.~{Kneer}.
\newblock {Analysis of a Wide Chromospheric Active Region}.
\newblock In D.~E. {Innes}, A.~{Lagg}, S.~A. {Solanki}, and D.~Danesy, editors, {\em
  Chromospheric and Coronal Magnetic Fields}, volume 596 of {\em ESA Special
  Publication}, Nov. 2005.

\item
\textbf{B.~{S{\'a}nchez-Andrade Nu{\~n}o}}, K.~G. {Puschmann}, M.~{S{\'a}nchez Cuberes},
  J.~{Blanco Rodr{\'{\i}}guez}, and F.~{Kneer}.
\newblock {Chromospheric Dynamics of a Solar Active Region}.
\newblock In {\em The Dynamic Sun: Challenges for Theory and Observations},
  volume 600 of {\em ESA Special Publication}, Dec. 2005.
\item
\textbf{B.~{S{\'a}nchez-Andrade Nu{\~n}o}}. 
\newblock {Study case: Solar Science Communication}.
\newblock  In L.~Lindberg Christensen, and M.~Zoulias, editors, {\em
  Communicating Astronomy with the Public 2007}. An IAU/Nat. Obs. of Athens/Eugenides Foundation Conference,  {\em } Oct. 2007.

\item
J.~{Blanco Rodr{\'{\i}}guez}, \textbf{B.~{S{\'a}nchez-Andrade Nu{\~n}o}}, K.~G.
  {Puschmann}, and F.~{Kneer}.
\newblock {Study of Polar Faculae}.
\newblock In {\em The Dynamic Sun: Challenges for Theory and Observations},
  volume 600 of {\em ESA Special Publication}, Dec. 2005.

\item
J.~{Blanco Rodr{\'{\i}}guez}, \textbf{B.~{S{\'a}nchez-Andrade Nu{\~n}o}}, K.~G.
  {Puschmann}, and F.~{Kneer}.
\newblock {Study of Polar Faculae}.
\newblock In D.~E. {Innes}, A.~{Lagg}, S.~A. {Solanki}, and D.~Danesy, editors, {\em
  Chromospheric and Coronal Magnetic Fields}, volume 596 of {\em ESA Special
  Publication}, Nov. 2005.

\item
F.~{Kneer}, K.~G. {Puschmann}, J.~{Blanco Rodr{\'{\i}}guez},
  \textbf{B.~{S{\'a}nchez-Andrade Nu{\~n}o}}, and A.~D. {Wittmann}.
\newblock {Magnetic Structures on the Sun: Observations with the New
  ''G{\"O}TTINGEN'' Two-Dimensional Spectrometer on Tenerife}.
\newblock In D.~E. {Innes}, A.~{Lagg}, and S.~A. {Solanki}, editors, {\em
  Chromospheric and Coronal Magnetic Fields}, volume 596 of {\em ESA Special
  Publication}, Nov. 2005.

\item
L.~{Valdivielso Casas}, N.~{Bello Gonz{\'a}lez}, K.~G. {Puschmann},
  \textbf{B.~{S{\'a}nchez-Andrade Nu{\~n}o}}, and F.~{Kneer}.
\newblock {Analysis of Polarimetric Sunspot Data from Tesos/vtt/tenerife}.
\newblock In D.~E. {Innes}, A.~{Lagg}, S.~A. {Solanki}, and D.~Danesy, editors, {\em
  Chromospheric and Coronal Magnetic Fields}, volume 596 of {\em ESA Special
  Publication}, Nov. 2005.
\item
C.~Denker, A.P.~Verdoni, F.~W\"oger, A.~Tritschler, T.R.~Rimmele, F.~Kneer,
K.~Reardon, \textbf{B.~S\'anchez-Andrade Nu\~no}, I.~Dom\'inguez Cerde\~na and K.G.~Puschmann
\newblock {Speckle Interferometry of Solar Adaptive Optics Imagery}
\newblock {DFG-NSF Astrophysics Research Conference ``Advanced Photonics in Application to Astrophysical Problems'', June 2007}
  
  \thispagestyle{empty}
\end{enumerate}

\chapter*{Acknowledgements\markboth{Acknowledgements}{Acknowledgements}}
\addcontentsline{toc}{chapter}{Acknowledgements}

I would like to acknowledge here the contributions from many people to the successful completion of this work. I want to express my gratitude to all those who helped me, from my supervisor this three last years to my teachers at my first university who encouraged me to start this adventure in astrophysics. During all this time I have been also surrounded and supported by many people to whom I want to thank within these lines. They all made this experience of doing a PhD, professionally and personally, one of the best times of my life.

My supervisor of this research work was Prof. Dr. Franz Kneer, or better \emph{franz}. He was the first person I saw when I came to the train station back three years ago (along with Julian). He was the guide to all my work, the observations, the interpretation, \dots basically, my formation as solar physicist. The amount of knowledge I learnt from him cannot be measured by any means, and for that I am professionally extremely grateful. In a personal way he was also very kind and helped me always with good advices whenever I needed it. He understood me perfectly when I had problems and also encouraged me to travel as much as I wanted (which was not a little). Working with him this time made the whole experience of PhD leaps better. 
Ich kann nicht diese Arbeit ohne Ihn zu vorstellen. Danke Franz.

Above all, I am and will be always grateful to my reference in life, my parents Conchita Nu\~no L\'opez and Julio S\'anchez-Andrade Fern\'andez, and my sister Deva S\'anchez Nu\~no. They always supported me, although  a bit far in physical distance during this last years. When I was 5 years old I was going home with my mother from the playground when I saw a poster announcing a conference about starts. I asked my mother to read it loud for me, and to her surprise I had to explained her: ``Don't you yet know that I want to be a \emph{researcher of stars}?''. The day after she gave me a children book about stars and explained me that if I wanted to be so I should know that they are called \emph{astronomers}. And that's how it all began.

My work was supported by two institutions: the \emph{Max Planck Institute f\"ur Sonnensystemforschung} (MPS) granted me the fellowship and the \emph{Institut f\"ur Astrophysik G\"ottingen} (IAG) provided me the facilities to work with Franz at G\"ottingen and support for the observations at Tenerife. Also, being part of the \emph{International Max Planck Research School on Physical Processes in the Solar System and Beyond at the Universities of Braunschweig and G\"ottingen} (IMPRS) I could attend seminars, retreats and courses on various astrophysical topics. I am very grateful to these institutions for providing such a broad curricula. I would like also to thank the coordinator of the MPS and IMPRS, Dr. Dieter Schmitt. 

The Vacuum Tower Telescope used for the observations is operated by the Kiepenheuer-Institut f\"ur Sonnenphysik, Freiburg, at the Spanish Observatorio del Teide of the Instituto de Astrof\'isica de Canarias.  

Here at the solar group of IAG we had a really stimulating environment with long discussions: Franz, Klaus, Markus, Nazaret, Julian and the various Erasmus people that went by (like Luisa, Manu, \dots ). We worked in the beginning at the beautiful historical \emph{Sternwarte} and then at the Faculty of Physics. I would like to thank specially Klaus for all the help and scientific discussions during all the time he was here.

Many other professional colleagues contributed to this work. Their input was extremely helpful. A short list with few names would include: (from the MPS) Andreas Lagg, Vasili Zakarov, (from the IAG) the solar group, Axel Wittman, Volker Bothmer, (from the IAC) Basilio Ruiz Cobos, Manolo Collados, Javier Trujillo, Rebecca Centeno. Thanks to Nurol Al and H. Schleicher for their computing codes. For the Blind Deconvolution section I got much help from Jaime and Mats L\"ofdahl at the Institute for Solar Physics, Stockholm.

From Asturias to here there is long way, specially passing through Canary Island. Without the advices and guidance from many people I would have been lost. In these sense I specially thank Cristina, Juanjo, Basilio and Fernando. 

Throughout all these years I had the luck to be surrounded by the best companion of friends. I would like to mention lastly some few friends that helped me to keep my feet on the ground. Here in G\"ottingen Thorsten, Gonzaga, Klaus, Miguel, Cristiano, Vladi, Cathi, Niko, Lorne, Diego, Iria, Teresa, Nora, Carlos, Olga, Julian, Nazaret, Benoit, Markus, Manu and Luisa. Thorsten in particular was not only a good friend but also my flat-mate, climbing-mate, snowboard-mate and of course party-mate. My new flat-mate, Richard, had to stand me this last months with my thesis mood, though. Thanks a lot!.  Mark, Lucas, Clementina, Pedro, Emre or Martin, all the guys from Lindau, always warmly welcomed his lost member of the family emigrated to the civilization.

During my short intense period in Tenerife I made good friends: Onti, Borja, Bendinat, Adriana, Manu, Miguel. Some time later I met Rebecca, with whom I also worked these years.

Whenever a crazy trip came up I had AEGEE people coming along. They would join me to any place in Europe. With them I had some of my best times: Luis, David, Neila, Saioa, Javiero, Marti, Marta and recently of course Carol. All those friends with crazy names: Annia, Annamary, Bir, Konstantina, Marios or Lutzn. I am sure I'll meet you again, somewhere.

Since I left Asturias, I have always missed the green and astounding landscapes, but also my old friends: Roberto, Raul, Estela, Flaci, Nacho, Cova, Lisa, David and of course my big and warm family.

Por ultimo me gustar\'ia terminar esta secci\'on, y con ello el fin de este trabajo, con la memoria de tres personas a quienes siempre echar\'e de menos. En mi primer a\~no en Alemania tuve que despedir a  mi abuela materna y el segundo a mi abuela paterna. Aunque sea ley de vida no deja de ser doloroso. Este \'ultimo a\~no, a mi primo Abraham, a quien tuve la suerte de ver una \'ultima vez. Con ellos he aprendido la lecci\'on m\'as importante:

\vspace{1cm}
{\emph{ \centering Disfruten de la vida.}}


\flushright{
G\"ottingen, January 2008} 


\chapter*{Lebenslauf\markboth{Lebenslauf}{Lebenslauf}}
\addcontentsline{toc}{chapter}{Lebenslauf}
\begin{table}[htdp]
\begin{tabular}{p{.23\linewidth}lp{.35\linewidth}}
\textbf{Name:} & \multicolumn{2}{l}{Bruno S\'anchez-Andrade Nu\~no}  \\
\\
\textbf{Geburtsdatum:} &   \multicolumn{2}{l}{6. Mai 1981}   \\
\\
\textbf{Geburstort:} &   \multicolumn{2}{l}{Oviedo, Asturias, Spanien}   \\
\\
\textbf{Familienstand:} &  \multicolumn{2}{l}{Ledig}  \\
\\
\textbf{Eltern:} &  \multicolumn{2}{l}{Julio Miguel S\'anchez-Andrade Fern\'andez}  \\
 &  \multicolumn{2}{l}{Concepci\'on Nu\~no L\'opez }  \\
 \\
 \\
 \textbf{Staatsangeh\"origkeit:} &   \multicolumn{2}{l}{Spanisch}   \\
 \\
 \\
  \textbf{Schulbildung:} &  September 1987 - Juni 1995 & Grundschule am \emph{Colegio P\'ublico ``R\'io Piles''} in {Gij\'on}, Asturias   \\
  & September 1995 - Juni 1999 & Weiter f\"uhrende Schule am \emph{I. B. ``R\'io Piles''} in {Gij\'on}, Asturias
  \\
  \textbf{Studium:} &  September 1999 - September 2003 & Physikalische Fakult\"at  der Universit\"at Oviedo, Asturias  \\
  &  September 2003 - September 2004 & Physikalische Fakult\"at  der Universit\"at La Laguna, Teneriffa (Fachrichtung Astrophysik)  \\
  \\
  
    \textbf{Promotion:} &  Januar 2005 - Januar 2008 & Promotion an der Universit\"ats-Sternwarte G\"ottingen (seit Juni 2005 Institut f\"ur Astrophysik  G\"ottingen, IAG)   \\
 &  Januar 2005 - Januar 2008 & Stipendium des Max-Planck-Instituts f\"ur Sonnensystemforschung    \\   
  \end{tabular}
\end{table}

\end{document}